\documentclass[12pt]{iopart}
\usepackage{iopams}

\usepackage{tikz}
\usetikzlibrary{arrows}

\usepackage{algorithm}

 \newtheorem{theorem}{\bf Theorem}[section]
 \newtheorem{lemma}{\bf Lemma}[section]
 
 \newtheorem{corollary}{\bf Corollary}[section]

 \newtheorem{@definition}{\bf Definition}[section]

 \newtheorem{@example}{\bf Example}[section]

\newtheorem{@nonexample}{\bf (Non)Example}[section]

 \newtheorem{@remark}{\bf Remark}[section]

 \newtheorem{alemma}{\bf Lemma}

\begin{document}

\title{Geometric Decompositions of Bell Polytopes with Practical Applications}

\author{Peter Bierhorst}

\address{Information Technology Laboratory, National Institute of Standards and Technology, Boulder, CO 80305}
\begin{abstract}
In the well-studied $(2,2,2)$ Bell experiment consisting of two parties, two measurement settings per party, and two possible outcomes per setting, it is known that if the experiment obeys no-signaling constraints, then the set of admissible experimental probability distributions is fully characterized as the convex hull of 24 distributions: 8 Popescu-Rohrlich (PR) boxes and 16 local deterministic distributions. Furthermore, it turns out that in the $(2,2,2)$ case, any nonlocal nonsignaling distribution can always be uniquely expressed as a convex combination of exactly one PR box and (up to) eight local deterministic distributions. In this representation each PR box will always occur only with a fixed set of eight local deterministic distributions with which it is affiliated. In this paper, we derive multiple practical applications of this result: we demonstrate an analytical proof that the minimum detection efficiency for which nonlocality can be observed is $\eta>2/3$ even for theories constrained only by the no-signaling principle, and we develop new algorithms that speed the calculation of important statistical functions of Bell test data. Finally, we enumerate the vertices of the no-signaling polytope for the $(2, n, 2)$ ``chained Bell'' scenario and find that similar decomposition results are possible in this general case. Here, our results allow us to prove the optimality of a bound, derived in Barrett \emph{et al.} \cite{barrett:2006}, on the proportion of local theories in a local/nonlocal mixture that can be inferred from the experimental violation of a chained Bell inequality.  
\end{abstract}

\maketitle

\section{Introduction}\label{s:introduction}

Ever since Bell's landmark 1965 work \cite{BELL}, it has been known that quantum mechanics can predict nonlocal behavior in certain experimental scenarios. The most widely studied scenario is the $(2,2,2)$ setting, in which there are two spatially separated parties with measuring devices where each measuring device can be put into one of two configurations (``measurement settings'') and has two possible measurement outcomes. In such a scenario, local realist theories must obey constraints, generally referred to as Bell inequalities, on the probabilities of certain outcomes. These constraints, which include the Clauser-Horne-Shimony-Holt (CHSH) inequality \cite{CHSH} and the CH/Eberhard inequalities \cite{CH74,eberhard:1993}, can be violated by quantum mechanics. In recent years, it has been shown that experimental violations of these inequalities can be exploited for practical purposes, such as device-independent quantum key distribution \cite{acin:2007} and device-independent quantum random number expansion \cite{pironio:2010}.

In a $(2,2,2)$ experiment, the relevant quantum predictions form just one example from a class of probability distributions collectively known as the no-signaling polytope. This polytope contains all possible quantum distributions for the experiment, as well as some additional distributions that cannot be realized by a quantum mechanical system. The no-signaling polytope can be expressed as the convex hull of 24 special extremal distributions -- 16 ``local deterministic'' (LD) distributions and 8 nonsignaling nonlocal distributions \cite{barrett:2005,tsirelson:1993}. Any convex combination of these 24 extremal distributions represents a probability distribution over possible experimental outcomes that is consistent with the ``no-signaling'' principle, a constraint imposed by special relativity; conversely, any nonsignaling distribution can be expressed as a convex combination of these 24 extremal distributions.

Any convex combination consisting only of the 16 LD distributions will be compatible with local realism and cannot violate any Bell inequalities; such a distribution cannot be used for device-independent quantum information applications. It is only distributions that contain some weight on the 8 nonlocal distributions, commonly referred to as Popescu-Rohrlich (PR) boxes following the work \cite{PRBOX}, that can display nonlocality. In particular, the CHSH inequality \cite{CHSH},
\begin{equation}\label{e:CHSH}
E_{ab}(D_AD_B)+E_{ab'}(D_AD_B)+E_{a'b}(D_AD_B)-E_{a'b'}(D_AD_B)\le 2,
\end{equation}
is maximally violated by one of the eight PR boxes. There are eight distinct versions of the inequality \eref{e:CHSH} that can be obtained by symmetry transformations (i.e., relabeling of settings and/or outcomes), which we can refer to as ``CHSH symmetry inequalities.'' As noted in \cite{barrett:2005}, each of the eight PR boxes is associated with a unique CHSH symmetry inequality that it maximally violates. 

In this paper, we highlight a useful fact related to the above results: any nonlocal nonsignaling distribution can always be written as a convex combination of exactly one PR box and at most eight LD distributions, where the LD distributions all saturate the CHSH symmetry inequality that is violated by the specific PR box occurring in the convex combination. (There are exactly eight such saturating LD distributions for each CHSH symmetry inequality.) This result refines what can be said about the no-signaling polytope using only standard results of convex geometry, such as Carath\'eodory's theorem as can be found in (for example) \cite{lauritzen:2013}.

While the existence of the decomposition into a single PR box and 8 LD distributions is a straightforward consequence of previous work on the $(2,2,2)$ setting \cite{barrett:2005,fine:1982,pironio:2003,BBP}, it offers valuable new insights into the nature of $(2,2,2)$ experiments. The decomposition is surprisingly easy to construct given nothing more than a table of nonsignaling empirical frequencies that violate a Bell inequality. This allows for useful applications, such as a new proof that detection efficiency must exceed $\eta=2/3$ in order for a $(2,2,2)$ experiment to display nonlocality even for a theory constrained only by the no-signaling principle.  Another application concerns the statistics of a Bell experiment. Specifically, finding the closest local distribution (in statistical distance) to a given nonlocal empirical distribution is important for quantifying the statistical evidence against local realism (LR) \cite{gillvandam:2005,zhang:2011,zhang:2013}, and distributions that have higher CHSH violations can have more information-theoretic potential. For two important measures of statistical distance, the calculation of the closest local distribution can be meaningfully simplified by using this decomposition: for the total variation distance (the $L_1$ norm), the calculation is rendered trivial and the statistical distance to the closest local distribution is shown to always be a constant multiple of the CHSH violation; for the Kullback-Leibler divergence, the calculation still requires a computer but the search space can be significantly reduced.

A natural question is whether this result can be extended to more general situations. We provide an example by extending the result to the scenario of the $(2, n, 2)$ chained Bell inequalities that were introduced in \cite{pearle:1970} and studied extensively in \cite{CHAINED}, and later shown to be relevant for quantum protocols for key distribution \cite{BHK} and randomness amplification \cite{colbeck:2012} where security can be guaranteed by the no-signaling conditions alone. To address the chained Bell scenario, we first classify the nonsignaling polytope for this situation, showing that the extremal points consist of $2^{2n}$ local deterministic distributions and $2^{2n-1}$ nonlocal nonsignaling distributions that generalize the concept of the PR box. Once the polytope is classified, we then prove the general version of our earlier result: that any nonsignaling nonlocal distribution in the $(2, n, 2)$ chained Bell scenario can be expressed as a convex combination of exactly one generalized PR box and $4n$ LD distributions. As an application of this result, we prove the optimality of a bound derived in Barrett \emph{et al.} \cite{barrett:2006} on the proportion of local theories in a local/nonlocal mixture that can be inferred from the experimental violation of a chained Bell inequality. 

The structure of the paper is as follows: in Section \ref{s:theorems}, we introduce the notation and definitions that we will use and prove the main result for the $(2,2,2)$ situation. In Section \ref{s:applications}, we demonstrate the applications of this result to statistical problems, and in Section \ref{s:chained}, we study the chained Bell scenario, followed by concluding remarks in Section \ref{s:conclusion}.

\section{Decomposition Theorems for the $(2,2,2)$ Scenario}\label{s:theorems}

\subsection{Defining the $(2,2,2)$ No-Signaling Polytope}

In the $(2,2,2)$ setting, there are two spatially separated parties which we call Alice and Bob. Alice and Bob each have a measuring apparatus, and each apparatus has two configurations which we call measurement settings. We label the two measurement settings for Alice with the symbols $\{a,a'\}$ and the two measurement settings for Bob with the symbols $\{b,b'\}$. Whatever the choice of setting, each apparatus always returns one of two outcomes ``+'' or ``0'' in an experimental trial.

In a given experiment, there are then four possible setting configurations $\{ab,ab',a'b,a'b'\}$ and for each there will be an associated probability distribution over the four possible outcomes $\{+_A+_B,\textnormal{ +}_A0_B,\textnormal{ 0}_A+_B,\textnormal{ }0_A0_B\}$, where $X_AY_B$ denotes outcome $X$ for Alice and outcome $Y$ for Bob. Henceforth, we will omit the $A$ and $B$ subscripts when possible, with the understanding if a pair of outcomes is written without subscripts, then the first outcome is Alice's, and the second is Bob's. We can conveniently represent the four settings-conditional distributions as a table with one row for each conditional distribution, such as the following example:

\begin{table}[h]\caption{An example of a $(2,2,2)$ distribution matrix}\label{t:anexample}
\begin{center}
\begin{tabular}{ r|c|c|c|c| }
 \multicolumn{1}{r}{}
  &  \multicolumn{1}{c}{++}
 &  \multicolumn{1}{c}{+0}
 &  \multicolumn{1}{c}{0+} 
 &  \multicolumn{1}{c}{00}\\
 \cline{2-5}
 $ab$ & 1/4 & 1/4 & 1/4  & 1/4 \\
 \cline{2-5}
 $ab'$ & 1/2  & 0 & 0 & 1/2 \\
 \cline{2-5}
 $a'b$ & 1/4  & 1/4 & 1/4 & 1/4  \\
 \cline{2-5}
 $a'b'$ & 0 & 1/2 & 1/2 & 0 \\
 \cline{2-5}
 \end{tabular}
\end{center}
\end{table}

\noindent So for an experiment obeying the distribution in Table \ref{t:anexample}, we would expect to see, for instance, + for Alice and 0 for Bob with probability 1/4 if the setting is $ab$. Bell inequalities like \eref{e:CHSH} are constraints on these outcome probabilities that can either be obeyed or violated. In particular, the terms $E_{ij}(D_AD_B)$ appearing in \eref{e:CHSH} refer to the expectation, conditioned on the setting $ij$, of the product of random variables $D_A$ and $D_B$, where $D_{A(B)}$ equals $+1$ or $-1$ if Alice's (Bob's) outcome is  + or 0, respectively.

The complete collection of all 16 entries in Table \ref{t:anexample} is an example of what we will refer to as a \emph{distribution matrix}. Mathematically, a valid distribution matrix is just an element of $\mathbb R^{16}$ satisfying certain linear constraints. To enumerate these constraints, first note that each row of Table \ref{t:anexample} must be a valid probability distribution. Referring to the entries of the table using the notation $P(XY\mid AB)$ for ``the probability of outcome $XY$ given that the setting is $AB$,'' the associated constraints can be written as follows: 
{\small\begin{equation}\label{e:probspos}
P(XY\mid AB)\ge 0\quad\textnormal{for all}\quad X,Y\in\{0,+\}, A\in \{a,a'\}, B\in \{b,b'\},
\end{equation}
\begin{eqnarray}
1 &=& P(\textnormal{++}\mid ab)+ P(\textnormal{+0}\mid ab) + P(\textnormal{0+}\mid ab)+ P(\textnormal{00}\mid ab) \label{e:prob1}\\
1 &=& P(\textnormal{++}\mid ab')+ P(\textnormal{+0}\mid ab') + P(\textnormal{0+}\mid ab')+ P(\textnormal{00}\mid ab')  \label{e:prob2}\\
1 &=& P(\textnormal{++}\mid a'b)+ P(\textnormal{+0}\mid a'b) + P(\textnormal{0+}\mid a'b)+ P(\textnormal{00}\mid a'b)  \label{e:prob3}\\
1 &=& P(\textnormal{++}\mid a'b')+ P(\textnormal{+0}\mid a'b') + P(\textnormal{0+}\mid a'b')+ P(\textnormal{00}\mid a'b')  \label{e:prob4}.
\end{eqnarray}
}
Furthermore, the example in Table \ref{t:anexample} is just one instance of a distribution matrix obeying the \emph{no-signaling conditions}:
{\small\begin{eqnarray}
P(\textnormal{++}\mid ab)+ P(\textnormal{+0}\mid ab) &=& P(\textnormal{++}\mid ab')+ P(\textnormal{+0}\mid ab')\label{e:nosig1}\\
P(\textnormal{++}\mid a'b)+ P(\textnormal{+0}\mid a'b) &=& P(\textnormal{++}\mid a'b')+ P(\textnormal{+0}\mid a'b')\label{e:nosig2}\\
P(\textnormal{++}\mid ab)+ P(\textnormal{0+}\mid ab) &=& P(\textnormal{++}\mid a'b)+ P(\textnormal{0+}\mid a'b)\label{e:nosig3}\\
P(\textnormal{++}\mid ab')+ P(\textnormal{0+}\mid ab') &=& P(\textnormal{++}\mid a'b')+ P(\textnormal{0+}\mid a'b')\label{e:nosig4}\\
P(\textnormal{0+}\mid ab)+ P(\textnormal{00}\mid ab) &=& P(\textnormal{0+}\mid ab')+ P(\textnormal{00}\mid ab')\label{e:nosig5}\\
P(\textnormal{0+}\mid a'b)+ P(\textnormal{00}\mid a'b) &=& P(\textnormal{0+}\mid a'b')+ P(\textnormal{00}\mid a'b')\label{e:nosig6}\\
P(\textnormal{+0}\mid ab)+ P(\textnormal{00}\mid ab) &=& P(\textnormal{+0}\mid a'b)+ P(\textnormal{00}\mid a'b)\label{e:nosig7}\\
P(\textnormal{+0}\mid ab')+ P(\textnormal{00}\mid ab') &=& P(\textnormal{+0}\mid a'b')+ P(\textnormal{00}\mid a'b')\label{e:nosig8}.
\end{eqnarray}
}
The no-signaling conditions capture the notion that Alice and Bob should not be able to exploit the experiment to send signals to each other, in the sense that Alice's marginal outcome distribution should not depend on Bob's measurement choice and vice versa. To illustrate, note that if the equality \eref{e:nosig1} did not hold, then if Alice were to choose setting $a$, she could gain some information about whether Bob chose $b$ or $b'$ because her probability of seeing ``+'' changes according to Bob's choice. Equalities \eref{e:nosig1}-\eref{e:nosig8} encapsulate all of the different versions of this scenario for different settings and outcomes. 

We are interested in the collection of distribution matrices that satisfy all of the constraints \eref{e:probspos}-\eref{e:nosig8}. These constraints are all linear equalities and inequalities so the collection will form a polyhedron in $\mathbb R^{16}$. In particular, we can say that it is a \emph{convex polytope} -- the convex hull of a finite set -- because it is bounded. The set of distribution matrices satisfying \eref{e:probspos}-\eref{e:nosig8} is known as the no-signaling polytope, and any element of the set is expressible as a convex combination of elements of a special finite collection of distribution matrices known as the extremal points or vertices. There are 24 extremal points \cite{barrett:2005} which we list in \ref{s:prsdets}. As noted earlier, 16 of these extremal points are local (i.e., satisfying all Bell inequalities) and 8 are nonlocal PR boxes.

We can observe a few things about the polytope by studying the constraints that define it. It is straightforward to check that equations \eref{e:nosig5}-\eref{e:nosig8} can be derived from \eref{e:prob1}-\eref{e:prob4} along with \eref{e:nosig1}-\eref{e:nosig4}. Removing \eref{e:nosig5}-\eref{e:nosig8} from the set of equalities, we are left with equalities  \eref{e:prob1}-\eref{e:prob4} and \eref{e:nosig1}-\eref{e:nosig4}, which are linearly independent. Thus these equalities reduce the dimension of the no-signaling polytope to 8 (down from the 16 dimensions of the ambient space $\mathbb R^{16}$). A theorem of convex geometry, due to Carath\'eodory, tells us that for a given element in the convex hull of a finite set of points $\mathcal S$, it is always possible to express it as a convex combination of an affinely independent subset of $\mathcal S$ \cite{lauritzen:2013}. For the $(2,2,2)$ no-signaling polytope, this indicates that any distribution matrix can always be expressed as a convex combination of no more than 9 extremal points. We can refine this result for our particular polytope by analyzing various properties belonging to this collection of 9 extremal points.

\subsection{Decomposing the Polytope}

We will later generalize these theorems to the $(2,n,2)$ chained Bell setting, but we include the following proofs of the special case for ease of understanding.

\begin{theorem}\label{t:onepr}
In the $(2,2,2)$ Bell scenario, any nonsignaling distribution matrix that is an equal mixture of two distinct PR boxes is expressible as an equal mixture of four local deterministic distributions. 
\end{theorem}
\emph{Proof.}
We prove this by inspection, referring to the table in \ref{s:prsdets}. We start by fixing the PR box labeled ``1.'' There are seven different possibilities for PR box 1 to be mixed with another PR box in equal amounts. For each of these seven pairs, we can list a group of four LD distributions that reproduce the equal mixture distribution. Since an arbitrary pair of PR boxes can be transformed into a pair containing PR box 1 by re-labelling the outcomes and/or settings, this is sufficient to show that the theorem is true for all possible distinct PR box pairs. 

For visualization purposes, we examine how this works for the equal mixture of PR box 1 and PR box 6, which results in the distribution matrix that was written earlier in Table \ref{t:anexample}. An equal mixture of LD distributions 9, 12, 14, and 15 will induce the same distribution. One way to see this is by proceeding in steps, first considering the equal mixture of LD distributions 9 and 14 (below left -- to reduce clutter, we leave blank the cells that contain zero), and then considering an equal mixture of LD distributions 12 and 15 (below right):

\medskip

\begin{center}
\begin{tabular}{ r|c|c|c|c| }
 \multicolumn{1}{r}{}
  &  \multicolumn{1}{c}{++}
 &  \multicolumn{1}{c}{+0}
 &  \multicolumn{1}{c}{0+} 
 &  \multicolumn{1}{c}{00}\\
 \cline{2-5}
 $ab$ & 1/2 & 1/2 & & \\
 \cline{2-5}
 $ab'$ & 1  &  &  & \\
 \cline{2-5}
 $a'b$ & & & 1/2 & 1/2  \\
 \cline{2-5}
 $a'b'$ & &  & 1 &  \\
 \cline{2-5}
 \end{tabular}
 \hspace{1cm}
 \begin{tabular}{ r|c|c|c|c| }
 \multicolumn{1}{r}{}
  &  \multicolumn{1}{c}{++}
 &  \multicolumn{1}{c}{+0}
 &  \multicolumn{1}{c}{0+} 
 &  \multicolumn{1}{c}{00}\\
 \cline{2-5}
 $ab$  & & & 1/2 & 1/2 \\
 \cline{2-5}
 $ab'$  &  &  & & 1  \\
 \cline{2-5}
 $a'b$ & 1/2 & 1/2 & &   \\
 \cline{2-5}
 $a'b'$ &  & 1 &  &  \\
 \cline{2-5}
 \end{tabular}
\end{center}

\medskip

\noindent It is easy to see that if we take an equal mixture of the above two tables -- which is an even mixture over all the four LD distributions 9, 12, 14, and 15 -- we get the same distribution matrix as the one in Table \ref{t:anexample}. The table below lists how to achieve this for each pairing of PR box 1 with another PR box, and thus completes the proof.
  
\begin{center}
\begin{tabular}{ c|c|c|c|c|c|c|c|c| }
PR box mixture & 1,2  & 1,3 & 1,4 & 1,5 & 1,6 & 1,7 & 1,8   \\
\hline
Det. collection     & 1,2,3,4 & 1,4,9,12  & 5,8,14,15 & 1,4,5,8 & 9,12,14,15 & 1,4,14,15  & 5,8,9,12 \\
 \end{tabular}
 \end{center}
 
$\hfill\Box$

\begin{corollary}\label{c:1}
In the $(2,2,2)$ scenario, any nonsignalling distribution matrix can always be expressed as a convex combination of extremal points that contains at most one PR box.
\end{corollary}
\emph{Proof.}
Any nonsignaling distribution matrix has a representation as a convex combination of the extremal points of the polytope. We can express this algebraically as 
\begin{equation}\label{e:coro1}
\sum_{i=1}^{8}p_iPR_i + \sum_{i=1}^{16}q_iD_i,
\end{equation}
where and $PR_i$ and $D_i$ refer to the distribution matrices given in Tables \ref{t:prboxes} and \ref{t:localdets} of \ref{s:prsdets} and the $p_i$ and $q_i$ are nonnegative numbers satisfying $\sum_{i=1}^{8}p_i+ \sum_{i=1}^{16}q_i= 1$. We can arrange the $p_iPR_i$ terms in decreasing order relative to the magnitude of $p_j$. Then if $p_{(8)}$ and $p_{(7)}$ are the respectively the smallest and second smallest $p_i$s,
\begin{eqnarray*}
p_{(7)}PR_{(7)}+p_{(8)}PR_{(8)}&=& (p_{(7)}-p_{(8)})PR_{(7)}+p_{(8)}(PR_{(7)}+PR_{(8)})\\
&=& (p_{(7)}-p_{(8)})PR_{(7)}+p_{(8)}(D_w+D_x+D_y+D_z)/2,
\end{eqnarray*}
where we have replaced $PR_{(7)}+PR_{(8)}$ with an equal mixture of deterministic strategies by using Theorem \ref{t:onepr}. We can then shift the $D_i$'s appearing above to the second sum of \eref{e:coro1}, and we have removed one PR box from the expression. This process can then be applied to the $p_{(6)}PR_{(6)}$ term and $(p_{(7)}-p_{(8)})PR_{(7)}$ to remove $PR_{(7)}$ from the expression, and similarly repeated until there is only one remaining PR box term. $\hfill\Box$

\begin{theorem}\label{t:8dets}
In the $(2,2,2)$ scenario, any convex combination of a fixed PR box and 16 local deterministic distributions can be re-expressed as either \emph{a)} a convex combination consisting only of local deterministic distributions, or \emph{b)} a convex combination of the same PR box and the eight local deterministic distributions that saturate the symmetry of the CHSH inequality maximally violated by the PR box.
\end{theorem}
\emph{Proof.}
We will explicitly prove the claim of the theorem for the particular case of PR box 1. Then a symmetry argument implies that the theorem holds for all 8 PR boxes, as the other PR boxes and their corresponding CHSH symmetries are equivalent to PR box 1 paired with inequality \eref{e:CHSH}, up to relabeling of settings and/or outcomes.

By assumption, we are given a distribution that can be written in the form $pPR_1 + \sum_{i=1}^{16}q_iD_i$ where $p,q_i\ge 0$ and $p + \sum_{i=1}^{16}q_i=1$. Now consider the subset of $D_i$ distributions listed in \ref{s:prsdets} that saturate the CHSH inequality \eref{e:CHSH}. By inspection, this is seen to be the eight distributions whose index falls in the set $\mathcal S = \{1, 4, 5, 8, 9, 12, 14, 15\}$; these are the LD distributions that ``go with'' PR box 1. The complement of $\mathcal S$ is $\mathcal S^C=\{2,3,6,7,10,11,13,16\}$, so we can re-write our distribution as
\begin{equation*}
pPR_1 + \sum_{i\in \mathcal S}q_iD_i + \sum_{i\in \mathcal S^C}q_iD_i.
\end{equation*}
Our goal is to show that the distribution above can be induced by a similar convex combination that either a) does not include the $pPR_1$ term, or b) does not include any of the terms in the $\mathcal S^C$ sum. To do this, we will show that for any deterministic distribution $D_i$ whose index $i$ is in $\mathcal S^C$, the linear combination $x\times D_s + 2x\times PR_1$ with $x> 0$ can be re-expressed equivalently by $x$ times a sum of three LD distributions whose indices are in the set $\mathcal S$. Then for each $i$ in $\mathcal S^C$, we can methodically replace $q_iD_i+2q_iPR_1$ with a collection of LD distributions in the set $\mathcal S$ until we either a) ``run out'' of $PR_1$ because $p$ was not big enough, or b) cast out all of the $\mathcal S^C$-indexed LD distributions and end up with a convex combination of $PR_1$ and LD distributions whose index is in $\mathcal S$. These two contingencies correspond to cases (a) and (b) in the statement of the theorem.

For visualization purposes, we again demonstrate one particular case. Suppose we have a linear combination consisting of $2x$ of $PR_1$ and $x$ of $D_{16}$, with $x>0$. Then the induced pseudo-distribution is given by the following table:

\medskip

\begin{center}
\begin{tabular}{ r|c|c|c|c| }
 \multicolumn{1}{r}{}
  &  \multicolumn{1}{c}{++}
 &  \multicolumn{1}{c}{+0}
 &  \multicolumn{1}{c}{0+} 
 &  \multicolumn{1}{c}{00}\\
  \cline{2-5}
   $ab$ & $x$ &  &  &  $2x$ \\
 \cline{2-5}
 $ab'$ & $x$ &  & $x$  & $x$\\
 \cline{2-5}
 $a'b$ & $x$ & $x$ &  & $x$ \\
 \cline{2-5}
 $a'b'$ & $x$ & $x$ & $x$ &  \\
 \cline{2-5}
 \end{tabular}
\end{center}

\medskip

\noindent By inspection, we can see that the above table is equivalent to the linear combination of LD distributions $x\times D_1 + x\times D_8 + x\times D_{12}$. Similarly, we can table off all the possibilities as follows:

\begin{center}
\begin{tabular}{ l|c|c|c|c|c|c|c|c|c| }
$PR_1$ mixed with: & 2 & 3 & 6 & 7 & 10 & 11 & 13 & 16   \\
\hline
Alt. Det. collection:     & 5,12,14 & 8,9,15 & 1,12,14 & 4,9,15 & 4,5,14 & 1,8,15 & 4,5,9 & 1,8,12 \\
 \end{tabular}
 \end{center}
 
 \medskip

\noindent This completes the proof. $\hfill \Box$

\medskip

Together, Corollary \ref{c:1} and Theorem \ref{t:8dets} tell us that any nonsignaling distribution matrix is either a) local (i.e., can be expressed as a convex combination of local deterministic distributions and thus violates no Bell inequalities), or b) can be expressed as a convex combination of exactly one PR box and (up to) 8 local deterministic distributions that saturate the specific CHSH symmetry violated by that PR box. The characterization (b), which we will call the 1 PR + 8 LD representation, offers a deeper understanding of nonlocal $(2,2,2)$ distributions, as illustrated in the following sections.

\subsection{Relationships between $(2,2,2)$ Bell inequalities}

Consider the following distribution matrix, which is a nonsignalling and approximates the empirical data reported in Table S-II of the supplementary material of \cite{shalm:2015}, a recent loophole-free Bell test:

 \begin{center}
 \begin{tabular}{ r|c|c|c|c| }
 \multicolumn{1}{r}{}
  &  \multicolumn{1}{c}{$++$}
 &  \multicolumn{1}{c}{$+$0}
 &  \multicolumn{1}{c}{0$+$} 
   &  \multicolumn{1}{c}{$00$}
\\
  \cline{2-5}
   $ab$&0.0001422 &0.0000743& 0.0000699& 0.9997136\\
 \cline{2-5}
   $ab'$&0.0001530& 0.0000635& 0.0005249& 0.9992586\\
 \cline{2-5}
   $a'b$&0.0001476&0.0004795& 0.0000644& 0.9993084\\
 \cline{2-5}
   $a'b'$&0.0000024&0.0006247&0.0006755&0.9986974\\
 \cline{2-5}
 \end{tabular}
 \end{center}

\noindent This distribution weakly violates the CHSH inequality \eref{e:CHSH}, and therefore it has a 1 PR + 8 LD representation. But at first glance, it is not at all obvious how to find the coefficients of this representation.  However, this task turns out to be rather easy. To see how, note that the 1 PR + 8 LD representation for a nonsignalling distribution violating the CHSH inequality is as follows:
\begin{equation}\label{e:convexexp}
\hspace{-2cm}p_{PR}PR_1 + p_1D_1+ p_4D_4+ p_5D_5+ p_8D_8+ p_9D_9+ p_{12}D_{12}+ p_{14}D_{14}+ p_{15}D_{15},
\end{equation}
where the nine $p$ coefficients are nonnegative and sum to 1. Referring to \ref{s:prsdets}, we can fill in a table for the distribution matrix based on the expression above by filling in a $p_i$ everywhere the corresponding $D_i$ has a ``1,'' putting a $\frac{1}{2}p_{PR}$ everywhere $PR_1$ has a ``1/2,'' and adding these entries together to get the following:

\begin{table}[h]\caption{General Expression for a Distribution Matrix Violating CHSH}\label{t:genexp}
\centering
{\footnotesize\begin{tabular}{ r|c|c|c|c| }
 \multicolumn{1}{r}{}
  &  \multicolumn{1}{c}{++}
 &  \multicolumn{1}{c}{+0}
 &  \multicolumn{1}{c}{0+} 
 &  \multicolumn{1}{c}{00}\\
  \cline{2-5}
$ab$ & $p_1+p_5+p_9+\frac{1}{2}p_{PR}$ & $p_{14}$ & $p_{15}$ & $p_4+p_8+p_{12}+\frac{1}{2}p_{PR}$\\
  \cline{2-5}
$ab'$ & $p_1+p_9+p_{14}+\frac{1}{2}p_{PR}$ & $p_{5}$ & $p_{8}$ & $p_4+p_{12}+p_{15}+\frac{1}{2}p_{PR}$\\
  \cline{2-5}
$a'b$ & $p_1+p_5+p_{15}+\frac{1}{2}p_{PR}$ & $p_{12}$ & $p_{9}$ & $p_4+p_{8}+p_{14}+\frac{1}{2}p_{PR}$\\
  \cline{2-5}
$a'b'$ & $p_{1}$ & $p_5+p_{12}+p_{15}+\frac{1}{2}p_{PR}$ & $p_8+p_{9}+p_{14}+\frac{1}{2}p_{PR}$ & $p_{4}$\\
  \cline{2-5}
\end{tabular}}
\end{table}

\noindent We notice immediately that there are eight cells in the table that are uniquely determined by the coefficient of a specific deterministic distribution in expression \eref{e:convexexp}. One can thus work backwards: presented with a nonsignaling distribution matrix that violates the CHSH inequality like the one at the beginning of the section, the construction of the representation \eref{e:convexexp} requires only one calculation to find $p_{PR}$ and the remaining $p_i$ coefficients can be copied directly from the table (so for instance $p_{14}=.0000743$ in the 1 PR + 8 LD representation of the distribution matrix given earlier.) The same process can also be applied to nonsignaling distributions violating other symmetries of the CHSH inequality, with appropriate changes to the collection of local distributions appearing in \eref{e:convexexp}.

Expression \eref{e:convexexp} and its counterpart, Table \ref{t:genexp}, offer a useful new way of visualizing Bell inequalities. For instance, consider the following Eberhard-type inequality \cite{eberhard:1993,giustina:2013,bierhorst:2015}, recently tested in loophole-free Bell experiments \cite{shalm:2015,giustina:2015}:
\begin{equation}\label{e:eberhard}
P(\textnormal{++}\mid ab) - P(\textnormal{+0}\mid ab') - P(\textnormal{0+}\mid a'b) - P(\textnormal{++}\mid a'b') \le 0.
\end{equation}
The left side of this Eberhard inequality is a linear combination of four of the entries in Table \ref{t:genexp}.  Furthermore, the value of this linear combination is $\frac{1}{2}p_{PR}$ -- so estimators of the Eberhard quantity are in fact estimators of the amount of PR box (up to a scale factor of $\frac{1}{2}$), which illustrates how positive values indicate nonlocality.

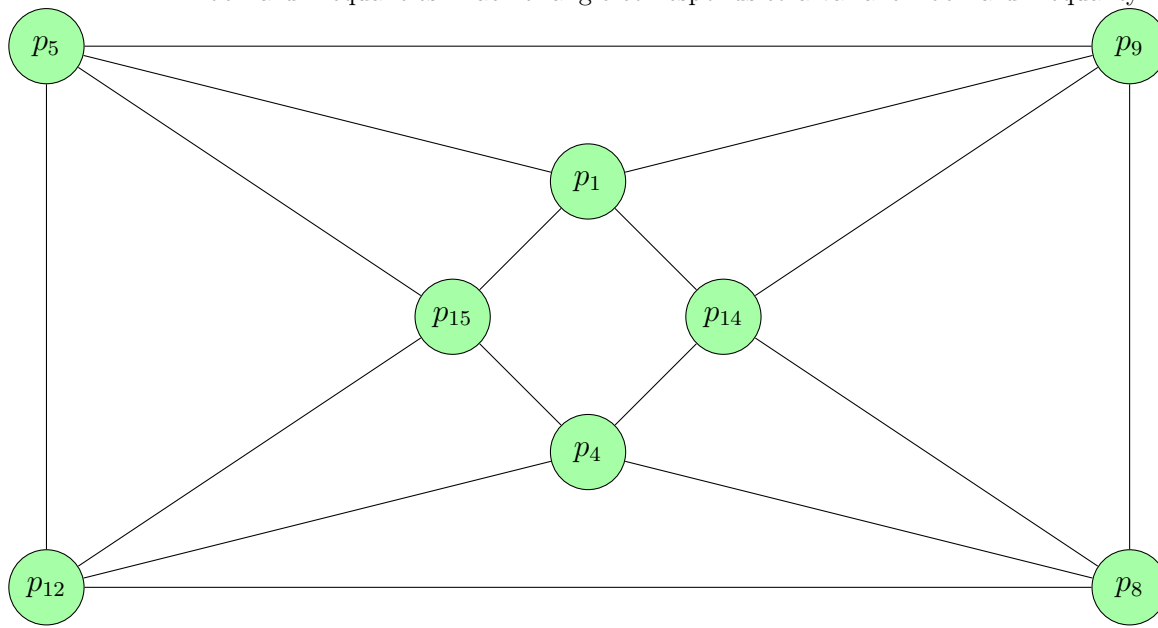
\begin{figure}[h]\centering\caption{Relationship between weights of deterministic distributions and variant-Eberhard Inequalities. Each triangle corresponds to a variant-Eberhard inequality.}\label{f:scheme}
\begin{tikzpicture}[scale=.9]

 \node [style=circle,draw,minimum size=1cm,fill=green!35,]  (1) at (8,-2)  {$p_1$};
 \node [style=circle,draw,minimum size=1cm,fill=green!35,] (15) at (6,-4) {$p_{15}$};
 \node [style=circle,draw,minimum size=1cm,fill=green!35,](4) at (8,-6) {$p_4$};
 \node [style=circle,draw,minimum size=1cm,fill=green!35,](14) at (10,-4) {$p_{14}$};
 \node [style=circle,draw,minimum size=1cm,fill=green!35,] (5) at (0,0)  {$p_5$};
 \node [style=circle,draw,minimum size=1cm,fill=green!35,] (12) at (0,-8)  {$p_{12}$};
 \node [style=circle,draw,minimum size=1cm,fill=green!35,](9) at (16,0)  {$p_9$};
 \node [style=circle,draw,minimum size=1cm,fill=green!35,] (8) at (16,-8)  {$p_{8}$};
  
  \draw[-] (1)--(14);
  \draw[-] (1)--(9);
  \draw[-] (14)--(9);
  \draw[-] (5)--(9);
  \draw[-] (8)--(9);
  \draw[-] (14)--(8);
  \draw[-] (4)--(8);
  \draw[-] (12)--(8);
  \draw[-] (12)--(4);
  \draw[-] (12)--(15);
  \draw[-] (12)--(5);  
  \draw[-] (1)--(15);
  \draw[-] (4)--(15);
  \draw[-] (4)--(14);
  \draw[-] (1)--(5);
  \draw[-] (5)--(15);
      
\end{tikzpicture} 
\end{figure}

Expression \eref{e:eberhard} singles out $p_{PR}$ by subtracting $p_1$, $p_5$, and $p_9$ from $P(\textnormal{++}\mid ab)$. We immediately see from Table \ref{t:genexp} that there are eight different ways to do this, one for each of the table entries that consists of $\frac{1}{2}p_{PR}$ plus three other $p_i$. Thus the Eberhard inequality is just one of a class of eight related inequalities associated with $PR_1$ and the CHSH inequality. Figure \ref{f:scheme} is a diagram that depicts these eight related inequalities. Each CHSH symmetry inequality will have its own version of Figure \ref{f:scheme}, and as there are 8 CHSH symmetries, this yields 64 total variant-Eberhard inequalities, but a given nonsignaling distribution can violate at most one CHSH symmetry.

For each triangle in Figure \ref{f:scheme}, one can generate a variant-Eberhard inequality by adding the unique cell of Table \ref{t:genexp} containing all three of the $p_i$ in the triangle's vertices, and subtracting the three cells of Table \ref{t:genexp} where these $p_i$ appear individually. While these variant inequalities can also be generated by adding various linear combinations of the no-signaling conditions \eref{e:nosig1}-\eref{e:nosig8} to inequality \eref{e:eberhard}, this does not reveal the way in which the variant-Eberhard inequalities all estimate the same parameter $p_{PR}$. Note also that if one adds all the variant-Eberhard inequalities together, one obtains the following inequality:
\begin{eqnarray*}
P(\textnormal{++}\mid ab)-3P(\textnormal{+0}\mid ab)-3P(\textnormal{0+}\mid ab)+P(\textnormal{00}\mid ab)\\
+P(\textnormal{++}\mid ab')-3P(\textnormal{+0}\mid ab')-3P(\textnormal{0+}\mid ab')+P(\textnormal{00}\mid ab')\\
+P(\textnormal{++}\mid a'b)-3P(\textnormal{+0}\mid a'b)-3P(\textnormal{0+}\mid a'b)+P(\textnormal{00}\mid a'b)\\
-3P(\textnormal{++}\mid a'b')+P(\textnormal{+0}\mid a'b')+P(\textnormal{0+}\mid a'b')-3P(\textnormal{00}\mid a'b')\le 0.
\end{eqnarray*}
The inequality above is equivalent to the CHSH inequality \eref{e:CHSH}, which can be seen by adding 2 to both sides and using the fact that $\sum_{x,y}P(xy\mid AB)=1$ for each of the four fixed choices $A,B\in\{ab,ab',a'b,a'b'\}$. Thus the amount of violation of the CHSH inequality is a constant multiple of the amount of PR box present in the 1 PR + 8 LD representation \eref{e:convexexp} of a nonlocal nonsignaling distribution matrix. Indeed, one can add variant-Eberhard inequalities together with different positive weightings to obtain a variety of Bell inequalities of the form $\sum_{i\in I} c_iP(xy_i\mid ab_i)\le0$, where $I$ indexes a subset of joint settings and outcomes.

This wide latitude in generating Bell inequalities can yield useful results. For instance, the left side of a variant-Eberhard inequality will take the value $p_{PR}/2$ when applied to a nonsignaling distribution that violates the CHSH inequality, and the same will be true for any linear combination of left sides of variant-Eberhard inequalities so long as the sum of the coefficients in the linear combination is 1. The resulting expression $\sum_{i\in I} c_iP(xy_i\mid ab_i)$ can thus induce an unbiased linear estimator of $p_{PR}/2$ by replacing the conditional probabilities with outcome tallies and multiplying by the setting probabilities: $\sum_{i\in I} c_iN(xy_i\mid ab_i)P(ab_i)$. So in an experimental situation where the setting probabilities are known, a minimum variance linear unbiased estimator of the ``amount of PR box'' can be generated by positing a best guess for the expected true distribution and then optimizing over linear combinations of variant-Eberhard inequalities using standard techniques. (Note this requires an assumption of independent, identically distributed experimental trials to be statistically valid.) This could yield smaller error bars for estimates of CHSH parameters in Bell experiments, especially for lossy experiments whose distributions lack in symmetry or otherwise deviate substantially from optimal quantum statistics.

\section{Applications}\label{s:applications}

The 1 PR + 8 LD representation of nonlocal nonsignaling distributions has useful applications, which we explore in this section.

\subsection{A proof that detection efficiency $\eta$ must exceed 2/3 in order to witness nonlocality}
 
Any real-world Bell test experiment will contain imperfections. In particular, an experimenter cannot always detect all of the particles being generated; some invariably evade detection. This can lead to the \emph{detection loophole}: if enough particles are not detected, the observed empirical probabilities can deviate from the ideal quantum probabilities to such an extent that the observed probabilities do not violate any Bell inequalities.

The 1 PR + 8 LD representation can help us understand how this issue causes problems in a ``one-channel'' experiment \cite{shalm:2015,giustina:2013,giustina:2015,christensen:2013}, which we briefly describe. In a single trial of a one-channel experiment, a particle flies through an obstacle which probabilistically either deflects the particle in an unmonitored direction, or allows the particle to pass where it then is (hopefully) registered by an awaiting particle detector. If the particle is deflected, the detector will not see anything and we assign this the outcome ``0.'' If the particle passes through and is detected, the detector registers a click, which we assign the outcome ``+.'' But if the imperfect detector fails to see a particle that is present, this will result in the outcome ``0.'' The quantum efficiency $\eta$ is the proportion of present particles that are actually detected by the detector.

\begin{figure}[h]\centering\caption{Effects of dropped signal on a PR box, with various probabilities}\label{f:PReta}

\begin{tikzpicture}[node distance=8cm, auto]
\node (1) {\begin{tabular}{ |c|c|c|c| }
\multicolumn{1}{c}{++}
 &  \multicolumn{1}{c}{+0}
 &  \multicolumn{1}{c}{0+} 
 &  \multicolumn{1}{c}{00}\\
\hline
$1/2$ &  &  & $1/2$ \\
\hline
$1/2$ &  &  & $1/2$ \\
\hline
$1/2$ &  &  & $1/2$ \\
\hline
&$1/2$ &  $1/2$&\\
\hline
\end{tabular}};

\node (2)[right of=1] {\begin{tabular}{ |p{.5cm}|c|p{.5cm}|c| }
\multicolumn{1}{c}{++}
 &  \multicolumn{1}{c}{+0}
 &  \multicolumn{1}{c}{0+} 
 &  \multicolumn{1}{c}{00}\\
\hline
 & $1/2$ &  & $1/2$ \\
\hline
  & $1/2$ & & $1/2$ \\
\hline
  &$1/2$  & & $1/2$ \\
\hline
 & $1/2$  & & $1/2$ \\
\hline
\end{tabular}};  

\node (3)[below of=1] {\begin{tabular}{ |p{.5cm}|p{.5cm}|c|c| }
\multicolumn{1}{c}{++}
 &  \multicolumn{1}{c}{+0}
 &  \multicolumn{1}{c}{0+} 
 &  \multicolumn{1}{c}{00}\\
\hline
& & $1/2$  & $1/2$ \\
\hline
&  & $1/2$  & $1/2$ \\
\hline
&  &$1/2$   & $1/2$ \\
\hline
& & $1/2$   & $1/2$ \\
\hline
\end{tabular}};  

\node (4)[below of=2] {\begin{tabular}{ |p{.5cm}|p{.5cm}|p{.5cm}|c| }
\multicolumn{1}{c}{++}
 &  \multicolumn{1}{c}{+0}
 &  \multicolumn{1}{c}{0+} 
 &  \multicolumn{1}{c}{00}\\
\hline
&& & $1$ \\
\hline
&& & $1$ \\
\hline
&& & $1$ \\
\hline
&& & $1$\\
\hline
\end{tabular}};  
  
  \path[->]  (1) edge node [above]{$\eta(1-\eta)$}(2)
  (1) edge node [left]{$\eta(1-\eta)$}(3)
  (1) edge node [above right]{$(1-\eta)^2$}(4)
  ;

\end{tikzpicture}
\end{figure}
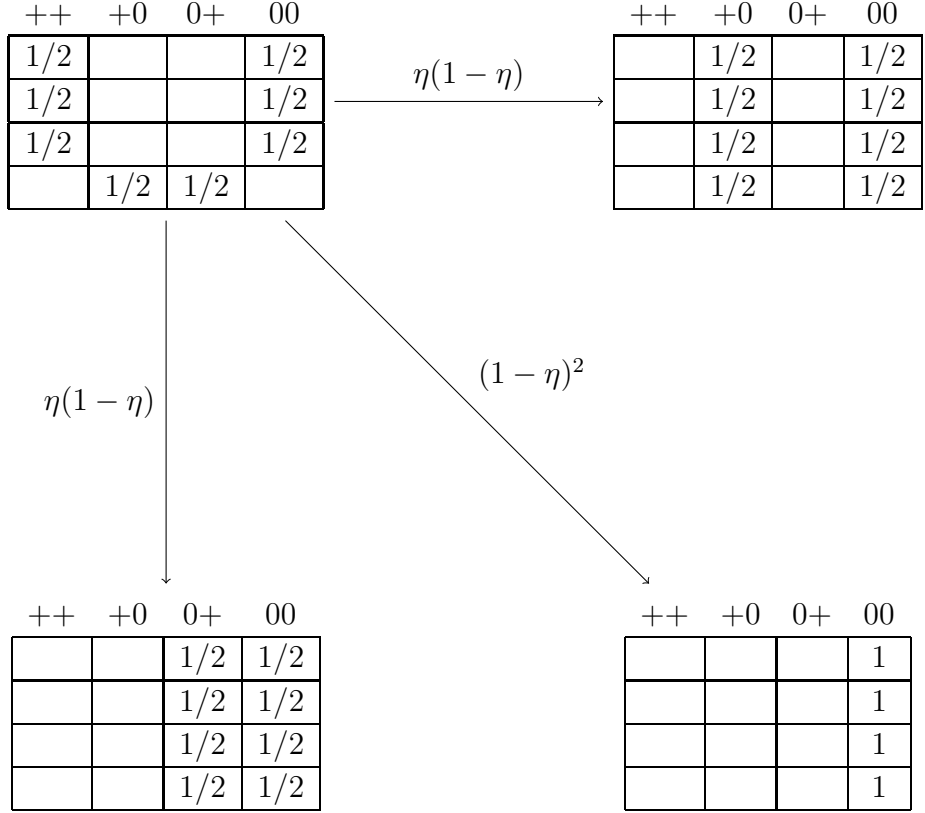

Thus the effect of a missed detection is to convert a ``+'' count to a ``0'' count with probability $1-\eta$. If the probability of a missed count is symmetrically $1-\eta$ for both Alice and Bob, the effect of this transition on a PR box is shown in Figure \ref{f:PReta}: the PR box remains a PR box with probability $\eta^2$ (all particles detected), becomes an equal mixture of $D_2$ and $D_4$ if Bob's detector fails (probability $(1-\eta)\eta$), becomes an equal mixture of $D_3$ and $D_4$ if Alice's detector fails (probability $(1-\eta)\eta$), and becomes $D_4$ if both detectors fail (probability $(1-\eta)^2$). 

With this understanding, we can now prove that a detection efficiency strictly exceeding $2/3$ is necessary to witness nonlocality. Suppose that we start with an ideal experimental system described by a nonsignaling nonlocal distribution. The distribution matrix can then be expressed in the form \eref{e:convexexp}. If $\eta$ is less then 1, the \emph{observed} distribution will be a transformation of \eref{e:convexexp}. Focusing on the term $p_{PR}PR_1$, the transformation replaces $PR_1$ with $\eta^2PR_1 + \eta(1-\eta)(D_2+D_4)/2 + \eta(1-\eta)(D_3+D_4)/2 + (1-\eta)^2(D_4)$, as illustrated by Figure \ref{f:PReta}. Note that we have picked up some weight on LD distributions $D_2$ and $D_3$, which do not saturate the CHSH inequality. As the $\eta$ transformation takes the local distributions in \eref{e:convexexp} to other local distributions (this can be easily checked), this means that the proportion of $PR_1$ post-transformation is at most $\eta^2$. However, from the proof of Theorem \ref{t:8dets}, we recall that for $i=2$ or $i=3$, $xD_i + 2x PR_1$ can be replaced by a mixture of LD distributions. Since under the transformation, the PR box itself generates some $D_2$ and $D_3$, this will decrease the subsequent amount of PR box below $\eta^2$. When $\eta$ is small enough so that $\eta^2=2\eta(1-\eta)$, the transformation will replace the entire coefficient of $PR_1$ with coefficients of LD distributions; this occurs when $\eta = 2/3$. Thus we can analytically prove the bound $\eta>2/3$ for a general nonsignaling distribution. This complements a numerical proof of the same fact in \cite{wilms:2008}, as well as related results in \cite{eberhard:1993,massar:2003,branciard:2011} where nondetection events are formulated as a third outcome.

We make one last point about the $\eta$ transformation. The transformation has an unequal effect on different $D_i$ distributions appearing in \eref{e:convexexp}: some of these ($D_4$, $D_8$, $D_{12}$) are always taken to other $D_i$ that saturate the CHSH inequality, but others can be mapped to ``bad'' LD distributions -- for instance, $D_1$ is mapped to $D_2$ with probability $\eta(1-\eta)$. Thus a distribution matrix with a fixed $p_{PR}$ is ``more hurt'' by the transformation if it has more weight on LD distributions such as $D_1$. So the search for partially entangled states that are noise tolerant, as introduced by Eberhard \cite{eberhard:1993}, is essentially the search for states whose representation \eref{e:convexexp} has less weight on LD distributions such as $D_1$ and more weight on the LD distributions $D_4$, $D_8$, and $D_{12}$.

\subsection{Calculating minimum-statistical-distance distributions}

In $(2,2,2)$ experiments and their applications, a fundamental concern is how to distinguish a given quantum distribution matrix from alternative local distributions. Thus we consider the problem of finding the local distribution that comes closest to approximating a given nonsignaling nonlocal distribution. Depending on which measure of statistical distance is being used, this can be a nontrivial calculation. Here we show that the 1 PR + 8 LD representation meaningfully simplifies this calculation for two important measures of statistical distance: the total variation distance and the Kullback-Leibler divergence.

The following lemma will be useful for this purpose. It states that if we draw a straight line connecting a local distribution $S$ to a CHSH-violating nonsignaling distribution $Q$, then at some point the straight line intersects the set of convex combinations of the eight local deterministic CHSH-saturating distributions. This result is useful because we expect that measures of statistical distance should decrease as we travel along lines that connect to the target distribution. So for any local distribution matrix that is not already a convex combination of the saturating distributions, there is always a ``better'' distribution matrix that is such a convex combination. 

\begin{lemma}\label{l:straightlinelemma}
Let $Q$ be a nonsignaling distribution matrix that violates the CHSH inequality \eref{e:CHSH}, and let $S$ be a local distribution. Then there is a $\lambda\in [0,1)$ for which the mixture distribution $\lambda Q + (1-\lambda)S$ is a convex combination of the eight local deterministic distributions that saturate the CHSH inequality.
\end{lemma}
\emph{Proof.}
$Q$ can be written in the form \eref{e:convexexp} so $Q=q_{PR}PR_1+\sum_{i\in \mathcal S}q_iD_i$ where the set $\mathcal S = \{1, 4, 5, 8, 9, 12, 14, 15\}$ enumerates the local deterministic distributions that saturate the CHSH inequality. $S$ can similarly be written in the form $\sum_{i=\in \mathcal S}s_iD_i+\sum_{i\in \mathcal S^C}s_iD_i$. Let $n=\sum_{i\in\mathcal S^C}s_i$ and let $\lambda = 2n/(2n+q_{PR})$. Then
\begin{eqnarray*}
\fl \lambda Q+ (1-\lambda) S &=& \left(\sum_{i \in \mathcal S} [\lambda q_i +(1-\lambda) s_i]D_i\right) + \lambda q_{PR} PR_1 + (1-\lambda)\sum_{i\in \mathcal S^C} s_iD_i\\
&=& \left(\sum_{i \in \mathcal S} [\lambda q_i +(1-\lambda) s_i]D_i\right) + \frac{q_{PR}}{2n+ q_{PR}} \sum_{i \in \mathcal S^C} \left[ 2s_i PR_1 + s_iD_i\right],
\end{eqnarray*}
and by Theorem \ref{t:8dets} each bracketed term in the second sum above is equivalent to a linear combination of $D_i$ distributions whose indices are in $\mathcal S$, so the above expression can be re-written as a convex combination of the eight local deterministic distributions satisfying the CHSH inequality. $\hfill\Box$


\medskip

If we start already on the CHSH-saturating face, then $\lambda$ is zero. Furthermore, it is clear that if we move any closer towards $Q$ past the CHSH-saturating face, we leave the set of local distributions. 

\medskip

\noindent \emph{Total Variation Distance}. For two distribution matrices $Q$ and $S$, we define the total variation distance between them as follows:
\begin{equation}\label{e:totvardist}
\delta(Q,S):=\frac{1}{2}\sum_{X,Y,A,B} \left|Q(XY\mid AB) - S(XY\mid AB)\right|,
\end{equation}
where $Q(XY\mid AB)$ is the probability that the distribution matrix $Q$ assigns to outcome $XY$ conditioned on setting $AB$, and similarly for $S(XY\mid AB)$. Note that this definition involves conditional probabilities, and thus is different from the usual definition of total variation distance. According to the usual definition, \eref{e:totvardist} is really a sum of four total variation distances, one for each of the setting configurations. (If we assume that all setting configurations are equiprobable, the conditional could be dropped, and the two definitions of total variation distance would align up to a scale factor.) 

We consider now the problem of finding the local distribution that is of minimum total variation distance from a given nonlocal nonsignaling distribution matrix. The following result gives a solution and shows that the distance to the closest local distribution is equal to the proportion $p_{PR}$ of PR box in the 1 PR + 8 LD representation of the nonlocal nonsignaling distribution matrix (and thus also equal to a constant multiple of the CHSH violation).
\begin{theorem}\label{t:vardistheorem}
Let a distribution $Q$ be nonsignaling and violate the CHSH inequality \eref{e:CHSH}, so that it can be expressed in the 1 PR + 8 LD form \eref{e:convexexp}. Then a local distribution $S$ achieves the minimum possible $\delta(Q,S)$ for local distributions if it is of the form 
\begin{equation}\label{e:Sexp}
s_1D_1+ s_4D_4+ s_5D_5+ s_8D_8+ s_9D_9+ s_{12}D_{12}+ s_{14}D_{14}+ s_{15}D_{15}
\end{equation}
where each $s_i$ is equal to $p_i+\frac{p_{PR}}{8}$, and $\min\delta(Q,S)=p_{PR}$.
\end{theorem}
We defer the straightforward but somewhat lengthy proof of Theorem \ref{t:vardistheorem} to \ref{s:vardistproof}.

Theorem \ref{t:vardistheorem} has intuitive appeal: starting with the nonsignaling nonlocal distribution given by \eref{e:convexexp}, you can get to the closest local distribution (with respect to total variation distance) by throwing away the $p_{PR}$ component and replacing it with an even mixture of the eight local distributions, which is then added to the original weight on these eight distributions. (Note that this is not in general the same throwing away the $p_{PR}$ coefficient and then renormalizing the local coefficients according to the formula $p_i/(1-p_{PR})$.) Interestingly, the distribution in the statement of Theorem \ref{t:vardistheorem} is not unique in minimizing the quantity \eref{e:totvardist} -- for instance, the distribution 
\begin{equation*}
\hspace{-2cm} (p_1+\frac{p_{PR}}{2})D_1+ (p_4+\frac{p_{PR}}{2})D_4+ p_5D_5+ p_8D_8+ p_9D_9+ p_{12}D_{12}+ p_{14}D_{14}+ p_{15}D_{15}
\end{equation*}
will also achieve the minimum -- the expression given in \eref{e:Sexp} is somewhat canonical in that it allocates the divergence from the quantum distribution matrix equally between all four measurement settings. 

Prior to knowing Theorem \ref{t:vardistheorem}, the problem of identifying a distribution that minimizes total variation distance from a nonsignaling nonlocal distribution would be a linear-programming-type problem likely requiring the use of a computer. With the result of Theorem \ref{t:vardistheorem}, the calculation is immediate: from the conditional probabilities in the distribution matrix, one can immediately find the $p_i$ coefficients by referring to Table \ref{t:genexp}, from which the $s_i$ coefficients in \eref{e:Sexp} can be obtained directly.

\medskip

\noindent \emph{Kullback-Leibler Divergence}. If one is conducting a hypothesis test against local realism according to the prediction-based-ratio (PBR) protocol of \cite{zhang:2011,zhang:2013}, calculations involving the Kullback-Leibler divergence are of central importance. Specifically, to define a test statistic that will be optimal for a given nonsignaling nonlocal distribution matrix, one must compute the closest local distribution with respect to the Kullback-Leibler divergence, which is in general \emph{not} a distribution that minimizes the total variation distance. It may not always be possible to find the closest local distribution by analytical methods, in which case a numerical optimization can be performed following the procedures outlined in \cite{zhang:2011}. This optimization problem can be simplified by using the results of this paper.

For two discrete probability distributions $q$ and $p$ over the same set $X$, the Kullback-Leibler divergence from $q$ to $p$ is defined as
\begin{equation}\label{e:KLdef}
D_{KL}(q\mid p) = \sum_{x\in X} q_x \log_2(q_x/p_x),
\end{equation}
where $q_x$ and $p_x$ denote the probability of outcome $x$ according to distribution $q$  and $p$, respectively. We cannot directly evaluate \eref{e:KLdef} for pairs of distribution matrices in the $(2,2,2)$ setting, because such a distribution matrix actually consists of four separate conditional probability distributions, one for each setting configuration. Thus to compute expression \eref{e:KLdef}, one multiplies by the known settings probabilities. So for a distribution matrix $Q$, let us define the induced probability distribution $Q'$ over 16 outcomes according to the rule
\begin{equation*}
Q'(XYAB):=Q(XY\mid AB)\times \textnormal{Prob}(AB).
\end{equation*}
Note that if we have two distribution matrices $Q$ and $S$, and a third distribution $P$ is a convex combination of the two such that $P = \lambda Q + (1-\lambda)S$ for some $\lambda\in(0,1)$, then equivalently $P'  = \lambda Q' + (1-\lambda)S'$. 

Now consider a nonsignaling nonlocal distribution matrix $Q$ violating the CHSH inequality. We assert that a local distribution $S$ of minimum-possible Kullback-Leibler divergence from $Q'$ to $S'$ must be expressible as a convex combination of only the eight local deterministic distributions that saturate the CHSH inequality. 

This is a consequence of Lemma \ref{l:straightlinelemma}; we just need to show that for $P = \lambda Q + (1-\lambda)S$, $D\big(Q'\mid P')$ is strictly less than $D(Q'\mid S')$ for $\lambda\in (0,1)$. This can be derived as a consequence of the well-known fact that $D(\cdot \mid \cdot)$ is convex in its second argument:
\begin{eqnarray*}
D_{KL}(Q'\mid P') &=& D_{KL}\left(Q'\mid\lambda Q' + (1-\lambda)S'\right)\\
&\le&  \lambda D_{KL}(Q'\mid Q') + (1-\lambda)D_{KL}(Q'\mid S')\\
&=&  0 + (1-\lambda)D_{KL}(Q'\mid S') \\
&<& D_{KL}(Q'\mid S').
\end{eqnarray*}
Thus if $S$ is not already a convex combination of the LD distributions that saturate \eref{e:CHSH}, Lemma \ref{l:straightlinelemma} tells us there is a choice of $\lambda\in(0,1)$ for which $P$ is such a convex combination and $D\big(Q'\mid P')<D(Q'\mid S')$. So any local distribution of minimum Kullback-Leibler divergence from $Q'$ is necessarily a convex combination of the eight LD distributions that saturate the CHSH inequality. The entire space of local distributions consists of convex combinations of all sixteen LD distributions, so this observation reduces the number of parameters of the search space by a factor of 2, thereby simplifying the optimization problem of finding the local distribution of minimum Kullback-Leibler divergence from $Q'$.

\section{Extension to the Chained Bell Setting}\label{s:chained}

Having studied the $(2,2,2)$ scenario extensively, it is natural to ask whether these results generalize to other Bell scenarios. In this section, we study one particular generalization, the $(2,n,2)$ ``chained Bell'' scenario of two parties, $n$ settings, and two outcomes per setting. We will find that analogs of Theorems \ref{t:onepr} and \ref{t:8dets} do indeed hold for this scenario, and will demonstrate an application.

\subsection{The Chained Bell Polytope}\label{s:chainedBellpoly}

To formulate the statements of the chained-Bell analogs of our previous theorems, it is necessary to enumerate the extremal points of the chained Bell polytope. To formulate the problem, recall that while the $n$-fold chained Bell scenario involves $n$ settings per party, it does not consider all $n^2$ possible setting configurations such as the situation studied in \cite{jones:2005}. Rather, the chained Bell scenario considers only a subset of all setting configurations of cardinality $2n$. Thus a distribution matrix will consist of $2n\times 4$ entries -- an element of $\mathbb R^{8n}$. We can organize distribution matrices into outcome tables as follows:

\vspace{.5cm}

\begin{center}
\begin{tabular}{ r|c|c|c|c| }
\multicolumn{1}{r}{}
&  \multicolumn{1}{c}{++}
&  \multicolumn{1}{c}{+0}
&  \multicolumn{1}{c}{0+} 
&  \multicolumn{1}{c}{00}\\
\cline{2-5}
$a_1b_1$ &  & & & \\
\cline{2-5}
$a_2b_1$ &  & & & \\
\cline{2-5}
$a_2b_2$ &  & & & \\
\cline{2-5}
$a_3b_2$ &  & & & \\
\cline{2-5}
\multicolumn{1}{c}{}&\multicolumn{1}{c}{$\vdots$}&\multicolumn{1}{c}{$\vdots$}&\multicolumn{1}{c}{$\vdots$}&\multicolumn{1}{c}{$\vdots$}\\
\cline{2-5}
$a_{n}b_{n-1}$ &  & & & \\
\cline{2-5}
$a_{n}b_{n}$ &  & & & \\
\cline{2-5}
$a_1b_n$  &  & & & \\
\cline{2-5}
\end{tabular}
\end{center}

\vspace{5mm}

To be a valid nonsignaling distribution matrix for a fixed $n$, an element of dimension $\mathbb R^{8n}$ must satisfy the probability constraints (all entries nonnegative and each row sums to one) and the following linearly independent collection of no-signaling conditions:
\begin{eqnarray}
\hspace{-2.5cm}P(++\mid a_ib_{i-1})+P(+0\mid a_ib_{i-1}) &=& P(++\mid a_ib_{i})+P(+0\mid a_ib_i), \quad i\in \{2,...,n\}\nonumber\\
\hspace{-2.5cm}P(++\mid a_1b_{1})+P(+0\mid a_1b_{1}) &=& P(++\mid a_1b_{n})+P(+0\mid a_1b_n)\nonumber\\
\hspace{-2.5cm}P(++\mid a_ib_{i})+P(0+\mid a_ib_{i}) &=& P(++\mid a_{i+1}b_{i})+P(0+\mid a_{i+1}b_i), \quad i\in \{1,...,n-1\}\nonumber\\
\hspace{-2.5cm}P(++\mid a_nb_{n})+P(0+\mid a_nb_{n}) &=& P(++\mid a_1b_{n})+P(0+\mid a_1b_n)\label{e:chainednosig}
\end{eqnarray}
We refer to the subset of $\mathbb R^{8n}$ satisfying these conditions as the \emph{$n$-fold chained Bell polytope}. This polytope can be thought of as a projection of the larger polytope studied in \cite{jones:2005} that considers all possible $n^2$ measurement configurations, but for our current purposes this viewpoint is not essential. Any quantum-induced distribution matrix for the $n$-fold chained Bell scenario will obey the no-signaling conditions \eref{e:chainednosig} and thus will be an element of the $n$-fold chained Bell polytope. 

Based on an understanding of the $(2,2,2)$ polytope, one would reasonably expect the extremal points for the $n$-fold chained Bell polytope to include analogs of the PR boxes as well as local deterministic distributions. In preparation for our classification of the extremal points, we define a \emph{generalized PR box} to be a $n$-fold chained Bell distribution matrix consisting of an odd number of rows of the form $\left(\frac{1}{2},0,0,\frac{1}{2}\right)$, an odd number of rows of the form $\left(0,\frac{1}{2},\frac{1}{2},0\right)$, and no other types of rows. Two examples of generalized PR boxes for $n=3$ are as follows:

\vspace{.5cm}

\begin{tabular}{ r|c|c|c|c| }
\multicolumn{1}{r}{}
&  \multicolumn{1}{c}{++}
&  \multicolumn{1}{c}{+0}
&  \multicolumn{1}{c}{0+} 
&  \multicolumn{1}{c}{00}\\
\cline{2-5}
$a_1b_1$ & 1/2 & & & 1/2 \\
\cline{2-5}
$a_2b_1$ & 1/2 & & & 1/2 \\
\cline{2-5}
$a_2b_2$ & 1/2 & & & 1/2 \\
\cline{2-5}
$a_3b_2$ & 1/2 & & & 1/2 \\
\cline{2-5}
$a_3b_3$ & 1/2 & & & 1/2 \\
\cline{2-5}
$a_1b_3$ &  & 1/2 & 1/2 & \\
\cline{2-5}
\end{tabular}
\hspace{2cm}
\begin{tabular}{ r|c|c|c|c| }
\multicolumn{1}{r}{}
&  \multicolumn{1}{c}{++}
&  \multicolumn{1}{c}{+0}
&  \multicolumn{1}{c}{0+} 
&  \multicolumn{1}{c}{00}\\
\cline{2-5}
$a_1b_1$ & 1/2 & & & 1/2 \\
\cline{2-5}
$a_2b_1$ &  & 1/2 & 1/2 & \\
\cline{2-5}
$a_2b_2$ &  & 1/2 & 1/2 & \\
\cline{2-5}
$a_3b_2$ & 1/2 & & & 1/2 \\
\cline{2-5}
$a_3b_3$ &  & 1/2 & 1/2 & \\
\cline{2-5}
$a_1b_3$ & 1/2 & & & 1/2\\
\cline{2-5}
\end{tabular}

\vspace{.5cm}

\noindent In the $n$-fold chained Bell setting, there are $2^{2n-1}$ generalized PR boxes, as there are $2n-1$ independent rows that can be assigned to one of two different formats; after these $2n-1$ choices are made, the format of the last row is fixed by the requirement that the number of rows of each type be odd.

As for the local deterministic distributions, it is no longer practical to list these explicitly as we did for the special case $n=2$. Instead, we enumerate them as the set of mappings that assign each $a_i$ and $b_i$ to either 0 or +. For example, here is one possible table of assignments for $n=4$:

\medskip

\begin{table}[h]\caption{Assignment Table for a Local Deterministic Distribution, $n=4$}\label{t:asstable}
\begin{center}
\begin{tabular}{c|c|c|c|c|c|c|c|}
$a_1$ & $b_1$ & $a_2$ & $b_2$ & $a_3$ & $b_3$ & $a_4$ & $b_4$ \\
+ & 0 & 0 & + & 0 & 0 &  + & + \\
\end{tabular}
\end{center}
\end{table}

\noindent In the above assignment table, we adopt a useful convention of separating all columns with a vertical line and adding an additional vertical line at the end. This way, each vertical line corresponds to a row of an outcome table: a vertical line between $a_x$ and $b_y$ corresponds to the $a_xb_y$ row, while the last vertical line corresponds to the $a_1b_n$ row. This will allow us to easily move from an assignment table like the one above to the corresponding outcome table of a distribution matrix -- for instance, as $a_1$ is mapped to + and $b_1$ is mapped to 0, then the $a_1b_1$ outcome row will have a ``1'' in the +0 cell and a zero in the other three cells, etc., and so the distribution matrix for the above assignment table is as follows:

\begin{center}
 \begin{tabular}{ r|c|c|c|c| }
\multicolumn{1}{r}{}
&  \multicolumn{1}{c}{++}
&  \multicolumn{1}{c}{+0}
&  \multicolumn{1}{c}{0+} 
&  \multicolumn{1}{c}{00}\\
\cline{2-5}
$a_1b_1$ & & 1 & &  \\
\cline{2-5}
$a_2b_1$ &  &  &  & 1\\
\cline{2-5}
$a_2b_2$ &  &  & 1& \\
\cline{2-5}
$a_3b_2$ &  & & 1 &  \\
\cline{2-5}
$a_3b_3$ &  &  &  &  1 \\
\cline{2-5}
$a_4b_3$ &  & 1 &  &   \\
\cline{2-5}
$a_4b_4$ & 1  &  &  &   \\
\cline{2-5}
$a_1b_4$ & 1 & &  & \\
\cline{2-5}
\end{tabular}
\end{center}

      We \emph{define} a local deterministic distribution to be a distribution matrix that can be generated from an assignment table like the one in Table \ref{t:asstable}. There are $2^{2n}$ distinct ways to fill an assignment table with 0 and +, and thus $2^{2n}$ LD distributions in the $n$-fold chained Bell scenario.

The following theorem classifies the extremal points of the $n$-fold chained Bell polytope. It is proved in \ref{s:chainedBellproof}.

\begin{theorem}\label{t:chainedBelltheorem}
The set of extremal points of the nonsignaling polytope corresponding to the $n$-fold chained Bell scenario consists of the $2^{2n}$ local deterministic distributions and the $2^{2n-1}$ ``generalized PR boxes.''
\end{theorem}

\subsection{Decomposition Theorems for the $n$-Fold Chained Bell Polytope}

Having enumerated the vertices of the $n$-fold chained Bell polytope, we can now proceed to state and prove generalized versions of Theorems \ref{t:onepr} and \ref{t:8dets}.

\begin{theorem}\label{t:genonepr}
In the $n$-fold chained Bell scenario, an equal mixture of any two (non-identical) generalized PR boxes is equivalent to a mixture of local deterministic distributions.
\end{theorem}
\emph{Proof.}
Consider two non-identical PR boxes called ``PR A'' and ``PR B.'' The equal mixture $\frac{1}{2}PR_A+\frac{1}{2}PR_B$ will yield a distribution matrix for which each row of the outcome table can take one of the three following forms: $\left(\frac{1}{2},0,0,\frac{1}{2}\right)$, $\left(0,\frac{1}{2},\frac{1}{2},0\right)$, and $\left(\frac{1}{4},\frac{1}{4},\frac{1}{4},\frac{1}{4}\right)$, where these vectors represent settings-conditional assignments of probabilities to the outcomes $(\textnormal{++},\textnormal{+0},\textnormal{0+},\textnormal{00})$. We call these three row types \emph{correlated}, \emph{anticorrelated}, and \emph{uniform}, respectively. We claim that the number of uniform rows is nonzero and even. First, note that to have zero uniform rows would require that PR A and PR B be identical, a possibility that we are excluding. To see that the number of uniform rows is even, recall that PR A and PR B will each have an odd number of correlated rows and an odd number of anticorrelated rows. Now, out of the complete set of $2n$ outcome rows, consider the collection of rows where PR A is anticorrelated and PR B is correlated. If the number of such rows is odd (even), then the number of rows where A is anticorrelated and B is anticorrelated must be even (odd), and then the number of rows where A is correlated and B is anticorrelated must be odd (even), so the total number of rows where A and B are inequivalent (either A anticorrelated/B correlated or A correlated/B anticorrelated) must be the sum of two odd numbers or the sum of two even numbers, and therefore even. Thus $\frac{1}{2}PR_A+\frac{1}{2}PR_B$ has a positive even number of uniform rows.

Now we will show how to select four LD distributions so that their mixture is equivalent to the equal mixture of PR A and PR B. To illustrate how this can work, first consider the case where PR A and PR B are perfectly mis-aligned so that their mixture consists entirely of uniform rows. Then the following suite of four LD distributions can replicate this distribution matrix:

\vspace{5mm}

\begin{center}
\begin{tabular}{cc|c|c|c|c|c|c|c|c|c|c|c|c|c|c|c|c|}
&$a_1$ & $b_1$ & $a_2$ & $b_2$ & $a_3$ & $b_3$ & $a_4$ & $b_4$ &$a_5$ & $b_5$ &$a_6$ & $b_6$ &$\cdots$&$a_{n-1}$ &$b_{n-1}$ & $a_n$ & $b_n$ \\
 1:& + & + & + & + & + & + & + & + & + & + & + & + &$\cdots$&+ & + & + & + \\
    2:& + & 0 & + & 0 & + & 0 & + & 0 & + & 0 & + & 0 &$\cdots$&+ & 0 & + & 0 \\
     3:& 0 & + & 0 & + & 0 & + & 0 & + & 0 & + & 0 & + &$\cdots$&0 & + & 0 & + \\
       4:& 0 & 0 & 0 & 0 & 0 & 0 & 0 & 0 & 0 & 0 & 0 & 0 &$\cdots$&0 & 0 & 0 & 0 \\
\end{tabular}
\end{center}

\vspace{5mm}

\noindent To see why, notice that each vertical line in the assignment table above is always straddled by ++ once, +0 once, 0+ once, and 00 once. Thus putting $1/4$th weight on each LD distribution generates the distribution $\left(\frac{1}{4},\frac{1}{4},\frac{1}{4},\frac{1}{4}\right)$ in each row of the outcome table.

Now consider the general case, where we have noted that the two PR boxes will be mis-aligned on a nonzero even number of rows. We describe an algorithm to generate four LD distributions whose mixture replicates the mixture of the two PR boxes. First, start with an empty assignment table like the one above, with the vertical lines labeled to indicate the nature of the corresponding row of the PR box mixture -- correlated, anticorrelated, or uniform. For example, this would look something like this:

\vspace{5mm}
{\scriptsize
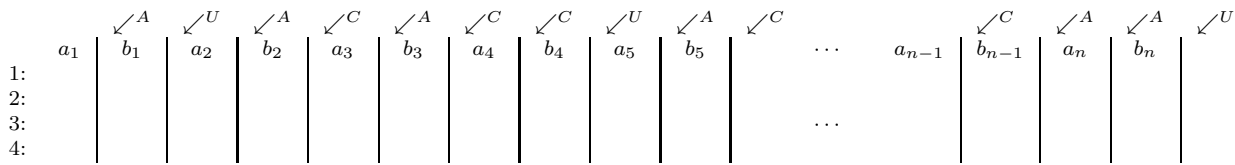
\begin{center}
\begin{tabular}{cc|c|c|c|c|c|c|c|c|c|cc|c|c|c|c}
\multicolumn{2}{c}{}&
\multicolumn{1}{l}{$\swarrow^A$}&
\multicolumn{1}{l}{$\swarrow^U$}&
\multicolumn{1}{l}{$\swarrow^A$}&
\multicolumn{1}{l}{$\swarrow^C$}&
\multicolumn{1}{l}{$\swarrow^A$}&
\multicolumn{1}{l}{$\swarrow^C$}&
\multicolumn{1}{l}{$\swarrow^C$}&
\multicolumn{1}{l}{$\swarrow^U$}&
\multicolumn{1}{l}{$\swarrow^A$}&
\multicolumn{1}{l}{$\swarrow^C$}&
\multicolumn{1}{l}{}&
\multicolumn{1}{l}{$\swarrow^C$}&
\multicolumn{1}{l}{$\swarrow^A$}&
\multicolumn{1}{l}{$\swarrow^A$}&
\multicolumn{1}{l}{$\swarrow^U$}\\
&$a_1$ & $b_1$ & $a_2$ & $b_2$ & $a_3$ & $b_3$ & $a_4$ & $b_4$ &$a_5$ & $b_5$  &$\quad\quad\quad\cdots\quad$&$a_{n-1}$ &$b_{n-1}$ & $a_n$ & $b_n$ &\\
 1:&  &  &    &   &  &    & &  &  &       &   &    &   &  &  &\\
 2:&  &  &    &   &  &    & &  &  &       &   &    &   &  &  &\\
 3:&  &  &    &   &  &    & &  &  &    &   $\quad\quad\quad\cdots\quad$ &   &   &  &  &\\
 4:&  &  &    &   &  &    & &  &  &     &  &    &   &  &  &\\
\end{tabular}
\end{center}
}
\vspace{5mm}

\noindent Now fill the table according to the following procedure:

\begin{enumerate}
\item[1.] Locate the leftmost line marked ``uniform'' and enter $\textnormal{+}\textnormal{+}00$ in descending order in the column to the right of this line.
\item[2.] Moving to the right, fill in the next blank column according to the following set of rules:

\medskip
\begin{tabular} {p{3.5cm}|p{5cm}|p{4cm}}
\emph{If the last filled column contained...} & \emph{... and the line label to the left of the blank column is...} & \emph{...then fill the blank column as follows}:\\
\hline
\multicolumn{1}{|c|} {$\textnormal{+}\textnormal{+}00$} & \multicolumn{1}{c} {$C$} & \multicolumn{1}{|c|} {$\textnormal{+}\textnormal{+}00$}\\
 \multicolumn{1}{|c|} {$00\textnormal{+}\textnormal{+}$} &    \multicolumn{1}{c} {$C$}  &  \multicolumn{1}{|c|} {$00\textnormal{+}\textnormal{+}$}   \\
\multicolumn{1}{|c|} {$\textnormal{+}0\textnormal{+}0$}  &    \multicolumn{1}{c} {$C$}   &  \multicolumn{1}{|c|} {$\textnormal{+}0\textnormal{+}0$}    \\
\multicolumn{1}{|c|} {$0\textnormal{+}0\textnormal{+}$}   &    \multicolumn{1}{c} {$C$}   &   \multicolumn{1}{|c|} {$0\textnormal{+}0\textnormal{+}$}    \\
\hline
\multicolumn{1}{|c|} {$\textnormal{+}\textnormal{+}00$} & \multicolumn{1}{c} {$A$} & \multicolumn{1}{|c|} {$00\textnormal{+}\textnormal{+}$}\\
\multicolumn{1}{|c|} {$00\textnormal{+}\textnormal{+}$} &    \multicolumn{1}{c} {$A$}  &   \multicolumn{1}{|c|} {$\textnormal{+}\textnormal{+}00$}  \\
\multicolumn{1}{|c|} {$\textnormal{+}0\textnormal{+}0$} &    \multicolumn{1}{c} {$A$}   &    \multicolumn{1}{|c|} {$0\textnormal{+}0\textnormal{+}$}   \\
\multicolumn{1}{|c|} {$0\textnormal{+}0\textnormal{+}$}  &    \multicolumn{1}{c} {$A$}   &   \multicolumn{1}{|c|} {$\textnormal{+}0\textnormal{+}0$}   \\
\hline
\multicolumn{1}{|c|} {$\textnormal{+}\textnormal{+}00$} & \multicolumn{1}{c} {$U$} & \multicolumn{1}{|c|} {$\textnormal{+}0\textnormal{+}0$}\\
 \multicolumn{1}{|c|} {$00\textnormal{+}\textnormal{+}$}  &    \multicolumn{1}{c} {$U$}  &  \multicolumn{1}{|c|} {$\textnormal{+}0\textnormal{+}0$}    \\
\multicolumn{1}{|c|} {$\textnormal{+}0\textnormal{+}0$}  &    \multicolumn{1}{c} {$U$}   &    \multicolumn{1}{|c|} {$\textnormal{+}\textnormal{+}00$}  \\
\multicolumn{1}{|c|} {$0\textnormal{+}0\textnormal{+}$}    &    \multicolumn{1}{c} {$U$}   &    \multicolumn{1}{|c|} {$\textnormal{+}\textnormal{+}00$}  \\
\hline
 \end{tabular}

\item[3.] Continue working to the right repeating step 2 until the assignment table is full, wrapping around from $b_n$ to $a_1$ if necessary.
\end{enumerate}

\noindent To illustrate, we demonstrate a few iterations of this algorithm below:

\vspace{5mm}
{\scriptsize
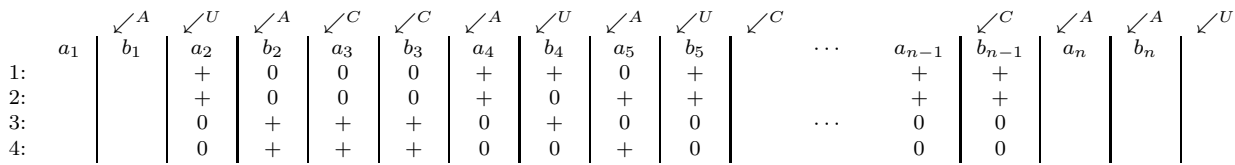
\begin{center}
\begin{tabular}{cc|c|c|c|c|c|c|c|c|c|cc|c|c|c|c}
\multicolumn{2}{c}{}&
\multicolumn{1}{l}{$\swarrow^A$}&
\multicolumn{1}{l}{$\swarrow^U$}&
\multicolumn{1}{l}{$\swarrow^A$}&
\multicolumn{1}{l}{$\swarrow^C$}&
\multicolumn{1}{l}{$\swarrow^C$}&
\multicolumn{1}{l}{$\swarrow^A$}&
\multicolumn{1}{l}{$\swarrow^U$}&
\multicolumn{1}{l}{$\swarrow^A$}&
\multicolumn{1}{l}{$\swarrow^U$}&
\multicolumn{1}{l}{$\swarrow^C$}&
\multicolumn{1}{l}{}&
\multicolumn{1}{l}{$\swarrow^C$}&
\multicolumn{1}{l}{$\swarrow^A$}&
\multicolumn{1}{l}{$\swarrow^A$}&
\multicolumn{1}{l}{$\swarrow^U$}\\
&$a_1$ & $b_1$ & $a_2$ & $b_2$ & $a_3$ & $b_3$ & $a_4$ & $b_4$ &$a_5$ & $b_5$  &$\quad\quad\quad\cdots\quad$&$a_{n-1}$ &$b_{n-1}$ & $a_n$ & $b_n$ &\\
 1:&  &  &  +  &  0 & 0  & 0   & +  & + & 0 &  +    &   &  +  & +  &  &  &\\
 2:&  &  &  +  &  0 & 0  & 0   & +  & 0 & + &  +    &   &  +  &  + &  &  &\\
 3:&  &  &  0  &  + & + & +   & 0  & + & 0 &  0  &   $\quad\quad\quad\cdots\quad$ & 0  & 0  &  &  &\\
 4:&  &  &  0  &  + & + & +   & 0  & 0 & + &  0   &  &  0  &  0 &  &  &\\
\end{tabular}
\end{center}
}
\vspace{5mm}

By inspection, we see that a new column is always filled in a manner so that an equiprobable distribution over the four LD distributions induces a row distribution that is consistent with the label of the vertical line to its left, and thus is consistent with the corresponding row of $\frac{1}{2}PR_A+\frac{1}{2}PR_B$. So the only question is what happens, after wrapping around $b_n$ to $a_1$, when we get to the final empty column (which is $b_1$ in the example) -- specifically, will the last-filled column also be consistent with the label of the vertical line to its \emph{right}? To see why this will work, note that the algorithm always fills columns with an entry from one of the following two sets: $S_1=\{\textnormal{+}\textnormal{+}00,00\textnormal{+}\textnormal{+}\}$ and $S_2=\{\textnormal{+}0\textnormal{+}0, 0\textnormal{+}0\textnormal{+}\}$. Each time a ``U'' line is crossed, we switch from being in set $S_1$ to being in set $S_2$ or vice-versa, whereas crossing ``A'' and ``C'' lines keeps us within whichever subset $S_i$ we are already in. Our earlier observation requires that there is an even total number of ``U'' labels, so we will have crossed an odd number of lines labeled ``U'' by the time we fill in the last column, with the last remaining ``U'' linking the last-filled column to the first-filled column. Starting with ++00 (in $S_1$) in the initial entry, the odd number of U-flips will thus always result in either $\textnormal{+}0\textnormal{+}0$ or $0\textnormal{+}0\textnormal{+}$ for the final entry. Either one of these results is consistent with the original ``U'' line, which has $\textnormal{+}\textnormal{+}00$ on its right, so we will always get a consistent final result such as what we see here for our illustrative example:

\vspace{5mm}
{\scriptsize
\begin{center}
\begin{tabular}{cc|c|c|c|c|c|c|c|c|c|cc|c|c|c|c}
\multicolumn{2}{c}{}&
\multicolumn{1}{l}{$\swarrow^A$}&
\multicolumn{1}{l}{{\color{red}$\swarrow^U$}}&
\multicolumn{1}{l}{$\swarrow^A$}&
\multicolumn{1}{l}{$\swarrow^C$}&
\multicolumn{1}{l}{$\swarrow^C$}&
\multicolumn{1}{l}{$\swarrow^A$}&
\multicolumn{1}{l}{$\swarrow^U$}&
\multicolumn{1}{l}{$\swarrow^A$}&
\multicolumn{1}{l}{$\swarrow^U$}&
\multicolumn{1}{l}{$\swarrow^C$}&
\multicolumn{1}{l}{}&
\multicolumn{1}{l}{$\swarrow^C$}&
\multicolumn{1}{l}{$\swarrow^A$}&
\multicolumn{1}{l}{$\swarrow^A$}&
\multicolumn{1}{l}{$\swarrow^U$}\\
&$a_1$ & $b_1$ & $a_2$ & $b_2$ & $a_3$ & $b_3$ & $a_4$ & $b_4$ &$a_5$ & $b_5$  &$\quad\quad\quad\cdots\quad$&$a_{n-1}$ &$b_{n-1}$ & $a_n$ & $b_n$ &\\
 1:& + & 0 &  +  &  0 & 0  & 0   & +  & + & 0 &  +    &   &  +  & +  & 0 & + &\\
 2:& 0 & + &  +  &  0 & 0  & 0   & +  & 0 & + &  +    &   &  +  &  + & 0 & + &\\
 3:& + & 0 &  0  &  + & + & +   & 0  & + & 0 &  0  &   $\quad\quad\quad\cdots\quad$ & 0  & 0  & + & 0 &\\
 4:& 0 & + &  0  &  + & + & +   & 0  & 0 & + &  0   &  &  0  &  0 & + & 0 &\\
\end{tabular}
\end{center}
}
\vspace{5mm}

\noindent Hence the algorithm always generates a collection of four LD distributions whose mixture induces the same distribution as the PR box mixture. $\hfill\Box$

\medskip

The previous result generalizes Theorem \ref{t:onepr}. An argument analogous to the proof of Corollary \ref{c:1} then shows that any convex combination of generalized PR boxes and LD distributions can be re-expressed as a convex combination that includes at most one generalized PR box. The next step is to generalize Theorem \ref{t:8dets}, which requires a few preliminary results.

\begin{lemma}\label{l:oddmismatch}
For a given local deterministic distribution and a given generalized PR box in the $n$-fold chained Bell scenario, there is always an odd number of outcome rows for which the local distribution has a ``1'' in an entry where the PR box has a ``0.''
\end{lemma}
\emph{Proof.} Each row of the distribution matrix of a LD distribution contains exactly one 1 and three 0s. Furthermore, we claim that the distribution matrix always has an even number of outcome rows where the 1 is in ++ or 00, and an even number of rows where the 1 is in +0 or 0+. To see why this is so, we can represent a general LD distribution with the following assignment table

\medskip

\begin{center}
\begin{tabular}{c|c|c|c|c|c|c|c|c|c|}
$a_1$ & $b_1$ & $a_2$ & $b_2$ & $a_3$ & $b_3$ & $\cdots$ &$b_{n-1}$ & $a_n$ & $b_n$ \\
$x_1$ & $y_1$ & $x_2$ & $y_2$ & $x_3$ & $y_3$ & $\cdots$ & $y_{n-1}$ & $x_n$ & $y_n$ \\
\end{tabular} 
\end{center}

\medskip

\noindent where each variable $x_i$ and $y_i$ takes a value in the two element set $\{+,0\}$ and each vertical line represents the row of the outcome table corresponding to the $a_i$ and $b_j$ that straddle the line. Note that the $a_ib_j$ row of the outcome table will have its support in +0 or 0+ if and only if $x_i\ne y_j$. Now, if we start at $x_1$ and move to the right through the above assignment table, the adjacent $x$ and $y$ values will either be equal or unequal as we cross each vertical line. If we go all the way to $y_n$ and cross the last vertical line, this returns us to $x_1$, and so consistency requires that we must cross an even number of lines for which $x_i\ne y_j$ across the line. This proves the claim.

Since a generalized PR box always has an odd number of each type of the two rows $\left(\frac{1}{2},0,0,\frac{1}{2}\right)$ and $\left(0,\frac{1}{2},\frac{1}{2},0\right)$, there must be an odd number of rows for which the LD distribution has a 1 where the generalized PR box has a 0. $\hfill\Box$

\medskip

For a given generalized PR box and LD distribution, Lemma \ref{l:oddmismatch} tells us that the LD distribution's outcome table must have at least one row where it has a ``1'' in a location where the generalized PR box has a ``0.'' The special case where the LD distribution has \emph{exactly} one such row is of particular interest. We will see that LD distributions satisfying this condition are the ones that are ``affiliated'' with a particular generalized PR box, in the same way that $(2,2,2)$ LD distributions saturating a CHSH symmetry inequality are affiliated with a PR box violating the same CHSH symmetry. For a given generalized PR box, we say that a LD distribution is a \emph{one-support-mismatch} if it has exactly one row where it has a ``1'' in a location where the generalized PR box has a ``0.'' Here are two examples for a specific generalized PR box:

{\footnotesize
\begin{center}
\begin{tabular}{ r|c|c|c|c| }
\multicolumn{1}{r}{}
&  \multicolumn{1}{c}{++}
&  \multicolumn{1}{c}{+0}
&  \multicolumn{1}{c}{0+} 
&  \multicolumn{1}{c}{00}\\
\cline{2-5}
$a_1b_1$ & 1/2 & & & 1/2 \\
\cline{2-5}
$a_2b_1$ & 1/2 & & & 1/2 \\
\cline{2-5}
$a_2b_2$ & 1/2 & & & 1/2 \\
\cline{2-5}
$a_3b_2$ & 1/2 & & & 1/2 \\
\cline{2-5}
$a_3b_3$ & 1/2 & & & 1/2 \\
\cline{2-5}
$a_1b_3$ &  & 1/2 & 1/2 & \\
\cline{2-5}
\end{tabular}
\hspace{1mm}
\begin{tabular}{ r|c|c|c|c| }
\multicolumn{1}{r}{}
&  \multicolumn{1}{c}{++}
&  \multicolumn{1}{c}{+0}
&  \multicolumn{1}{c}{0+} 
&  \multicolumn{1}{c}{00}\\
\cline{2-5}
$a_1b_1$ & 1 & & & \\
\cline{2-5}
$a_2b_1$ & 1 & & & \\
\cline{2-5}
$a_2b_2$ & 1 & & & \\
\cline{2-5}
$a_3b_2$ & 1 & & & \\
\cline{2-5}
$a_3b_3$ & 1 & & & \\
\cline{2-5}
$a_1b_3$ & {\color{red}1} & & & \\
\cline{2-5}
\end{tabular}
\hspace{1mm}\begin{tabular}{ r|c|c|c|c| }
\multicolumn{1}{r}{}
&  \multicolumn{1}{c}{++}
&  \multicolumn{1}{c}{+0}
&  \multicolumn{1}{c}{0+} 
&  \multicolumn{1}{c}{00}\\
\cline{2-5}
$a_1b_1$ & 1 & & &  \\
\cline{2-5}
$a_2b_1$ &  &  & {\color{red} 1}&  \\
\cline{2-5}
$a_2b_2$ & &  & & 1 \\
\cline{2-5}
$a_3b_2$ & & & & 1 \\
\cline{2-5}
$a_3b_3$ & & & & 1 \\
\cline{2-5}
$a_1b_3$ &  & 1 & &  \\
\cline{2-5}
\end{tabular}
\end{center}
}

\medskip

\begin{lemma}\label{l:castout}
Consider a generalized PR box and a local deterministic distribution for which there are $2m+1$ (with $m \ne 0$) outcome rows in which there is an outcome that the local deterministic distribution assigns probability one and the PR box assigns zero. Then the mixture distribution induced by $\frac{2m}{2m+1}$ times the PR box and $\frac{1}{2m+1}$ times the local deterministic distribution can also be induced by a uniform mixture of $2m+1$ local deterministic distributions that are one-support-mismatches with respect to the generalized PR box.
\end{lemma}
\emph{Proof.}
We represent the initial local deterministic distribution with the now-familiar assignment table:

\medskip

\begin{center}
\begin{tabular}{c|c|c|c|c|c|c|c|c|c|c|c|c|c|c|c|}
$a_1$ & $b_1$ & $a_2$ & $b_2$ & $a_3$ & $b_3$ & $a_4$ & $b_4$ &$a_5$ & $b_5$ &$a_6$ & $b_6$ &$\cdots$ &$b_{n-1}$ & $a_n$ & $b_n$ \\
 $x_1$ & $y_1$ & $x_2$ & $y_2$ & $x_3$ & $y_3$ & $x_4$ & $y_4$ & $x_5$ & $y_5$ & $x_6$ & $y_6$ &$\cdots$ & $y_{n-1}$ & $x_n$ & $y_n$ \\
\end{tabular}
\end{center}

\medskip

\noindent By assumption, there are exactly $2m+1$ outcome rows where the LD distribution assigns probability one and the PR box assigns probability zero to an outcome. If we mark the corresponding $2m+1$ lines in the assignment table with arrows, it will look something like this: 

\medskip

\begin{center}
\begin{tabular}{c|c|c|c|c|c|c|c|c|c|c|c|c|c|c|c|}
\multicolumn{2}{c}{} & \multicolumn{1}{c}{$\swarrow$}& \multicolumn{2}{c}{}& \multicolumn{1}{c}{$\swarrow$}& \multicolumn{1}{c}{}&\multicolumn{1}{c}{$\swarrow$}&\multicolumn{1}{c}{$\swarrow$}&\multicolumn{6}{c}{} & \multicolumn{1}{c}{$\searrow$}\\
$a_1$ & $b_1$ & $a_2$ & $b_2$ & $a_3$ & $b_3$ & $a_4$ & $b_4$ &$a_5$ & $b_5$ &$a_6$ & $b_6$ &$\cdots$ &$b_{n-1}$ & $a_n$ & $b_n$ \\
 $x_1$ & $y_1$ & $x_2$ & $y_2$ & $x_3$ & $y_3$ & $x_4$ & $y_4$ & $x_5$ & $y_5$ & $x_6$ & $y_6$ &$\cdots$ & $y_{n-1}$ & $x_n$ & $y_n$ \\
\end{tabular}
\end{center}

\medskip

The presence of an arrow marking a line between $x_i$ and $y_j$ means that the PR box assigns probability 0 to the event $x_iy_j$ when the setting is $a_ib_j$. If we use -$x_i$ to represent the opposite outcome of $x_i$ (i.e., -$x_i=0$ if $x_i=\textnormal{+}$ and -$x_i=$+ if $x_i=0$), then we can say that at a marked line, the PR box assigns probability 1/2 to the two events (-$x_i)y_j$ and $x_i$(-$y_j)$ when the setting is $a_ib_j$. So if an LD distribution is to be a one-support-mismatch from the original PR box, its assignment table should contain the entries (-$x_i)y_j$ or $x_i$(-$y_j)$ on either side of all but one of the marked lines above. For unmarked lines, a one-support-mismatch must contain either the entry $x_iy_i$ or (-$x_i)($-$y_i)$.

In light of these observations, there is an algorithm to take a marked table like the one above and generate a collection of $2m+1$ one-support-mismatch LD distributions that satisfy the statement of the theorem. Consider the following collection of $2m+1$ LD distributions, generated according to rules enumerated below. 

\vspace{5mm}

{\footnotesize
\hspace{-3cm}\begin{tabular}{ccc|c|c|c|c|c|c|c|c|c|c|c|c|c|c|c|}
\multicolumn{4}{c}{} & \multicolumn{1}{l}{$\swarrow$}& \multicolumn{2}{c}{}& \multicolumn{1}{c}{$\swarrow$}& \multicolumn{1}{c}{}&\multicolumn{1}{l}{$\swarrow$}&\multicolumn{1}{c}{$\swarrow$}&\multicolumn{6}{c}{} & \multicolumn{1}{r}{$\searrow$}\\
&Index&$a_1$ & $b_1$ & $a_2$ & $b_2$ & $a_3$ & $b_3$ & $a_4$ & $b_4$ &$a_5$ & $b_5$ &$a_6$ & $b_6$ &$\cdots$ &$b_{n-1}$ & $a_n$ & $b_n$ \\
\hline
& 1 & $x_1$ & $y_1$ & -$x_2$ & -$y_2$ & -$x_3$ & $y_3$ & $x_4$ & -$y_4$ & $x_5$ & $y_5$ & $x_6$ & $y_6$ &$\cdots$ & $y_{n-1}$ & $x_n$ & \Large{$y_n$} \\
&2 & $x_1$ & $y_1$ & -$x_2$ & -$y_2$ & -$x_3$ & $y_3$ & $x_4$ & -$y_4$ & $x_5$ & $y_5$ & $x_6$ & $y_6$ &$\cdots$ & -$y_{n-1}$ & -$x_n$ & -$y_n$ \\
&3 & $x_1$ & $y_1$ & -$x_2$ & -$y_2$ & -$x_3$ & $y_3$ & $x_4$ & -$y_4$ & $x_5$ & $y_5$ & $x_6$ & $y_6$ &$\cdots$ & -$y_{n-1}$ & -$x_n$ & -$y_n$ \\
Org. aligned&$\vdots$&\multicolumn{16}{c|}{$\vdots$}\\
&$m-2$ & $x_1$ & $y_1$ & -$x_2$ & -$y_2$ & -$x_3$ & $y_3$ & $x_4$ & -$y_4$ & $x_5$ & $y_5$ & $x_6$ & $y_6$ &$\cdots$ & -$y_{n-1}$ & -$x_n$ & -$y_n$ \\
&$m-1$ & $x_1$ & $y_1$ & -$x_2$ & -$y_2$ & -$x_3$ & $y_3$ & $x_4$ & -$y_4$ & $x_5$ & $y_5$ & $x_6$ & $y_6$ &$\cdots$ & -$y_{n-1}$ & -$x_n$ & -$y_n$ \\
&$m$ & $x_1$ & $y_1$ & -$x_2$ & -$y_2$ & -$x_3$ & $y_3$ & \Large{$x_4$} & \Large{$y_4$} & -$x_5$ & -$y_5$ & -$x_6$ & -$y_6$ &$\cdots$ & -$y_{n-1}$ & -$x_n$ & -$y_n$ \\
&$m+1$ & $x_1$ & \Large{$y_1$} & \Large{$x_2$} & $y_2$ & $x_3$ & -$y_3$ & -$x_4$ & $y_4$ & -$x_5$ & -$y_5$ & -$x_6$ & -$y_6$ &$\cdots$ & -$y_{n-1}$ & -$x_n$ & -$y_n$ \\
\hline
&$m+2$ & -$x_1$ & -$y_1$ & $x_2$ & $y_2$ & \Large{$x_3$} & \Large{$y_3$} & $x_4$ & -$y_4$ & $x_5$ & $y_5$ & $x_6$ & $y_6$ &$\cdots$ & $y_{n-1}$ & $x_n$ & $y_n$ \\
&$m+3$ & -$x_1$ & -$y_1$ & $x_2$ & $y_2$ & $x_3$ & -$y_3$ & -$x_4$ & \Large{$y_4$} & \Large{$x_5$} & $y_5$ & $x_6$ & $y_6$ &$\cdots$ & $y_{n-1}$ & $x_n$ & $y_n$ \\
&$m+4$ & -$x_1$ & -$y_1$ & $x_2$ & $y_2$ & $x_3$ & -$y_3$ & -$x_4$ & $y_4$ & -$x_5$ & -$y_5$ & -$x_6$ & -$y_6$ &$\cdots$ & $y_{n-1}$ & $x_n$ & $y_n$ \\
Orig. anti-a.&$m+5$ & -$x_1$ & -$y_1$ & $x_2$ & $y_2$ & $x_3$ & -$y_3$ & -$x_4$ & $y_4$ & -$x_5$ & -$y_5$ & -$x_6$ & -$y_6$ &$\cdots$ & $y_{n-1}$ & $x_n$ & $y_n$ \\
&$\vdots$&\multicolumn{16}{c|}{$\vdots$}\\
&$2m-1$ & -$x_1$ & -$y_1$ & $x_2$ & $y_2$ & $x_3$ & -$y_3$ & -$x_4$ & $y_4$ & -$x_5$ & -$y_5$ & -$x_6$ & -$y_6$ &$\cdots$ & $y_{n-1}$ & $x_n$ & $y_n$ \\
&$2m$ & -$x_1$ & -$y_1$ & $x_2$ & $y_2$ & $x_3$ & -$y_3$ & -$x_4$ & $y_4$ & -$x_5$ & -$y_5$ & -$x_6$ & -$y_6$ &$\cdots$ & $y_{n-1}$ & $x_n$ & $y_n$ \\
&$2m+1$ & -$x_1$ & -$y_1$ & $x_2$ & $y_2$ & $x_3$ & -$y_3$ & -$x_4$ & $y_4$ & -$x_5$ & -$y_5$ & -$x_6$ & -$y_6$ &$\cdots$ & $y_{n-1}$ & $x_n$ & $y_n$ \\
\end{tabular}
}
\vspace{5mm}

The table above is filled according to the following general procedure:

\begin{enumerate}
\item[1.] Starting with an empty table, fill the first column with $m+1$ instances of $x_1$ (``aligned'' with the initial LD distribution) followed by $m$ instances of -$x_1$ (``anti-aligned'' with the initial LD distribution).
\item[2.] Moving to the right, fill in the next blank column according to the following rules:
\begin{enumerate}
\item If the line to the left of the blank column is not marked with an arrow, then fill in each entry with the corresponding $x_i$ or $y_i$ value from the original assignment table, with a ``-'' prefix if and only if the previous column's entry on the same row had a ``-'' prefix.
\item If the line to the left of the blank column is marked with an arrow, then with one exception fill in each entry with the corresponding $x_i$ or $y_i$ value from the original assignment table, with a ``-'' in front of it if and only if the previous column's entry on the same row did \emph{not} have  a ``-'' in front of it. The exception entry is filled with the $x_i$ or $y_i$ value from the original assignment table always without a ``-'', and the location of the exception row follows this rule: for the $k$th marked line, if $k$ is odd the exception row is $m+1- \frac{k-1}{2}$ and if $k$ is even the exception row is $m+1+ \frac{k}{2}$. 
\end{enumerate}
\item[3.] Repeat step 2 until the table is full.
\end{enumerate}

Each of the $2m+1$ LD distributions created by the above algorithm is a one-support-mismatch to the original PR box. To see why, note that the algorithm ensures that each row is an ``exception row'' exactly once, and at the location of the exception, it aligns with the initial LD distribution on both sides of a marked line. So at this location, it must assign probability 1 to an outcome that is assigned zero probability in the corresponding row of the original PR box. But outside of the ``exception'' location, a given row always 1) across a marked line, anti-aligns with the initial LD distribution on exactly one side of the line, or 2) across an unmarked line, either aligns with the initial LD distribution on both sides of the line or anti-aligns with it on both sides of the line. In either case, the LD distribution assigns probability 1 to an outcome that the PR box assigns probability 1/2.

Now we examine how an equal mixture of these $2m+1$ LD distributions compares to a mixture of $2m/(2m+1)$ of the original PR box and $1/(2m+1)$ of the original LD distribution. Consider first an outcome row corresponding to an unmarked line. Here, if the original LD distribution assigns probability 1 to outcome $x_iy_j$, then the PR box assigns 1/2 probability to $x_iy_j$ and 1/2 probability to (-$x_i$)(-$y_j$). If we consider this outcome row for the equal mixture of $2m+1$ LD distributions, one of the aligned distributions will simulate the contribution of the original LD distribution, and the remaining $2m$ distributions simulate the original PR box, because the algorithm ensures that half of the remaining LD distributions cross the line as $x_iy_i$ and the other half of these cross the line as (-$x_i)($-$y_i)$. Now consider an outcome row corresponding to a marked line. In this circumstance, if the original local distribution assigns probability 1 to outcome $x_iy_j$, then the PR box assigns 1/2 probability to $x_i$(-$y_j$) and 1/2 probability to (-$x_i$)$y_j$. Once again, the equal mixture of $2m+1$ LD distributions simulates this behavior: one LD distribution aligns with the original local distribution, and the remaining $2m$ LD distributions simulate the PR box with $m$ LD distributions crossing the line as $x_i$(-$y_j$) and $m$ LD distributions crossing the line as (-$x_i$)$y_j$. $\hfill\Box$

\medskip

If we are presented with a mixture of a single generalized PR box and an arbitrary collection of LD distributions, Lemma \ref{l:castout} allows us to methodically cast out LD distributions that deviate from the support of the PR box in more than one location and replace them with one-support-mismatch LD distributions. This process ``uses up'' a portion of the PR box: if an LD distribution with $2m+1$ support mismatches has weighting $x$, then we must use $2mx$ of the weighting of the PR box to cast out this local distribution. If there is insufficient weight on the PR box in the initial mixture to cast out all of the mismatched LD distributions, this will lead to clause (a) of the following generalization of Theorem \ref{t:8dets}.
\begin{theorem}\label{t:gen8dets}
In the $n$-fold chained Bell scenario, any convex combination of a fixed generalized PR box and an arbitrary collection of local deterministic distributions can be re-expressed as either \emph{a)} a convex combination consisting only of local deterministic distributions, or \emph{b)} a convex combination of the same generalized PR box and the $4n$ local deterministic distributions that differ from the support of the PR box in exactly one location.
\end{theorem}

Theorems \ref{t:genonepr} and \ref{t:gen8dets} show us that in the $n$-fold chained Bell inequality scenario, any nonsignaling distribution matrix is either local (i.e., a convex combination of only local deterministic distributions) or can be expressed as a convex combination of a single generalized PR box and (up to) $4n$ one-support-mismatch LD distributions. 

Equation \eref{e:convexexp} and Table \ref{t:genexp} now have counterparts for the chained Bell scenario. This novel point of view makes new insights possible, as exemplified by the application presented in the next subsection.

\subsection{Bounding the Proportion of Local Theories in a Chained Bell Experiment}

In a 1992 paper \cite{elitzur:1992}, Elitzur, Popescu and Rohrlich asked the following question: if measurements on an ensemble of particle pairs manifest a violation of a Bell inequality, does this imply that each pair was behaving nonlocally, or only the ensemble as a whole? That is, could there be a model reproducing the quantum predictions in which some of the pairs behave locally? The authors concluded that for particles in a singlet state, each particle must indeed behave nonlocally.

This question can be re-formulated as the question of whether quantum distributions can be written as convex combinations of local and nonlocal distributions. In a 2006 paper \cite{barrett:2006}, Barrett, Kent, and Pironio demonstrated that the violation of a chained Bell inequality implies an upper bound on the possible proportion of local distributions in such a convex combination. Specifically, if we define $I_n$ be the expression on the left side of the chained Bell inequality
\begin{eqnarray}\label{e:barrettchained}
\sum_{x=1}^{n-1}\left[P(X_A\ne Y_B \mid a_xb_x)+P(X_A\ne Y_B \mid a_xb_{x+1})\right] \nonumber \\
+ P(X_A\ne Y_B \mid a_nb_n)+P(X_A = Y_B \mid a_nb_{1})\ge 1
\end{eqnarray}
(equation (5) in \cite{barrett:2006}), then for any distribution matrix yielding a value of $I_n$ less than 1, the proportion $p$ of local theories must obey the following bound:
\begin{equation}\label{e:barrettbound}
p\le I_n.
\end{equation}
In an ideal experiment, a maximally entangled state of two qubits can achieve values of $I_n$ arbitrarily close to zero as $n\to\infty$. As experimental imperfections will generally prevent this asymptotic behavior as $n$ increases, \cite{barrett:2006} suggests an experimental program to find the best (lowest) possible empirical values for this upper bound, which has subsequently been undertaken by numerous tests of chained Bell inequalities  \cite{pomarico:2011}, \cite{stuart:2012}, \cite{christensen:2015}.

One can ask if the bound \eref{e:barrettbound} is tight. That is, given a nonsignaling distribution matrix for an $n$-fold chained Bell outcome table that yields a value $I_n$ less than 1, is it ever possible to find a lower bound on $p$ than $I_n$, or will there always be a convex combination of local and nonlocal distributions with $p=I_n$ that can replicate the distribution matrix? Our results in the previous subsection allow us to resolve this question: the bound is indeed tight.

To see why, note first that all local deterministic distributions have to satisfy the inequality \eref{e:barrettchained}. Thus to violate \eref{e:barrettchained}, a nonsignaling nonlocal distribution -- which can always be expressed as a convex combination of local deterministic distributions and generalized PR boxes -- must contain some weight on a generalized PR box. Theorem \ref{t:genonepr} tells us that a nonlocal nonsignaling distribution containing weight on generalized PR boxes can always be expressed as a convex combination with weight on only one PR box. Furthermore it is straightforward to verify that the only generalized PR box that violates \eref{e:barrettchained} is the one with that has a single anticorrelated outcome row in $a_1b_n$:

\medskip

\begin{center}
\begin{tabular}{ r|c|c|c|c| }
\multicolumn{1}{r}{}
&  \multicolumn{1}{c}{++}
&  \multicolumn{1}{c}{+0}
&  \multicolumn{1}{c}{0+} 
&  \multicolumn{1}{c}{00}\\
\cline{2-5}
$a_1b_1$ & 1/2 & & & 1/2 \\
\cline{2-5}
$a_2b_1$ & 1/2 & & & 1/2 \\
\cline{2-5}
$a_2b_2$ & 1/2 & & & 1/2 \\
\cline{2-5}
$a_3b_2$ & 1/2 & & & 1/2 \\
\cline{2-5}
\multicolumn{1}{c}{}&\multicolumn{1}{c}{$\vdots$}&\multicolumn{1}{c}{$\vdots$}&\multicolumn{1}{c}{$\vdots$}&\multicolumn{1}{c}{$\vdots$}\\
\cline{2-5}
$a_{n}b_{n-1}$ & 1/2 & & & 1/2 \\
\cline{2-5}
$a_{n}b_{n}$ & 1/2 & & & 1/2 \\
\cline{2-5}
$a_1b_n$ & & 1/2 & 1/2 & \\
\cline{2-5}
\end{tabular}
\end{center}

\medskip

\noindent This is in fact the only generalized PR box that assigns zero probability to every term in $I_n$. This observation, combined with the results of the previous subsection, tells us that a nonsignaling distribution violating \eref{e:barrettchained} will necessarily be representable as a convex combination of the above generalized PR box and (up to) $4n$ local deterministic distributions. Indexing these local deterministic distributions with $i$, such a distribution matrix can be represented as 
\begin{equation}\label{e:localportion}
p_{PR}PR+\sum_{i=1}^{4n}p_iD_i.
\end{equation}
Now if we form the chained-Bell equivalent of Table \ref{t:genexp}, a judicious choice of indices in \eref{e:localportion} gives us the following outcome table:

\begin{center}
\begin{tabular}{ r|c|c|c|c| }
\multicolumn{1}{r}{}
&  \multicolumn{1}{c}{++}
&  \multicolumn{1}{c}{+0}
&  \multicolumn{1}{c}{0+} 
&  \multicolumn{1}{c}{00}\\
\cline{2-5}
$a_1b_1$ & $\ast$ & $p_1$ & $p_2$ & $\ast$ \\
\cline{2-5}
$a_2b_1$ & $\ast$ & $p_3$ & $p_4$ & $\ast$ \\
\cline{2-5}
$a_2b_2$ & $\ast$ & $p_5$ & $p_6$ & $\ast$ \\
\cline{2-5}
$a_3b_2$ &$\ast$ & $p_7$ & $p_8$ & $\ast$ \\
\cline{2-5}
\multicolumn{1}{c}{}&\multicolumn{1}{c}{$\vdots$}&\multicolumn{1}{c}{$\vdots$}&\multicolumn{1}{c}{$\vdots$}&\multicolumn{1}{c}{$\vdots$}\\
\cline{2-5}
$a_{n}b_{n-1}$ &$\ast$ & $p_{4n-5}$ & $p_{4n-4}$ & $\ast$ \\
\cline{2-5}
$a_{n}b_{n}$  &$\ast$ &  $p_{4n-3}$ & $p_{4n-2}$ & $\ast$ \\
\cline{2-5}
$a_1b_n$ & $p_{4n-1}$& $\ast$& $\ast$  & $p_{4n}$  \\
\cline{2-5}
\end{tabular}
\end{center}
Each ``$\ast$'' entry above represents a sum of a different subset of the $p_i$ terms and $p_{PR}/2$. For a distribution given by the above table, the value of $I_n$ is $\sum_{i=1}^{4n}p_i$. This is precisely equal to the weight of the local portion in the expression \eref{e:localportion}. We have thus proved a ``converse'' of the bound \eref{e:barrettbound}: if a nonsignaling distribution matrix violates \eref{e:barrettchained} with a $I_n$ value less than 1, then it is always possible to express the distribution as a convex combination of local and nonlocal distributions with weight $p=I_n$ on the local portion. As quantum distributions necessarily belong to the nonsignaling polytope, this result applies to all quantum distributions.

\section{Conclusion}\label{s:conclusion}

In this paper, the structure of non-signaling polytopes was studied in the $(2,2,2)$ scenario and the $(2,n,2)$ chained Bell scenario. A method for decomposing nonlocal nonsignaling distributions into convex combinations of local distributions and maximally nonlocal nonsignaling distributions was introduced, and the decomposition was shown to have useful practical applications. For the $(2,2,2)$ scenario, the decomposition allowed for a new method to analyze the effect of detection efficiency on nonlocality and extend the $\eta>2/3$ efficiency bound to general nonlocal nonsignaling distributions. The decomposition was also shown to be computationally useful for studying the statistical distance from a given nonlocal nonsignaling distribution to the set of local distributions for two commonly used distance measures: the total variation distance and Kullback-Leibler divergence. For the $(2,n,2)$ distribution, the nonsignaling polytope was classified and the resulting decomposition was used to demonstrate the optimality of the bound \eref{e:barrettbound} (from \cite{barrett:2006}) on the proportion of local theories in a local/nonlocal mixture that can be inferred from the experimental violation of a chained Bell inequality, provided that the violating distribution is non-signaling. 

In light of recent development of device-independent quantum protocols involving the $(2,2,2)$ scenario and $(2,n,2)$ chained Bell scenario (\cite{acin:2007}, \cite{pironio:2010}, \cite{barrett:2005}), the decomposition results presented here may prove useful for examining the information-theoretic potential for various quantum distributions. As it is the nonlocal component of a distribution that allows for the certification that a device-independent protocol is secure, the methods of this paper that isolate and reveal the nonlocal component may generate new ways to assess the information-theoretic utility of nonlocal distributions.

\ack{The author graciously thanks Scott Glancy and Manny Knill for helpful discussions of the material and detailed comments on the manuscript, and also thanks Yi-Kai Liu and Yanbao Zhang for additional helpful suggestions. This work is a contribution of the National Institute of Standards and Technology and is not subject to U.S. copyright.}

\section*{References}
\bibliographystyle{unsrt}
\bibliography{c:/Users/plb/Desktop/Everything/metabib}

\begin{thebibliography}{10}

\bibitem{barrett:2006}
J.~Barrett, A.~Kent, and S.~Pironio.
\newblock Maximally nonlocal and monogamous quantum correlations.
\newblock {\em Phys. Rev. Lett.}, 97:170409, Oct 2006.

\bibitem{BELL}
J.~Bell.
\newblock On the {E}instein {P}odolsky {R}osen paradox.
\newblock {\em Physics}, 1:195--200, 1964.

\bibitem{CHSH}
J.~Clauser, A.~Horne, A.~Shimony, and R.~Holt.
\newblock Proposed experiment to test local hidden-variable theories.
\newblock {\em Phys. Rev. Lett.}, 23(15):880--884, 1969.

\bibitem{CH74}
J.~Clauser and M.~Horne.
\newblock Experimental consequences of objective local theories.
\newblock {\em Phys. Rev. D}, 10(2):526--535, 1974.

\bibitem{eberhard:1993}
P.~H. Eberhard.
\newblock Background level and counter efficiencies required for a
  loophole-free {E}instein-{P}odolsky-{R}osen experiment.
\newblock {\em Phys. Rev. A}, 47:R747--R750, Feb 1993.

\bibitem{acin:2007}
Antonio Ac{\'\i}n, Nicolas Brunner, Nicolas Gisin, Serge Massar, Stefano
  Pironio, and Valerio Scarani.
\newblock Device-independent security of quantum cryptography against
  collective attacks.
\newblock {\em Phys. Rev. Lett.}, 98:230501, Jun 2007.

\bibitem{pironio:2010}
S.~Pironio et~al.
\newblock Random numbers certified by {B}ell's theorem.
\newblock {\em Nature}, 464:1021--4, 2010.

\bibitem{barrett:2005}
J.~Barrrett, N.~Linden, S.~Massar, S.~Pironio, S.~Popescu, and D.~Roberts.
\newblock Nonlocal correlations as an information-theoretic resource.
\newblock {\em Phys. Rev. A}, 71:022101, Feb 2005.

\bibitem{tsirelson:1993}
B.~S. Tsirelson.
\newblock Some results and problems on quantum {B}ell-type inequalities.
\newblock {\em Hadronic J. Suppl.}, 8:329, 1993.

\bibitem{PRBOX}
S.~Popescu and D.~Rohrlich.
\newblock Quantum nonlocality as an axiom.
\newblock {\em Found. Phys.}, 24(3):379--85, 1994.

\bibitem{lauritzen:2013}
N.~Lauritzen.
\newblock {\em Undergraduate Convexity}.
\newblock World Scientific, Singapore, 2013.

\bibitem{fine:1982}
Arthur Fine.
\newblock Hidden variables, joint probability, and the {B}ell inequalities.
\newblock {\em Phys. Rev. Lett.}, 48:291--295, Feb 1982.

\bibitem{pironio:2003}
S.~Pironio.
\newblock Violations of {B}ell inequalities as lower bounds on the
  communication cost of nonlocal correlations.
\newblock {\em Phys. Rev. A}, 68:062102, Dec 2003.

\bibitem{BBP}
Nicolas Brunner, Daniel Cavalcanti, Stefano Pironio, Valerio Scarani, and
  Stephanie Wehner.
\newblock Bell nonlocality.
\newblock {\em Rev. Mod. Phys.}, 86:419--78, Apr 2014.

\bibitem{gillvandam:2005}
W.~van Dam, R.~D. Gill, and P.~D. Grunwald.
\newblock The statistical strength of nonlocality proofs.
\newblock {\em IEEE T. Inform. Theory}, 51:2812--35, 2005.

\bibitem{zhang:2011}
Y.~Zhang, S.~Glancy, and E.~Knill.
\newblock Asymptotically optimal data analysis for rejecting local realism.
\newblock {\em Phys. Rev. A}, 84:062118, 2011.

\bibitem{zhang:2013}
Y.~Zhang, S.~Glancy, and E.~Knill.
\newblock Efficient quantification of experimental evidence against local
  realism.
\newblock {\em Phys. Rev. A}, 88:052119, Nov 2013.

\bibitem{pearle:1970}
Philip~M. Pearle.
\newblock Hidden-variable example based upon data rejection.
\newblock {\em Phys. Rev. D}, 2:1418--1425, Oct 1970.

\bibitem{CHAINED}
S.~Braunstein and C.~Caves.
\newblock Wringing out better {B}ell inequalities.
\newblock {\em Ann. Phys. (Berlin)}, 202:22--56, Aug 1990.

\bibitem{BHK}
J.~Barrett, L.~Hardy, and A.~Kent.
\newblock No signaling and quantum key distribution.
\newblock {\em Phys. Rev. Lett.}, 95:010503, 2005.

\bibitem{colbeck:2012}
R.~Colbeck and R.~Renner.
\newblock Free randomness can be amplified.
\newblock {\em Nature Physics}, 8:450--453, 2012.

\bibitem{shalm:2015}
L.~K. Shalm et~al.
\newblock Strong loophole-free test of local realism.
\newblock {\em Phys. Rev. Lett.}, 115:250402, Dec 2015.

\bibitem{giustina:2013}
M.~Giustina et~al.
\newblock Bell violation using entangled photons without the fair-sampling
  assumption.
\newblock {\em Nature}, 497:227--30, 2013.

\bibitem{bierhorst:2015}
P.~Bierhorst.
\newblock A robust mathematical model for a loophole-free {C}lauser–{H}orne
  experiment.
\newblock {\em J. Phys. A: Math. Theor.}, 48(19):195302, 2015.

\bibitem{giustina:2015}
M.~Giustina et~al.
\newblock Significant-loophole-free test of bell's theorem with entangled
  photons.
\newblock {\em Phys. Rev. Lett.}, 115:250401, Dec 2015.

\bibitem{christensen:2013}
B.~G. Christensen et~al.
\newblock Detection-loophole-free test of quantum nonlocality, and
  applications.
\newblock {\em Phys. Rev. Lett.}, 111:130406, Sep 2013.

\bibitem{wilms:2008}
J.~Wilms, Y.~Disser, G.~Alber, and I.~C. Percival.
\newblock Local realism, detection efficiencies, and probability polytopes.
\newblock {\em Phys. Rev. A}, 78:032116, Sep 2008.

\bibitem{massar:2003}
S.~Massar and S.~Pironio.
\newblock Violation of local realism versus detection efficiency.
\newblock {\em Phys. Rev. A}, 68:062109, Dec 2003.

\bibitem{branciard:2011}
C.~Branciard.
\newblock Detection loophole in {B}ell experiments: how postselection modifies
  the requirements to observe nonlocality.
\newblock {\em Phys. Rev. A}, 83:032123, Mar 2011.

\bibitem{jones:2005}
N.~Jones and L.~Masanes.
\newblock Interconversion of nonlocal correlations.
\newblock {\em Phys. Rev. A}, 72:052312, Nov 2005.

\bibitem{elitzur:1992}
A.~Elitzur, S.~Popescu, and D.~Rohrlich.
\newblock Quantum nonlocality for each pair in an ensemble.
\newblock {\em Phys. Lett. A}, 162:25--8, January 1992.

\bibitem{pomarico:2011}
E.~Pomarico, J.-D. Bancal, B.~Sanguinetti, A.~Rochdi, and N.~Gisin.
\newblock Various quantum nonlocality tests with a commercial two-photon
  entanglement source.
\newblock {\em Phys. Rev. A}, 83:052104, May 2011.

\bibitem{stuart:2012}
T.~E. Stuart, J.~A. Slater, R.~Colbeck, R.~Renner, and W.~Tittel.
\newblock Experimental bound on the maximum predictive power of physical
  theories.
\newblock {\em Phys. Rev. Lett.}, 109:020402, Jul 2012.

\bibitem{christensen:2015}
B.~G. Christensen, Y.-C. Liang, N.~Brunner, N.~Gisin, and P.~G. Kwiat.
\newblock Exploring the limits of quantum nonlocality with entangled photons.
\newblock {\em Phys. Rev. X}, 5:041052, Dec 2015.

\end{thebibliography}

\appendix

\newpage

\section{Extremal Points of the $(2,2,2)$ Polytope}\label{s:prsdets}

\begin{table}[h!]\caption{The PR boxes. Empty cells contain 0 probability.}\label{t:prboxes}
\begin{center}
\begin{tabular}{ r|c|c|c|c| }
\multicolumn{1}{c}{} & \multicolumn{4}{c}{1}\\
 \multicolumn{1}{r}{}
  &  \multicolumn{1}{c}{++}
 &  \multicolumn{1}{c}{+0}
 &  \multicolumn{1}{c}{0+} 
 &  \multicolumn{1}{c}{00}\\
 \cline{2-5}
 $ab$ & 1/2 &  &  & 1/2 \\
 \cline{2-5}
 $ab'$ & 1/2  &  &  & 1/2 \\
 \cline{2-5}
 $a'b$ & 1/2  &  &  & 1/2  \\
 \cline{2-5}
 $a'b'$ &  & 1/2 & 1/2 &  \\
 \cline{2-5}
 \end{tabular}
 \hspace{1cm}
 \begin{tabular}{ r|c|c|c|c| }
 \multicolumn{1}{c}{} & \multicolumn{4}{c}{2}\\
 \multicolumn{1}{r}{}
  &  \multicolumn{1}{c}{++}
 &  \multicolumn{1}{c}{+0}
 &  \multicolumn{1}{c}{0+} 
 &  \multicolumn{1}{c}{00}\\
 \cline{2-5}
 $ab$  &   & 1/2 & 1/2 & \\
 \cline{2-5}
 $ab'$ &  & 1/2 & 1/2  & \\
 \cline{2-5}
 $a'b$ &  & 1/2 & 1/2 &  \\
 \cline{2-5}
 $a'b'$ & 1/2 & &  & 1/2 \\
 \cline{2-5}
 \end{tabular}
 
 \vspace{1cm}

\begin{tabular}{ r|c|c|c|c| }
 \multicolumn{1}{c}{} & \multicolumn{4}{c}{3}\\
 \multicolumn{1}{r}{}
  &  \multicolumn{1}{c}{++}
 &  \multicolumn{1}{c}{+0}
 &  \multicolumn{1}{c}{0+} 
 &  \multicolumn{1}{c}{00}\\
 \cline{2-5}
 $ab$ & 1/2 &  &  & 1/2 \\
 \cline{2-5}
 $ab'$ & 1/2  &  &  & 1/2 \\
 \cline{2-5}
 $a'b$ &  & 1/2 & 1/2 &  \\
 \cline{2-5}
 $a'b'$ & 1/2 &  &  & 1/2 \\
 \cline{2-5}
 \end{tabular}
 \hspace{1cm}
 \begin{tabular}{ r|c|c|c|c| }
  \multicolumn{1}{c}{} & \multicolumn{4}{c}{4}\\
 \multicolumn{1}{r}{}
  &  \multicolumn{1}{c}{++}
 &  \multicolumn{1}{c}{+0}
 &  \multicolumn{1}{c}{0+} 
 &  \multicolumn{1}{c}{00}\\
 \cline{2-5}
 $ab$  &   & 1/2 & 1/2 & \\
 \cline{2-5}
 $ab'$ &  & 1/2 & 1/2  & \\
 \cline{2-5}
 $a'b$ & 1/2 &  &  & 1/2 \\
 \cline{2-5}
 $a'b'$ & & 1/2 & 1/2 &  \\
 \cline{2-5}
 \end{tabular}

\vspace{1cm}

\begin{tabular}{ r|c|c|c|c| }
 \multicolumn{1}{c}{} & \multicolumn{4}{c}{5}\\
 \multicolumn{1}{r}{}
  &  \multicolumn{1}{c}{++}
 &  \multicolumn{1}{c}{+0}
 &  \multicolumn{1}{c}{0+} 
 &  \multicolumn{1}{c}{00}\\
 \cline{2-5}
 $ab$ & 1/2 &  &  & 1/2 \\
 \cline{2-5}
 $ab'$ &  & 1/2 & 1/2 &  \\
 \cline{2-5}
 $a'b$ & 1/2  &  &  & 1/2  \\
 \cline{2-5}
 $a'b'$ & 1/2 &  &  & 1/2 \\
 \cline{2-5}
 \end{tabular}
 \hspace{1cm}
 \begin{tabular}{ r|c|c|c|c| }
  \multicolumn{1}{c}{} & \multicolumn{4}{c}{6}\\
 \multicolumn{1}{r}{}
  &  \multicolumn{1}{c}{++}
 &  \multicolumn{1}{c}{+0}
 &  \multicolumn{1}{c}{0+} 
 &  \multicolumn{1}{c}{00}\\
 \cline{2-5}
 $ab$  &   & 1/2 & 1/2 &  \\
 \cline{2-5}
 $ab'$ & 1/2 &  &   & 1/2 \\
 \cline{2-5}
 $a'b$ &  & 1/2 & 1/2 &  \\
 \cline{2-5}
 $a'b'$ &  & 1/2 & 1/2 & \\
 \cline{2-5}
 \end{tabular}

\vspace{1cm}

\begin{tabular}{ r|c|c|c|c| }
 \multicolumn{1}{c}{} & \multicolumn{4}{c}{7}\\
 \multicolumn{1}{r}{}
  &  \multicolumn{1}{c}{++}
 &  \multicolumn{1}{c}{+0}
 &  \multicolumn{1}{c}{0+} 
 &  \multicolumn{1}{c}{00}\\
 \cline{2-5}
 $ab$ &  & 1/2 & 1/2 &  \\
 \cline{2-5}
 $ab'$ & 1/2  &  &  & 1/2 \\
 \cline{2-5}
 $a'b$ & 1/2  &  &  & 1/2  \\
 \cline{2-5}
 $a'b'$ & 1/2 & & & 1/2 \\
 \cline{2-5}
 \end{tabular}
 \hspace{1cm}
 \begin{tabular}{ r|c|c|c|c| }
  \multicolumn{1}{c}{} & \multicolumn{4}{c}{8}\\
 \multicolumn{1}{r}{}
  &  \multicolumn{1}{c}{++}
 &  \multicolumn{1}{c}{+0}
 &  \multicolumn{1}{c}{0+} 
 &  \multicolumn{1}{c}{00}\\
 \cline{2-5}
 $ab$  & 1/2  &  &  & 1/2 \\
 \cline{2-5}
 $ab'$ &  & 1/2 & 1/2  & \\
 \cline{2-5}
 $a'b$ &  & 1/2 & 1/2 &  \\
 \cline{2-5}
 $a'b'$ &  & 1/2 & 1/2 & \\
 \cline{2-5}
 \end{tabular}
\end{center}
\end{table}

\begin{table}[p]\caption{The Local Deterministic Distributions}\label{t:localdets}
\begin{center}
{\scriptsize
\begin{tabular}{ r|c|c|c|c| }
\multicolumn{1}{c}{} & \multicolumn{4}{c}{1}\\
 \multicolumn{1}{r}{}
  &  \multicolumn{1}{c}{++}
 &  \multicolumn{1}{c}{+0}
 &  \multicolumn{1}{c}{0+} 
 &  \multicolumn{1}{c}{00}\\
 \cline{2-5}
 $ab$ & 1 &  &  & \\
 \cline{2-5}
 $ab'$ & 1 &  &  & \\
 \cline{2-5}
 $a'b$ & 1 &  &  &  \\
 \cline{2-5}
 $a'b'$ & 1 &  &  &   \\
 \cline{2-5}
 \end{tabular}
 \hspace{1cm}
\begin{tabular}{ r|c|c|c|c| }
\multicolumn{1}{c}{} & \multicolumn{4}{c}{2}\\
 \multicolumn{1}{r}{}
  &  \multicolumn{1}{c}{++}
 &  \multicolumn{1}{c}{+0}
 &  \multicolumn{1}{c}{0+} 
 &  \multicolumn{1}{c}{00}\\
 \cline{2-5}
 $ab$ &  & 1 &  & \\
 \cline{2-5}
 $ab'$ &  & 1 &  & \\
 \cline{2-5}
 $a'b$ &  & 1 &  &  \\
 \cline{2-5}
 $a'b'$ &  & 1 &  &   \\
 \cline{2-5}
 \end{tabular}

\medskip\medskip

\begin{tabular}{ r|c|c|c|c| }
\multicolumn{1}{c}{} & \multicolumn{4}{c}{3}\\
 \multicolumn{1}{r}{}
  &  \multicolumn{1}{c}{++}
 &  \multicolumn{1}{c}{+0}
 &  \multicolumn{1}{c}{0+} 
 &  \multicolumn{1}{c}{00}\\
 \cline{2-5}
 $ab$ &  &  & 1 & \\
 \cline{2-5}
 $ab'$ &  &  & 1 & \\
 \cline{2-5}
 $a'b$ &  &  & 1 &  \\
 \cline{2-5}
 $a'b'$ &  &  & 1  &   \\
 \cline{2-5}
 \end{tabular}
 \hspace{1cm}
 \begin{tabular}{ r|c|c|c|c| }
\multicolumn{1}{c}{} & \multicolumn{4}{c}{4}\\
 \multicolumn{1}{r}{}
  &  \multicolumn{1}{c}{++}
 &  \multicolumn{1}{c}{+0}
 &  \multicolumn{1}{c}{0+} 
 &  \multicolumn{1}{c}{00}\\
 \cline{2-5}
 $ab$ &  &  &  & 1\\
 \cline{2-5}
 $ab'$ &  &  &  & 1\\
 \cline{2-5}
 $a'b$ &  &  &  &  1\\
 \cline{2-5}
 $a'b'$ &  &  &  &  1 \\
 \cline{2-5}
 \end{tabular}

\medskip\medskip

\begin{tabular}{ r|c|c|c|c| }
\multicolumn{1}{c}{} & \multicolumn{4}{c}{5}\\
 \multicolumn{1}{r}{}
  &  \multicolumn{1}{c}{++}
 &  \multicolumn{1}{c}{+0}
 &  \multicolumn{1}{c}{0+} 
 &  \multicolumn{1}{c}{00}\\
 \cline{2-5}
 $ab$ & 1 &  &  & \\
 \cline{2-5}
 $ab'$ &  & 1 &  & \\
 \cline{2-5}
 $a'b$ & 1 &  &  &  \\
 \cline{2-5}
 $a'b'$ &  & 1 &  &   \\
 \cline{2-5}
 \end{tabular}
 \hspace{1cm}
 \begin{tabular}{ r|c|c|c|c| }
\multicolumn{1}{c}{} & \multicolumn{4}{c}{6}\\
 \multicolumn{1}{r}{}
  &  \multicolumn{1}{c}{++}
 &  \multicolumn{1}{c}{+0}
 &  \multicolumn{1}{c}{0+} 
 &  \multicolumn{1}{c}{00}\\
 \cline{2-5}
 $ab$ &  & 1 &  & \\
 \cline{2-5}
 $ab'$ & 1 &  &  & \\
 \cline{2-5}
 $a'b$ &  & 1 &  &  \\
 \cline{2-5}
 $a'b'$ & 1 &  &  &   \\
 \cline{2-5}
 \end{tabular}

\medskip\medskip

\begin{tabular}{ r|c|c|c|c| }
\multicolumn{1}{c}{} & \multicolumn{4}{c}{7}\\
 \multicolumn{1}{r}{}
  &  \multicolumn{1}{c}{++}
 &  \multicolumn{1}{c}{+0}
 &  \multicolumn{1}{c}{0+} 
 &  \multicolumn{1}{c}{00}\\
 \cline{2-5}
 $ab$ &  &  & 1 & \\
 \cline{2-5}
 $ab'$ &  &  &  & 1\\
 \cline{2-5}
 $a'b$ &  &  & 1 &  \\
 \cline{2-5}
 $a'b'$ &  &  &  &  1 \\
 \cline{2-5}
 \end{tabular}
 \hspace{1cm}
 \begin{tabular}{ r|c|c|c|c| }
\multicolumn{1}{c}{} & \multicolumn{4}{c}{8}\\
 \multicolumn{1}{r}{}
  &  \multicolumn{1}{c}{++}
 &  \multicolumn{1}{c}{+0}
 &  \multicolumn{1}{c}{0+} 
 &  \multicolumn{1}{c}{00}\\
 \cline{2-5}
 $ab$ &  &  &  &1 \\
 \cline{2-5}
 $ab'$ &  &  & 1 & \\
 \cline{2-5}
 $a'b$ &  &  &  & 1 \\
 \cline{2-5}
 $a'b'$ &  &  & 1 &   \\
 \cline{2-5}
 \end{tabular}
 
\medskip\medskip
 
\begin{tabular}{ r|c|c|c|c| }
\multicolumn{1}{c}{} & \multicolumn{4}{c}{9}\\
 \multicolumn{1}{r}{}
  &  \multicolumn{1}{c}{++}
 &  \multicolumn{1}{c}{+0}
 &  \multicolumn{1}{c}{0+} 
 &  \multicolumn{1}{c}{00}\\
 \cline{2-5}
 $ab$ & 1 &  &  & \\
 \cline{2-5}
 $ab'$ & 1 &  &  & \\
 \cline{2-5}
 $a'b$ &  &  & 1 &  \\
 \cline{2-5}
 $a'b'$ &  &  & 1 &   \\
 \cline{2-5}
 \end{tabular}
 \hspace{1cm}
 \begin{tabular}{ r|c|c|c|c| }
\multicolumn{1}{c}{} & \multicolumn{4}{c}{10}\\
 \multicolumn{1}{r}{}
  &  \multicolumn{1}{c}{++}
 &  \multicolumn{1}{c}{+0}
 &  \multicolumn{1}{c}{0+} 
 &  \multicolumn{1}{c}{00}\\
 \cline{2-5}
 $ab$ &  & 1 &  & \\
 \cline{2-5}
 $ab'$ &  & 1 &  & \\
 \cline{2-5}
 $a'b$ &  &  &  & 1 \\
 \cline{2-5}
 $a'b'$ &  &  &  &  1 \\
 \cline{2-5}
 \end{tabular}

\medskip\medskip

\begin{tabular}{ r|c|c|c|c| }
\multicolumn{1}{c}{} & \multicolumn{4}{c}{11}\\
 \multicolumn{1}{r}{}
  &  \multicolumn{1}{c}{++}
 &  \multicolumn{1}{c}{+0}
 &  \multicolumn{1}{c}{0+} 
 &  \multicolumn{1}{c}{00}\\
 \cline{2-5}
 $ab$ &  &  & 1 & \\
 \cline{2-5}
 $ab'$ &  &  & 1 & \\
 \cline{2-5}
 $a'b$ & 1 &  &  &  \\
 \cline{2-5}
 $a'b'$ & 1 &  &  &   \\
 \cline{2-5}
 \end{tabular}
 \hspace{1cm}
 \begin{tabular}{ r|c|c|c|c| }
\multicolumn{1}{c}{} & \multicolumn{4}{c}{12}\\
 \multicolumn{1}{r}{}
  &  \multicolumn{1}{c}{++}
 &  \multicolumn{1}{c}{+0}
 &  \multicolumn{1}{c}{0+} 
 &  \multicolumn{1}{c}{00}\\
 \cline{2-5}
 $ab$ &  &  &  & 1\\
 \cline{2-5}
 $ab'$ &  &  &  & 1\\
 \cline{2-5}
 $a'b$ &  & 1 &  &  \\
 \cline{2-5}
 $a'b'$ &  & 1 &  &   \\
 \cline{2-5}
 \end{tabular}

\medskip\medskip

\begin{tabular}{ r|c|c|c|c| }
\multicolumn{1}{c}{} & \multicolumn{4}{c}{13}\\
 \multicolumn{1}{r}{}
  &  \multicolumn{1}{c}{++}
 &  \multicolumn{1}{c}{+0}
 &  \multicolumn{1}{c}{0+} 
 &  \multicolumn{1}{c}{00}\\
 \cline{2-5}
 $ab$ & 1 &  &  & \\
 \cline{2-5}
 $ab'$ &  & 1 &  & \\
 \cline{2-5}
 $a'b$ &  &  & 1 &  \\
 \cline{2-5}
 $a'b'$ &  &  &  & 1  \\
 \cline{2-5}
 \end{tabular}
 \hspace{1cm}
 \begin{tabular}{ r|c|c|c|c| }
\multicolumn{1}{c}{} & \multicolumn{4}{c}{14}\\
 \multicolumn{1}{r}{}
  &  \multicolumn{1}{c}{++}
 &  \multicolumn{1}{c}{+0}
 &  \multicolumn{1}{c}{0+} 
 &  \multicolumn{1}{c}{00}\\
 \cline{2-5}
 $ab$ &  & 1 &  & \\
 \cline{2-5}
 $ab'$ & 1 &  &  & \\
 \cline{2-5}
 $a'b$ &  &  &  & 1 \\
 \cline{2-5}
 $a'b'$ &  &  & 1 &   \\
 \cline{2-5}
 \end{tabular}

\medskip\medskip

 \begin{tabular}{ r|c|c|c|c| }
\multicolumn{1}{c}{} & \multicolumn{4}{c}{15}\\
 \multicolumn{1}{r}{}
  &  \multicolumn{1}{c}{++}
 &  \multicolumn{1}{c}{+0}
 &  \multicolumn{1}{c}{0+} 
 &  \multicolumn{1}{c}{00}\\
 \cline{2-5}
 $ab$ &  &  & 1 & \\
 \cline{2-5}
 $ab'$ &  &  &  & 1\\
 \cline{2-5}
 $a'b$ & 1  &  &  &  \\
 \cline{2-5}
 $a'b'$ &  & 1 &  &   \\
 \cline{2-5}
 \end{tabular}
 \hspace{1cm}
  \begin{tabular}{ r|c|c|c|c| }
\multicolumn{1}{c}{} & \multicolumn{4}{c}{16}\\
 \multicolumn{1}{r}{}
  &  \multicolumn{1}{c}{++}
 &  \multicolumn{1}{c}{+0}
 &  \multicolumn{1}{c}{0+} 
 &  \multicolumn{1}{c}{00}\\
 \cline{2-5}
 $ab$ &  &  &  & 1 \\
 \cline{2-5}
 $ab'$ &  &  & 1 & \\
 \cline{2-5}
 $a'b$ &  & 1 &  &  \\
 \cline{2-5}
 $a'b'$ & 1 &  &  &   \\
 \cline{2-5}
 \end{tabular}
 }
\end{center}
\end{table}

\newpage

\section{Proof of Theorem \ref{t:vardistheorem}}\label{s:vardistproof}

{\bf Theorem \ref{t:vardistheorem}} \emph{Let a distribution $Q$ be nonsignaling and violate the CHSH inequality \eref{e:CHSH}, so that it can be expressed in the 1 PR + 8 LD form \eref{e:convexexp}. Then a local distribution $S$ achieves the minimum possible $\delta(Q,S)$ for local distributions if it is of the form} 

\medskip

$
\hspace{1cm} 
s_1D_1+ s_4D_4+ s_5D_5+ s_8D_8+ s_9D_9+ s_{12}D_{12}+ s_{14}D_{14}+ s_{15}D_{15}\hfill \eref{e:Sexp}
$

\medskip

\noindent\emph{where each $s_i$ is equal to $p_i+\frac{p_{PR}}{8}$.}

\medskip

\noindent \emph{Proof.} We note that by Lemma \ref{l:straightlinelemma}, it is sufficient to consider only distributions that are convex combinations of the CHSH-saturating LD distributions. The lemma applies here because total variation distance to a target distribution $Q$, as defined in \eref{e:totvardist}, clearly decreases if we follow a straight line path from any fixed distribution $S$ towards the target distribution. Hence our problem reduces to the problem of minimizing $\delta(Q,S)$ where $Q$ is a fixed distribution in 1 PR + 8 LD form and $S$ is any convex combination of CHSH-saturating distributions: 
\begin{eqnarray}
\fl  Q= p_1D_1+ p_4D_4+ p_5D_5+ p_8D_8+ p_9D_9+ p_{12}D_{12}+ p_{14}D_{14}+ p_{15}D_{15}+p_{PR}PR_1\nonumber\\
\fl S= s_1D_1+ s_4D_4+ s_5D_5+ s_8D_8+ s_9D_9+ s_{12}D_{12}+ s_{14}D_{14}+ s_{15}D_{15},\label{e:Sdefi}
\end{eqnarray}
where the $p_i,s_j$ coefficients are nonnegative and $p_{PR}+\sum_ip_i=\sum_js_j=1$. Recalling Table \ref{t:genexp}, we can see that the distribution matrices of $Q$ and $S$ are given respectively by the following tables:
\begin{center}
{\footnotesize\begin{tabular}{ r|c|c|c|c| }
 \multicolumn{1}{r}{}
  &  \multicolumn{1}{c}{++}
 &  \multicolumn{1}{c}{+0}
 &  \multicolumn{1}{c}{0+} 
 &  \multicolumn{1}{c}{00}\\
  \cline{2-5}
$ab$ & $p_1+p_5+p_9+\frac{1}{2}p_{PR}$ & $p_{14}$ & $p_{15}$ & $p_4+p_8+p_{12}+\frac{1}{2}p_{PR}$\\
  \cline{2-5}
$ab'$ & $p_1+p_9+p_{14}+\frac{1}{2}p_{PR}$ & $p_{5}$ & $p_{8}$ & $p_4+p_{12}+p_{15}+\frac{1}{2}p_{PR}$\\
  \cline{2-5}
$a'b$ & $p_1+p_5+p_{15}+\frac{1}{2}p_{PR}$ & $p_{12}$ & $p_{9}$ & $p_4+p_{8}+p_{14}+\frac{1}{2}p_{PR}$\\
  \cline{2-5}
$a'b'$ & $p_{1}$ & $p_5+p_{12}+p_{15}+\frac{1}{2}p_{PR}$ & $p_8+p_{9}+p_{14}+\frac{1}{2}p_{PR}$ & $p_{4}$\\
  \cline{2-5}
\end{tabular}

\medskip

\begin{tabular}{ r|c|c|c|c| }
 \multicolumn{1}{r}{}
  &  \multicolumn{1}{c}{++}
 &  \multicolumn{1}{c}{+0}
 &  \multicolumn{1}{c}{0+} 
 &  \multicolumn{1}{c}{00}\\
  \cline{2-5}
$ab$ & $s_1+s_5+s_9$ & $s_{14}$ & $s_{15}$ & $s_4+s_8+s_{12}$\\
  \cline{2-5}
$ab'$ & $s_1+s_9+s_{14}$ & $s_{5}$ & $s_{8}$ & $s_4+s_{12}+s_{15}$\\
  \cline{2-5}
$a'b$ & $s_1+s_5+s_{15}$ & $s_{12}$ & $s_{9}$ & $s_4+s_{8}+s_{14}$\\
  \cline{2-5}
$a'b'$ & $s_{1}$ & $s_5+s_{12}+s_{15}$ & $s_8+s_{9}+s_{14}$ & $s_{4}$\\
  \cline{2-5}
\end{tabular}}
\end{center}
  
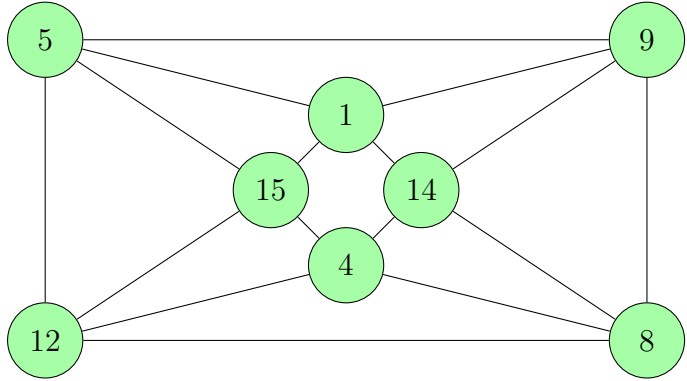
\begin{figure}[h]\centering\caption{Scheme to organize terms of $\delta(Q,S)$}\label{f:scheme2}
\begin{tikzpicture}[scale=.5]

 \node [style=circle,draw,minimum size=1cm,fill=green!35,]  (1) at (8,-2)  {$1$};
 \node [style=circle,draw,minimum size=1cm,fill=green!35,] (15) at (6,-4) {$15$};
 \node [style=circle,draw,minimum size=1cm,fill=green!35,](4) at (8,-6) {$4$};
 \node [style=circle,draw,minimum size=1cm,fill=green!35,](14) at (10,-4) {$14$};
 \node [style=circle,draw,minimum size=1cm,fill=green!35,] (5) at (0,0)  {$5$};
 \node [style=circle,draw,minimum size=1cm,fill=green!35,] (12) at (0,-8)  {$12$};
 \node [style=circle,draw,minimum size=1cm,fill=green!35,](9) at (16,0)  {$9$};
 \node [style=circle,draw,minimum size=1cm,fill=green!35,] (8) at (16,-8)  {$8$};
  
  \draw[-] (1)--(14);
  \draw[-] (1)--(9);
  \draw[-] (14)--(9);
  \draw[-] (5)--(9);
  \draw[-] (8)--(9);
  \draw[-] (14)--(8);
  \draw[-] (4)--(8);
  \draw[-] (12)--(8);
  \draw[-] (12)--(4);
  \draw[-] (12)--(15);
  \draw[-] (12)--(5);  
  \draw[-] (1)--(15);
  \draw[-] (4)--(15);
  \draw[-] (4)--(14);
  \draw[-] (1)--(5);
  \draw[-] (5)--(15);
      
\end{tikzpicture} 
\end{figure}

\noindent Thus we can see that $\delta(Q,S)$ as defined in \eref{e:totvardist} consists of a sum of eight terms of the form $|p_1-s_1|, |p_{14}-s_{14}|,...$ and eight more-complicated terms of the form $|p_1+p_5+p_9+\frac{1}{2}p_{PR} -(s_1+s_5+s_9)|$,... . The 16 terms have interrelationships captured by Figure \ref{f:scheme2} (which is similar to Figure \ref{f:scheme}).
Figure \ref{f:scheme2} organizes the sixteen terms that make up $\delta(Q,S)$ in the following manner: each simple term of the form $|p_i-s_i|$ corresponds to a single circular node of Figure \ref{f:scheme2}, and each of the more complicated terms corresponds to a fixed triangle of Figure \ref{f:scheme2}, where the three corners of the triangle contain the three indexes occurring in the corresponding term of $\delta(Q,S)$. So for instance, the term $|p_1+p_5+p_9+\frac{1}{2}p_{PR} -(s_1+s_5+s_9)|$ is represented by the top triangle. We will notate the triangle terms like $|p_1+p_5+p_9+\frac{1}{2}p_{PR} -(s_1+s_5+s_9)|$ as $\triangle_{1,5,9}$, etc., and refer to the simpler terms $|p_1-s_1|$ as $\delta_1$, etc. This will make it easier to keep track of the overlapping codependencies between all of the terms.

Our first step is to justify the following claim: 

\noindent \emph{Claim}: Given an $S$ of the form \eref{e:Sdefi}, if $s_k<p_k$ for a particular choice of $k$, we can always find a new distribution $S'$ that a) is at least as close to $Q$ as $S$ is, and b) has $s'_k=p_k$ and no $j$ for which $s'_j<p_j$ but $s_j\ge p_j$. 

This will show that it is sufficient to consider only the distributions \eref{e:Sdefi} for which each $s_i$ is at least as large as its corresponding $p_i$. We prove the claim explicitly only for $s_k=s_1$, but it should become clear as we work through the argument that it would work for any choice of $s_k$ due to the symmetry of Figure \ref{f:scheme2}. So suppose we have a distribution $S$ induced by a convex combination \eref{e:Sdefi} for which $s_1<p_1$. This implies that there must be some $j$ for which $s_j > p_j$, because the coefficients are nonnegative and $\sum_i s_i = p_{PR}+ \sum_i p_i = 1$. Based on a case analysis of which $s_j > p_j$, we now provide an algorithm that results in a modified distribution $S'$ satisfying the conditions outlined above.

First, suppose that $s_5 > p_5$. Then if we increase $s_1$ and decrease $s_5$ while leaving all the other $s_i$ unchanged, this will decrease $\delta_1$ and $\delta_5$, this will leave unchanged $\triangle_{1,5,9}$ and $\triangle_{1,5,15}$, and this will have an effect (possibly positive, possibly negative) on $\triangle_{1,9,14}$ and $\triangle_{5,12,15}$. No other terms of $\delta(Q,S)$ will change. At worst, the effect of this action for $\triangle_{1,9,14}$ and $\triangle_{5,12,15}$ will be to increase these terms, but even in this worst-case scenario, $\delta(Q,S)$ remains unchanged because the magnitude of the increase is equal to the magnitude of the decrease of $\delta_1$ and $\delta_5$ (this is easily checked).

So if $s_5 > p_5$, we will take this action until either $s_1=p_1$ or $s_5=p_5$, whichever comes first. If $s_5=p_5$ occurs first, we then test to see if $s_9 > p_9$; if this is true, we can continue to increase $s_1$ by decreasing $s_9$; this will have (at worst) a neutral effect on $\delta(Q,S)$ by an argument symmetric to the one given for $s_5$.

If taking these actions is sufficient to modify $S$ so that $s_1=p_1$, we are done. So suppose instead that we find ourselves in a situation where $s_1<p_1$ but $s_5\le p_5$ and $s_9\le p_9$, either because we have taken all available action to increase $s_5$ and $s_9$ and $s_1$ is still less than $p_1$, or because neither $s_5$ nor $s_9$ were bigger than $p_5$ and $p_9$ to begin with. 

Suppose now that $s_{14}>p_{14}$. (Note that this is similar to considering $s_{15}>p_{15}$). If we increase $s_1$ while decreasing $s_{14}$, we will decrease $\delta_1$ and $\delta_{14}$, we will also decrease $\triangle_{1,5,9}$ (recall our current assumption that $s_5\le p_5$ and $s_9\le p_9$), we will leave unchanged $\triangle_{1,9,14}$, and we will have an unknown effect on $\triangle_{4,8,14}$, $\triangle_{8,9,14}$, and $\triangle_{1,5,15}$. Hence three terms will decrease and at most three will increase. As the absolute values of each of the six changes will all be equal (easily checked), the net effect will be to either decrease $\delta(Q,S)$ or to leave it unchanged.

So if $s_{14}>p_{14}$ or $s_{15}>p_{15}$, we can increase $s_1$ by decreasing these terms. Suppose now that a) we do this until we run out of excess and that $s_1$ is still less than $p_1$, or b) from the start, no element of the set $\{s_5,s_9,s_{14},s_{15}\}$ exceeded its $p_i$ counterpart. We then find ourselves in the situation where at least one of the coefficients $s_4$, $s_8$, or $s_{12}$ must exceed its $p_i$ counterpart (because, recall, all of the coefficients sum to 1, and $s_1<p_1$). Now, decreasing any of these three coefficients will decrease its $\delta_i$ term while potentially increasing three adjacent ``$\triangle$'' terms. However, we now can say that increasing $s_1$ will decrease \emph{all} of the constraints $\delta_1$, $\triangle_{1,5,9}$, $\triangle_{1,5,15}$, and $\triangle_{1,9,14}$, because by assumption no element of the set $\{s_5,s_9,s_{14},s_{15}\}$ exceeds its $p_i$ counterpart. So in this circumstance, we would always improve $\delta(Q,S)$ by increasing $s_1$ while decreasing whichever of $s_4$, $s_8$, or $s_{12}$ exceeds the corresponding $p_4$, $p_8$, or $p_{12}$. 

This covers all of the $s_j>p_j$ contingencies for the algorithm, so we have proved the claim. Since the claim is true, we see that it is sufficient to consider only $S$ for which all $s_i\ge p_i$ while searching for a distribution that minimizes $\delta(Q,S)$. Let us consider the set of all such $S$ satisfying $s_i \ge p_i$ and denote it by $\mathcal S$. Our arguments so far show that to find $\min_{S\in local}\delta(Q,S)$, it is sufficient to find $\min_{S\in \mathcal S}\delta(Q,S)$.

To complete the proof, consider now the sub-distribution $S^s$ that is obtained by setting $s^s_i=p_i$ (i.e., just removing the $p_{PR}$ coefficient and keeping what remains). While $S^s$ is not a probability distribution, we can still consider the value of the function $\delta(S^s,P)$ (which will be, we note, $2p_{PR}$). Now for any $S\in \mathcal S$, we can lower-bound $\delta(Q,S)$ by first calculating $\delta(Q,S^s)$ and then seeing how $\delta(Q,S^s)$ changes as we transform $S^s$ into $S$, if we consider the following observations. We first note that $s_1$ must be greater than or equal to $s^s_1$, so if we modify $S^*$ by replacing $s^s_1=p_1$ by $s_1$, then the value of $\delta(Q,S^s)$ changes in four ways: $\delta_1$ always increases (``bad'' for $\delta(Q,S^s)$), and three $\triangle$ components (initially) decrease (``good'' for $\delta(Q,S^s)$). So, increasing $s_1$ by $\epsilon_1$ decreases $\delta(Q,S^s)$ by at most $\frac{1}{2}\times 2\epsilon_1 = \epsilon_1$. Now, if we move on to $s^s_4$, we know that increasing it to $s_4$ will again result in one definite worsening (of $\delta_4$), and (up to) three (potential) improvements of $\triangle$-terms. So now, having already increased $s^s_1$ to $s_1$, a further increasing of $s^s_4$ to $s_4$ now results in a further improvement to $\delta(Q,S^s)$ of (at most) $\epsilon_4$, where $\epsilon_4$ is the difference between $s^s_4$ and $s_4$. We can repeat this argument as we successively increase each $s_i$, and eventually we will have turned $S^s$ into $S$, and we know that $\delta(Q,S)$ can be no smaller than $\delta(Q,S^s)$ minus the sum of all of the $\epsilon_i$ terms. Furthermore, it is clear that the $\epsilon_i$ all have to sum to $p_{PR}$. Hence the minimum possible value for $\delta(Q,S)$ is $p_{PR}$. 

The above arguments provide a lower bound for $\min_{S\in local}\delta(Q,S)$ of $p_{PR}$. By inspection, we can see that the distribution cited in the statement of Theorem \ref{t:vardistheorem} achieves this bound. $\hfill\Box$

\section{Classification of the Extremal Points of the Chained Bell Polytope}\label{s:chainedBellproof}

Theorem \ref{t:chainedBelltheorem} enumerates the extremal points of the polytope characterized by the no-signaling equations \eref{e:chainednosig} in conjunction with the probability constraints that require nonnegative entries and row sums of one. Proving this theorem requires solving the well-known problem of moving from the halfspace representation of a convex polytope to the vertex representation. For a given fixed polytope, there is an algorithm for solving this problem, but we seek to perform the calculation in a way that will apply to all choices of $n$ in the $n$-fold chained Bell scenario.

In the canonical formulation of the vertex enumeration problem, the halfspace representation of a polytope $P$ in $\mathbb R^n$ consists of a $m\times n$ matrix $A$ and a $m$-dimensional vector $b$ for which $P=\{x\in \mathbb R^n \mid Ax\le b\}$. For the $n$-fold chained Bell inequality, we will be working in $\mathbb R^{8n}$, and some of the constraints will take  the form of equalities. Each equality constraint (the nosignaling constraints \eref{e:chainednosig} and the condition that the rows sum to one) can be turned into a pair of inequality constraints to match the canonical formulation of the problem. The inequality constraints (that stipulate that each entry must be nonnegative) can be included in $A$ directly.

For a given $z\in P$, define the submatrix $A_z$ to be the matrix whose rows are exactly the rows of $A$ for which $z$ saturates the inequality constraint. That is, the $i$th row of A is included in $A_z$ if and only if $(Az)_i=b_i$. Then a theorem of convex polytopes tells us that $z$ is an extremal point if and only if $A_z$ is a matrix of rank $8n$ \cite{lauritzen:2013}. With this result, we are now ready to prove Theorem \ref{t:chainedBelltheorem}, which we reproduce here:

\medskip

\noindent {\bf Theorem \ref{t:chainedBelltheorem}} \emph{The set of extremal points of the nonsignaling polytope corresponding to the $n$-fold chained Bell scenario consists of the $2^{2n}$ local deterministic distributions and the $2^{2n-1}$ generalized PR boxes.}

\medskip

\noindent \emph{Proof.} Let $P$ denote the $n$-fold chained Bell polytope for a fixed $n$, so $P$ can be described as the set of elements $x$ satisfying the equation $Ax\le b$, where $A$ and $b$ are derived from the probability conditions and no-signaling constraints as discussed earlier. Now consider the conditions that some $z\in\mathbb R^{8n}$ must satisfy in order to be an extremal point of $P$. First, $z$ of course must be an element of $P$, and membership in $P$ requires $z$ to saturate the all inequality constraints in $A$ that were generated in pairs from equality constraints. This automatically brings the rank of $A_z$ up to $4n$ (out of $8n$ total possible) for any $z\in P$ (even if $z$ is not an extremal point). Thus we can think of the ``baseline'' $A_z$ matrix as follows (blank cells contain ``0''):

\medskip

{\tiny

\hspace{-2.2cm}\begin{tabular}{ |cccc|cccc|cccc|cccc|cccc|cccc|cccc| }
\multicolumn{4}{c}{$a_1b_1$} 
&\multicolumn{4}{c}{$a_2b_1$} 
&\multicolumn{4}{c}{$a_2b_2$} 
&\multicolumn{4}{c}{$a_3b_2$} 
&\multicolumn{4}{c}{$\cdots$}
&\multicolumn{4}{c}{$a_nb_n$}
&\multicolumn{4}{c}{$a_1b_n$}  \\

 \multicolumn{1}{c}{++}
 &  \multicolumn{1}{c}{+0}
 &  \multicolumn{1}{c}{0+} 
 &  \multicolumn{1}{c}{00}
  & \multicolumn{1}{c}{++}
 &  \multicolumn{1}{c}{+0}
 &  \multicolumn{1}{c}{0+} 
 &  \multicolumn{1}{c}{00}
  & \multicolumn{1}{c}{++}
 &  \multicolumn{1}{c}{+0}
 &  \multicolumn{1}{c}{0+} 
 &  \multicolumn{1}{c}{00}
  & \multicolumn{1}{c}{++}
 &  \multicolumn{1}{c}{+0}
 &  \multicolumn{1}{c}{0+} 
 &  \multicolumn{1}{c}{00}
 &\multicolumn{4}{c}{$\cdots$}
     & \multicolumn{1}{c}{++}
 &  \multicolumn{1}{c}{+0}
 &  \multicolumn{1}{c}{0+} 
 &  \multicolumn{1}{c}{00}
   & \multicolumn{1}{c}{++}
 &  \multicolumn{1}{c}{+0}
 &  \multicolumn{1}{c}{0+} 
 &  \multicolumn{1}{c}{00} 
\\
\cline{1-17}\cline{20-28}
 -1  & -1  &      &      &     &       &      &      &      &      &      &      &      &      &      &   &&&&  &&&& &1&1&&   \\ 
$ 1$&$ 1$&$ 1$& $ 1$&     &      &      &      &      &      &      &       &      &      &      &   && && &&&& &&&&    \\ 
$ 1$&      &$ 1$&      &    -1&      &-1&      &      &      &         &      &      &      &    & &&&& &&&& &&&& \\
      &      &      &      &$ 1$&$ 1$&$ 1$&$ 1$&      &      &      &       &      &      &      &   & \multicolumn{4}{c|}{$\cdots$} &&&& &&&&  \\ 
      &      &      &      &$ 1$&$ 1$&      &      &  -1  &  -1  &     &      &      &      &      &    &&&& &&&& &&&&    \\ 
      &      &      &      &      &      &      &      &$ 1$&$ 1$&$ 1$&$ 1$ &      &      &      &   &&&& &&&& &&&&  \\ 
      &      &      &      &       &      &      &      &$ 1$&      &$ 1$&     &  -1  &      & -1   &    &&&& &&&& &&&&   \\
      &      &      &      &      &      &      &      &      &      &      &     &$ 1$&$ 1$&$ 1$&$ 1$ &&&& &&&& &&&& \\
      &      &      &      &      &      &      &      &      &     &      &      &   1 &   1  &      &     &  & && &&&& &&&& \\ 

\multicolumn{4}{c}{$\vdots$} & \multicolumn{4}{c}{$\vdots$} & \multicolumn{4}{c}{$\vdots$} & \multicolumn{4}{c}{$\vdots$} & \multicolumn{4}{c}{$\ddots$} & \multicolumn{4}{c}{$\vdots$}& \multicolumn{4}{c}{$\vdots$} \\

   &   &      &      &     &       &      &      &      &      &      &      &      &      &      &   &&&&  &-1&-1&& &&&&   \\ 
   &   &      &      &     &       &      &      &      &      &      &      &      &      &      &   &&&&  &1&1&1&1 &&&&   \\ 
   &   &      &      &     &       &      &      &      &      &      &      &      &      &      &   &\multicolumn{4}{c|}{$\cdots$}  &1&&1& &-1&&-1&   \\ 
      &   &      &      &     &       &      &      &      &      &      &      &      &      &      &   &&&&  &&&& &1&1&1&1   \\ 
      \cline{1-17}\cline{20-28}
 \end{tabular}
}

\noindent In the above table, the first row comes from the inequality
\begin{equation*}
-P(\textnormal{++}\mid a_1b_1)  -P(\textnormal{+0}\mid a_1b_1)+P(\textnormal{++}\mid a_1b_n) + P(\textnormal{+0}\mid a_1b_n)\le 0,
\end{equation*}
which is one of two inequality constraints induced by a no-signaling equality in \eref{e:chainednosig}. The second row comes from the inequality 
\begin{equation*}
P(\textnormal{++}\mid a_1b_1)+P(\textnormal{+0}\mid a_1b_1)+P(\textnormal{0+}\mid a_1b_1)+P(\textnormal{00}\mid a_1b_1)\le 1,
\end{equation*}
which is one of the two inequality constraints induced by the probability equality stipulating that the first outcome row sums to one. There are technically two rows for each equality constraint, but we have suppressed these duplicate rows in the table above because the rank of $A_z$ is the same if we only include one row per constraint. The rest of the rows in the above table are similarly derived from no-signaling and probability constraints.

The key question now becomes this: how can we add additional rows of $A$ to the above matrix so that we end up with a matrix $A_z$ of rank $8n$ with a solution $z$ so that $A_zz=b$, with $z$ also satisfying $Az\le b$. The pool from which we can draw additional rows consists of the rows generated by the nonnegative-probability constraints. Such rows have a ``-1'' in exactly one column and zeros in all other columns. $A_z$ contains such a row whenever $z$ has ``0'' as one of its entries -- for example, if $z$ assigns zero probability to the +$0|a_2b_2$ outcome, then $A_z$ will contain the row with a $-1$ in the $a_2b_2$,+$0$ column and zeros elsewhere.

We answer the question with a series of observations/lemmas that together imply Theorem \ref{t:chainedBelltheorem}. We will use certain expressions repeatedly: for an element $z$ of $P\subseteq \mathbb R^{8n}$, recall that its \emph{outcome table} is formed by organizing its entries into rows labeled $a_ib_j$, each with four elements. We refer to the corresponding four columns of $A_z$ as the $a_ib_j$ \emph{cell} of the $A_z$ matrix.
 
\medskip

\begin{alemma}\label{l:allinout}
Suppose that $z$ is an extremal point of the $n$-fold chained Bell polytope $P$, and that the outcome table of $z$ contains a row $a_xb_y$ that has support in exactly two entries. Then all rows must be supported in exactly two entries, and every row must be in either form $(p,0,0,q)$ with all probability on the ++ and 00 outcomes, or $(0,p,q,0)$ with all probability being on the +0 and 0+ outcomes.
\end{alemma}
\emph{Proof.} We can break this down into 12 cases based on different possible formats for the $a_xb_y$ row that has support in two entries. The row can be either of the form $a_{i}b_{i}$ or of form $a_{i+1}b_{i}$ (we include $a_1b_n$ in this second category), which determines how the corresponding cell of $A_z$ relates to the cells to its left and right. For each of these two cell types, there are six ways for the outcome row to have support in two locations: The $a_xb_y$-conditional probabilities $($++$,$+$0,0$+$,00)$ can be $(p,q,0,0)$, $(p,0,q,0)$, $(p,0,0,q)$, $(0,p,q,0)$, $(0,p,0,q)$, or $(0,0,p,q)$. The statement of the lemma indicates that the cases $(p,0,0,q)$ and $(0,p,q,0)$ are of particular importance, and we start with these. We will sometimes refer to a row of the form $(p,0,0,q)$ as an ``in-row'' and a row of the form $(0,p,q,0)$ as an ``out-row;'' or an ``in-cell'' and ``out-cell'' if we are discussing the corresponding cell of the $A_z$ matrix.

{\bf Case 1}: $a_ib_i$-style cell, $(p,0,0,q)$ outcome row. In the outcome table of $z$, the entries +0$|a_ib_i$ and 0+$|a_ib_i$ are equal to zero, so the $a_ib_i$ cell of $A_z$ will contain the rows that have ``$-1$'' in these two entries. Only three other rows of $A_z$ differ from zero in the $a_ib_i$ cell: two rows corresponding to no-signaling equalities, and one row corresponding to the probability sum-to-one equality. So we can represent the nonzero portion of this cell of the $A_z$ matrix as follows:

\vspace{5mm}
{\scriptsize\begin{center}
\begin{tabular}{ cccc|c|c|c|c|cccc }
 \multicolumn{4}{c}{$a_{i}b_{i-1}$}& \multicolumn{4}{|c|}{$a_{i}b_{i}$} &\multicolumn{4}{c}{$a_{i+1}b_{i}$} \\
&&&&  \multicolumn{1}{|c}{++}
&  \multicolumn{1}{c}{+0}
&  \multicolumn{1}{c}{0+} 
&  \multicolumn{1}{c|}{00}&&&&\\
\cline{5-8}
1&1&0&0& $-1$ &$-1$& 0 &0 &&&& \\
&&&& 1&1&1&1&&&&\\
&&&& 1&0&1&0&$-1$&0&$-1$&0\\
&&&& 0&$-1$&0&0&&&&\\
&&&& 0&0&$-1$&0&&&&\\
\end{tabular} 
\end{center}
}
\vspace{5mm}

If $A_z$ is to have full rank, we must be able to generate, through row operations, a row containing a 1 in any given column with 0s everywhere else. Looking at our $a_ib_i$ cell above, let's refer to the top three rows as $A$, $X$, and $B$, and the bottom two rows as the 0100 and 0010 ``pivots.'' Now let us try to form a linear combination of $A_z$ rows that generates the row containing a 1 in the $a_ib_i$00 column and 0s everywhere else. This linear combination clearly must include $1\times X$, because this is the only row supported in the $a_ib_i$00 column. The inclusion of $X$ in the linear combination also generates some ``slack'' in three other columns: $a_ib_i$++, $a_ib_i$+0, and $a_ib_i$0+. Reducing this ``slack'' to zero necessitates the inclusion of other rows in the linear combination. Inclusion of the pivot rows reduces the $a_ib_i$+0 and $a_ib_i$0+ entries, but reducing the $a_ib_i$++ entry requires including at least one of the two rows $A$ and $B$ in the linear combination. These rows both contain nonzero entries in other cells, so their inclusion in the linear combination creates some new slack in adjacent cells which cannot be avoided. Moving to the next cell sets off a sort of cascade effect from cell to cell. This effect merits a lemma of its own, whose proof we temporarily defer:

\begin{alemma}``Domino Lemma'': Suppose that the $a_xb_y$ cell of an extremal point's $A_z$ matrix is either an ``in-cell'' containing the the two pivots 0100 and 0010 or an ``out-cell'' containing the two pivots 1000 and 0001. Suppose that a linear combination of the rows of $A_z$ contains a nonzero multiple of the row that contains support in both $a_xb_y$ and the next cell to its right (left). Then in order for this linear combination of rows to contain only zeros in the entries in the cell to the right (left) of $a_xb_y$, that cell must also be either an in-cell or an out-cell, and furthermore the linear combination must also contain a nonzero multiple of the $A_z$ row that contains support in both the cell to the right (left) and the next subsequent further cell to the right (left).
\end{alemma}

For the present case, suppose that we use row $B$ in our linear combination of $A_z$ rows. Then the domino lemma tells us that the $a_{i+1}b_i$ cell of $A_z$ must be an in-cell or an out-cell, and that the linear combination of $A_z$ rows must also contain the row that links $a_{i+1}b_i$ and $a_{i+1}b_{i+1}$. Thus inductive application of the domino lemma implies that all cells must be in-cells or out-cells. This would also apply -- just moving to the left -- if we instead used row $A$ in the linear combination of $A_z$ rows. This proves the statement of Lemma \ref{l:allinout} for case 1. 

\medskip

{\bf Case 2}: $a_ib_i$-style cell, $(0,p,q,0)$ outcome row.
\vspace{5mm}
{\scriptsize
\begin{center}
\begin{tabular}{ cccc|c|c|c|c|cccc }
 \multicolumn{4}{c}{$a_{i}b_{i-1}$}& \multicolumn{4}{|c|}{$a_{i}b_{i}$} &\multicolumn{4}{c}{$a_{i+1}b_{i}$} \\
&&&&  \multicolumn{1}{|c}{++}
&  \multicolumn{1}{c}{+0}
&  \multicolumn{1}{c}{0+} 
&  \multicolumn{1}{c|}{00}&&&&\\
\cline{5-8}
1&1&0&0& $-1$ &$-1$& 0 &0 &&&& \\
&&&& 1&1&1&1&&&&\\
&&&& 1&0&1&0&$-1$&0&$-1$&0\\
&&&& $-1$&0&0&0&&&&\\
&&&& 0&0&0&$-1$&&&&\\
\end{tabular} 
\end{center}
}
\vspace{5mm}

\noindent Here, we can see by inspection that if a linear combination of the rows of $A_z$ is to generate a row containing a 1 in the $a_ib_i$+0 column and zeros elsewhere, then it necessarily must include a nonzero multiple of at least one of the two rows containing nonzero entries in an adjacent cell. Since $a_ib_i$ is an out-cell, the domino lemma applies, which then by induction tells us that all cells must be in-cells or out-cells.

\medskip

{\bf Case 3}: $a_{i+1}b_i$-style cell, $(0,p,q,0)$ outcome row. Similar to cases 1 and 2.

\medskip

{\bf Case 4}: $a_{i+1}b_i$-style cell, $(p,0,0,q)$ outcome row. Similar to cases 1 and 2.

\medskip

\noindent We note that the first four cases show us that if any one cell of $A_z$ is an in-cell or an out-cell, then every other cell must also be either an in-cell or an out-cell. This observation simplifies the treatment of the remaining 8 cases. 

\medskip

{\bf Case 5}: $a_{i+1}b_i$-style cell, $(0,0,p,q)$. (Recalling the statement of Lemma \ref{l:allinout}, we expect this case to lead to a contradiction.)
\vspace{5mm}
{\scriptsize
\begin{center}
\begin{tabular}{ cccc|c|c|c|c|cccc }
 \multicolumn{4}{c}{$a_{i}b_{i}$}& \multicolumn{4}{|c|}{$a_{i+1}b_{i}$} &\multicolumn{4}{c}{$a_{i+1}b_{i+1}$} \\
&&&&  \multicolumn{1}{|c}{++}
&  \multicolumn{1}{c}{+0}
&  \multicolumn{1}{c}{0+} 
&  \multicolumn{1}{c|}{00}&&&&\\
\cline{5-8}
1&0&1&0& $-1$ &0& $-1$ &0 &&&& \\
&&&& 1&1&1&1&&&&\\
&&&& 1&1&0&0&$-1$&$-1$&0&0\\
&&&& $-1$&0&0&0&&&&\\
&&&& 0&$-1$&0&0&&&&\\
\end{tabular} 
\end{center}
}
\vspace{5mm}

We refer to the first three rows depicted above as $B$, $X$, and $A$. For a linear combination of the $A_z$ rows to yield 1 in the $a_{i+1}b_i$0+ column and 0 elsewhere, it must contain $-1\times B$. ($X$ cannot be used because of its entry in the $a_{i+1}b_i$00 column.) Moving one cell to the left to $a_ib_i$, we now find ourselves with a ``slack'' of $(-1,0,-1,0)$ that must be turned into $(0,0,0,0)$ by adding a further linear combination of rows supported in the $a_ib_i$ cell of $A_z$:

{\scriptsize
\begin{center}
\begin{tabular}{ cccc|c|c|c|c| }
 \multicolumn{4}{c}{$a_{i}b_{i-1}$} &\multicolumn{4}{c}{$a_{i}b_{i}$} \\
&&&&  \multicolumn{1}{|c}{++}
&  \multicolumn{1}{c}{+0}
&  \multicolumn{1}{c}{0+} 
&  \multicolumn{1}{c|}{00}\\
\cline{5-8}
1&1&0&0&$-1$ & $-1$ &0&0 \\
&&&&1&1&1&1\\
&&&&{\color{red}$-1$}&0&{\color{red}$-1$}&0\\
\end{tabular} 
\end{center}
}

The addition of various pivot rows to this cell can allow for the elimination of the slack. However, we must add them in a consistent manner, so that $A_z$ actually has a solution $z$ for which $A_zz=b$. As the $z$-entries for the $a_{i+1}b_i$ row are of the form $(0,0,p,q)$, if we denote the $a_{i}b_i$ row of $z$ as $(s,t,u,v)$, then $s+u=p$ and $t+v=q$ must hold -- this is a consequence of the no-signaling equalities \eref{e:chainednosig}. This means that $(s,t,u,v)$ can contain at most two zeros, and additionally $(s,0,u,0)$ and $(0,t,0,v)$ are forbidden. This means that the only allowable pivot collections for the $a_{i}b_i$ cell are $\{1000\}$, $\{0100\}$, $\{0010\}$, $\{0001\}$, $\{1000,0100\}$, $\{1000,0001\}$, $\{0100,0010\}$, and $\{0010, 0001\}$. If the pivot 0010 is not present, then the use of the $X$ row is required to reduce the $a_ib_i$0+ column to zero, and this creates new slack in $a_ib_i$00 that can only be reduced if the 0001 pivot is present. Similarly, if the 1000 pivot is not present, then any choice of row used to reduce the $a_ib_i$++ column creates slack in the $a_ib_i$+0 entry, which then necessitates the presence of the 0100 pivot. These observations ensure that only two of the eight mentioned pivot collections are allowable: $\{1000,0001\}$ or $\{0100,0010\}$. So the $a_ib_i$ cell must be an out-cell or an in-cell.  But by our study of the earlier cases, this would imply that all cells must be out-cells or in-cells, which contradicts the case assumption that the original $a_{i+1}b_i$ outcome row was of the form $(0,0,p,q)$. Hence we have the anticipated contradiction.

\medskip

{\bf Case 6}: $a_{i+1}b_i$-style cell, $(0,p,0,q)$. The argument for this case and the remaining cases are all quite similar to the argument for case 5, and lead to similar contradictions. Hence we only list the remaining cases.

 \medskip
 
{\bf Case 7}: $a_{i+1}b_i$-style cell, $(p,0,q,0)$. 

{\bf Case 8}: $a_{i+1}b_i$-style cell, $(p,q,0,0)$. 

{\bf Case 9}: $a_{i}b_i$-style cell, $(0,0,p,q)$. 
 
{\bf Case 10}: $a_{i}b_i$-style cell, $(0,p,0,q)$. 

{\bf Case 11}: $a_{i}b_i$-style cell, $(p,0,q,0)$. 
 
{\bf Case 12}: $a_{i}b_i$-style cell, $(p,q,0,0)$.  $\hfill \Box$

\medskip

\emph{Proof of ``Domino Lemma.''} We divide the proof into four cases: For $a_xb_y$, either $x=y$ or $x=y+1$ (we include $x=1$, $y=n$ in the category $x=y+1$), and for each case we must consider that we can be moving to the adjacent cell either to the right or to the left. 

{\bf Case 1}: $x=y$, movement to the right. By assumption, $a_ib_i$ is either an in-cell or an out-cell, and we are considering a linear combination of $A_z$ rows that contains a nonzero multiple of the unique row that is supported in both $a_ib_i$ and $a_{i+1}b_i$. Let $-c$ denote the coefficient of this row in the linear combination. Now the goal is for the linear combination to have zeros in the columns corresponding to the $a_{i+1}b_i$ cell:

\vspace{5mm}
\begin{center}
{\scriptsize
\begin{tabular}{ rcccc|c|c|c|c|cccc }
& \multicolumn{4}{c|}{$a_{i}b_{i}$} & \multicolumn{4}{|c|}{$a_{i+1}b_{i}$} &\multicolumn{4}{c}{$a_{i+1}b_{i+1}$} \\
&&&&&  \multicolumn{1}{|c}{++}
&  \multicolumn{1}{c}{+0}
&  \multicolumn{1}{c}{0+} 
&  \multicolumn{1}{c|}{00}&&&&\\
\cline{6-9}
$-c\quad\times$& 1 & 0 & 1 & 0 & $-1$ & 0 & -1 &0 &&&& \\
?$\quad\times$&&&&&1&1&1&1&&&&\\
+\quad?$\quad\times$&&&&&1&1&0&0&$-1$&$-1$&0&0\\
\hline
  \multicolumn{1}{l}{=}&&&&\multicolumn{1}{c}{}&\multicolumn{1}{c}{0}&\multicolumn{1}{c}{0}&\multicolumn{1}{c}{0}&\multicolumn{1}{c}{0}&&&&
\end{tabular} 
}\end{center}
\vspace{5mm}

\noindent Clearly this is impossible if the only other rows supported in the $a_{i+1}b_i$ cell are the middle and bottom rows in the above figure, which we call $X$ and $A$. So some pivots must also be present in the $a_{i+1}b_i$ cell. The assumption that $A_zz=b$ can be satisfied imposes constraints on which pivots can be present. Specifically, the assumption that the $a_ib_i$ outcome row of $z$ is of the form $(p,0,0,q)$ or $(0,p,q,0)$ means, by no-signaling, that the $a_{i+1}b_i$ row can contain at most two zeros, and furthermore the forms $(s,0,u,0)$ and $(0,t,0,v)$ are also excluded. (Recall case 5 of the previous lemma, which also used this type of reasoning.) Thus the permissible pivot collections are $\{1000\}$, $\{0100\}$, $\{0010\}$, $\{0001\}$, $\{1000,0100\}$, $\{1000,0001\}$, $\{0100,0010\}$, and $\{0010, 0001\}$. Now, note that if the pivot 0010 is not present, then use of the row $X$ is required to eliminate the $c$ occurring in the $a_{i+1}b_i$0+ column, which creates new slack in $a_{i+1}b_i$00 that can only be eliminated if the pivot 0001 is present. Furthermore, we note that if the pivot 1000 is absent, the use of row $X$ or $A$ (or some combination of both) is required to eliminate the other $c$, which always creates slack in $a_{i+1}b_i$+0 that necessitates the pivot 0100. The only pivot collections satisfying both of these requirements are $\{1000,0001\}$ and $\{0100,0010\}$.

So we have proved that $a_{i+1}b_i$ is either an out-cell or an in-cell, but we still need to show that there is a cascade effect: that a nonzero coefficient on the next ``link'' row $A$ is necessary to obtain zeros in all four entries of the $a_{i+1}b_i$ cell. If $a_{i+1}b_i$ is an out-cell (pivots 1000 and 0001), the use of $X$ is necessary to reduce the $c$ quantity in the $a_{i+1}b_i$0+ column. This creates slack in the $a_{i+1}b_i$+0 column that can only be addressed with the inclusion of row $A$ in the linear combination. On the other hand, if $a_{i+1}b_i$ is an in-cell (pivots 0100 and 0010), then we might use either $X$ or $A$ to reduce the $c$ in the $a_{i+1}b_i$++ column, but if we use $X$, we create slack in $a_{i+1}b_i$00 that cannot be reduced, and so the link row must be used instead.

The remaining cases all follow by similar arguments and so we only list them.

{\bf Case 2}: $x=y$, movement to the left.

{\bf Case 3}: $x=y+1$, movement to the right.

{\bf Case 4}: $x=y+1$, movement to the left. $\hfill \Box$

\begin{alemma}
If a distribution matrix $z$ is an extremal point of $P$, then any given row of its outcome table can have support in at most two entries. 
\end{alemma}
\emph{Proof.} We first dispose of the possibility of an outcome row with support in all four entries. The existence of such an outcome row would result in there being no pivots in the corresponding cell of the polytope matrix $A_z$. By inspecting the cell diagrams that we have been drawing, one can see that without pivots it is impossible to form a linear combination of the baseline rows alone that yields a 1 in one column and 0 in the remaining three columns of the cell, so $A_z$ cannot have rank $8n$.

Now, suppose $z$ were to have an outcome row with support in exactly three entries. The no-signaling conditions rule out the possibility that this three-support row borders a row supported in exactly one location. Furthermore, the three-support row cannot border a row that is supported in exactly two locations, because our previous results only allow exact-two-support in a row if \emph{all} rows are exact-two-supported. Since support in all four entries has also been ruled out, the only option is for the bordering rows of $z$ to have support in exactly three locations. By induction, this implies that all of the rows must be exact-three-supported. Thus each cell of the polytope matrix $A_z$ has only one pivot row. Since there are $2n$ cells, $A_z$ consists of $2n$ pivot rows, $2n$ probability-sums-to-one constraint rows, and $2n$ no-signaling constraint rows. Thus the rank of $A_z$ is at most $6n$, and so $z$ cannot be an extremal point. $\hfill \Box$

\begin{alemma}\label{l:extremal}
If a distribution $z$ is an extremal point, then either A) the rows of $z$ are all either of the form $(p,0,0,q)$ and $(0,p,q,0)$, or B) the rows of $z$ consist entirely of rows with one 1 and three 0s.
\end{alemma}
\emph{Proof.} By no-signaling, a row consisting of one 1 and three 0s can only border another row that has one 1 and three 0s, or a row with support in exactly two locations. However, Lemma \ref{l:allinout} tells us that if one row of an extremal distribution $z$ is two-supported, all rows must be two-supported. Therefore the bordering rows, and thus inductively all the rows, must consist of exactly one 1 and three 0s. $\hfill \Box$

\medskip

\begin{alemma}\label{l:mustbeLD}
Any element of the $n$-fold chained Bell polytope whose rows each consist of one 1 and three 0s must be a local deterministic distribution.
\end{alemma}
\emph{Proof.} As discussed in Section \ref{s:chainedBellpoly}, any given LD distribution can be represented by an assignment table like the one below, where each vertical line corresponds to a row in the outcome table:

\medskip

\begin{tabular}{c|c|c|c|c|c|c|c|c|c|}
$a_1$ & $b_1$ & $a_2$ & $b_2$ & $a_3$ & $b_3$ & $\cdots$ &$b_{n-1}$ & $a_n$ & $b_n$ \\
+ & 0 & 0 & + & 0 & + & $\cdots$ & 0 & 0 & + \\
\end{tabular}

\medskip

\noindent Given a valid distribution matrix consisting entirely of rows of one 1 and three 0s, we want to show that we can consistently construct an assignment table like the one above that induces the distribution matrix. The fact that a valid distribution matrix must satisfy the no-signaling conditions \eref{e:chainednosig} is what makes this possible.

To construct such an assignment table, consider a given $a_i$ (the argument also works if looking at a $b_i$). There are two rows of the given distribution matrix that include this $a_i$: $a_ib_i$ and $a_ib_{i+1}$, which by assumption both consist of one 1 and three 0s assigned to the four possibilities ++, +0, 0+, 00. Let $xy,zw\in \{++, +0, 0+, 00\}$ denote the outcome with a ``1'' for $a_ib_i$ and $a_ib_{i+1}$, respectively. Then the no-signaling condition
\begin{equation*}
P(++|a_ib_i)+P(+0|a_ib_i) = P(++|a_ib_{i+1}) +P(+0|a_ib_{i+1}) 
\end{equation*}
is only satisfied if $x$ and $z$ are the same element of $\{$+$,0\}$. This gives us an unambiguous rule for how to fill in the $a_i$ cell in the assignment table. If we fill in the entire assignment table according to this method, then if we chose any vertical line and consider the induced outcome row distribution corresponding to this line, the method of construction ensures that this will be consistent with the original distribution matrix.$\hfill\Box$

\begin{alemma}\label{l:eachodd} Suppose $z$ is an distribution matrix for which each outcome row is either an in-row or an out-row. Then $z$ can be a extremal point of the $n$-fold chained Bell polytope only if it has an odd number of each type of row.
\end{alemma} 
 \emph{Proof.} Recall the baseline polytope matrix, whose rows are always part of $A_z$:

\medskip
{\tiny
\hspace{-2.2cm}\begin{tabular}{ |cccc|cccc|cccc|cccc|cccc|cccc|cccc| }
\multicolumn{4}{c}{$a_1b_1$} 
&\multicolumn{4}{c}{$a_2b_1$} 
&\multicolumn{4}{c}{$a_2b_2$} 
&\multicolumn{4}{c}{$a_3b_2$} 
&\multicolumn{4}{c}{$\cdots$}
&\multicolumn{4}{c}{$a_nb_n$}
&\multicolumn{4}{c}{$a_1b_n$}  \\

 \multicolumn{1}{c}{++}
 &  \multicolumn{1}{c}{+0}
 &  \multicolumn{1}{c}{0+} 
 &  \multicolumn{1}{c}{00}
  & \multicolumn{1}{c}{++}
 &  \multicolumn{1}{c}{+0}
 &  \multicolumn{1}{c}{0+} 
 &  \multicolumn{1}{c}{00}
  & \multicolumn{1}{c}{++}
 &  \multicolumn{1}{c}{+0}
 &  \multicolumn{1}{c}{0+} 
 &  \multicolumn{1}{c}{00}
  & \multicolumn{1}{c}{++}
 &  \multicolumn{1}{c}{+0}
 &  \multicolumn{1}{c}{0+} 
 &  \multicolumn{1}{c}{00}
 &\multicolumn{4}{c}{$\cdots$}
     & \multicolumn{1}{c}{++}
 &  \multicolumn{1}{c}{+0}
 &  \multicolumn{1}{c}{0+} 
 &  \multicolumn{1}{c}{00}
   & \multicolumn{1}{c}{++}
 &  \multicolumn{1}{c}{+0}
 &  \multicolumn{1}{c}{0+} 
 &  \multicolumn{1}{c}{00} 
\\
\cline{1-17}\cline{20-28}

 -1  & -1  &      &      &     &       &      &      &      &      &      &      &      &      &      &   &&&&  &&&& &1&1&&   \\ 
$ 1$&$ 1$&$ 1$& $ 1$&     &      &      &      &      &      &      &       &      &      &      &   && && &&&& &&&&    \\ 
$ 1$&      &$ 1$&      &    -1&      &-1&      &      &      &         &      &      &      &    & &&&& &&&& &&&& \\
      &      &      &      &$ 1$&$ 1$&$ 1$&$ 1$&      &      &      &       &      &      &      &   & \multicolumn{4}{c|}{$\cdots$} &&&& &&&&  \\ 
      &      &      &      &$ 1$&$ 1$&      &      &  -1  &  -1  &     &      &      &      &      &    &&&& &&&& &&&&    \\ 
      &      &      &      &      &      &      &      &$ 1$&$ 1$&$ 1$&$ 1$ &      &      &      &   &&&& &&&& &&&&  \\ 
      &      &      &      &       &      &      &      &$ 1$&      &$ 1$&     &  -1  &      & -1   &    &&&& &&&& &&&&   \\
      &      &      &      &      &      &      &      &      &      &      &     &$ 1$&$ 1$&$ 1$&$ 1$ &&&& &&&& &&&& \\
      &      &      &      &      &      &      &      &      &     &      &      &   1 &   1  &      &     &  & && &&&& &&&& \\ 

\multicolumn{4}{c}{$\vdots$} & \multicolumn{4}{c}{$\vdots$} & \multicolumn{4}{c}{$\vdots$} & \multicolumn{4}{c}{$\vdots$} & \multicolumn{4}{c}{$\ddots$} & \multicolumn{4}{c}{$\vdots$}& \multicolumn{4}{c}{$\vdots$} \\

   &   &      &      &     &       &      &      &      &      &      &      &      &      &      &   &&&&  &-1&-1&& &&&&   \\ 
   &   &      &      &     &       &      &      &      &      &      &      &      &      &      &   &&&&  &1&1&1&1 &&&&   \\ 
   &   &      &      &     &       &      &      &      &      &      &      &      &      &      &   &\multicolumn{4}{c|}{$\cdots$}  &1&&1& &-1&&-1&   \\ 
      &   &      &      &     &       &      &      &      &      &      &      &      &      &      &   &&&&  &&&& &1&1&1&1   \\ 
      \cline{1-17}\cline{20-28}
 \end{tabular}
}
\medskip

\noindent $A_z$ will contain the above rows plus a collection of pivot rows determined by whether each cell of $A_z$ corresponds to an in-row or out-row of $z$. If $z$ is to be a extremal point, then it must be possible to generate, as a linear combination of the rows of $A_z$, a row with a 1 in any particular column and zeros elsewhere. We will show that this is only possible if the condition in Lemma \ref{l:eachodd} is satisfied.

In the $a_1b_1$ cell, we know that by assumption this cell will also contain either the pivots $\{1000,0001\}$ or the pivots $\{0100,0010\}$. Let us start by examining the first situation, where $a_1b_1$ is an out-cell, and try to construct a linear combination of $A_z$ rows that generates the row with 1 in $a_1b_1$+0 and zeros elsewhere. We first note by inspection that any linear combination of $A_z$ rows containing a 1 in $a_1b_1$+0 and a 0 in $a_1b_1$0+ must contain the expression $qX-qB-pA$ where $q$ and $p$ are constants satisfying $q=1-p$ and $A$, $X$ and $B$ are the first, second and third rows of the cell. The zeros in $a_1b_1$00 and $a_1b_1$++ can then be obtained by including appropriate weights of the 1000 and 0001 pivots. Now, deleting the rows supported in $a_1b_1$ and replacing them with the linear combination that we have constructed so far, we have:

\medskip

\hspace{-2.2cm}{\tiny\begin{tabular}{ |cccc|cccc|cccc|cccc|cccc|cccc|cccc| }
\multicolumn{4}{c}{$a_1b_1$} 
&\multicolumn{4}{c}{$a_2b_1$} 
&\multicolumn{4}{c}{$a_2b_2$} 
&\multicolumn{4}{c}{$a_3b_2$} 
&\multicolumn{4}{c}{$\cdots$}
&\multicolumn{4}{c}{$a_nb_n$}
&\multicolumn{4}{c}{$a_1b_n$}  \\

 \multicolumn{1}{c}{++}
 &  \multicolumn{1}{c}{+0}
 &  \multicolumn{1}{c}{0+} 
 &  \multicolumn{1}{c}{00}
  & \multicolumn{1}{c}{++}
 &  \multicolumn{1}{c}{+0}
 &  \multicolumn{1}{c}{0+} 
 &  \multicolumn{1}{c}{00}
  & \multicolumn{1}{c}{++}
 &  \multicolumn{1}{c}{+0}
 &  \multicolumn{1}{c}{0+} 
 &  \multicolumn{1}{c}{00}
  & \multicolumn{1}{c}{++}
 &  \multicolumn{1}{c}{+0}
 &  \multicolumn{1}{c}{0+} 
 &  \multicolumn{1}{c}{00}
 &\multicolumn{4}{c}{$\cdots$}
     & \multicolumn{1}{c}{++}
 &  \multicolumn{1}{c}{+0}
 &  \multicolumn{1}{c}{0+} 
 &  \multicolumn{1}{c}{00}
   & \multicolumn{1}{c}{++}
 &  \multicolumn{1}{c}{+0}
 &  \multicolumn{1}{c}{0+} 
 &  \multicolumn{1}{c}{00} 
\\
\cline{1-17}\cline{20-28}

   & 1  &      &      &  q   &       &  q    &      &      &      &      &      &      &      &      &   &&&&  &&&& &-p&-p&&   \\ 

      &      &      &      &$ 1$&$ 1$&$ 1$&$ 1$&      &      &      &       &      &      &      &   & \multicolumn{4}{c|}{$\cdots$} &&&& &&&&  \\ 
      &      &      &      &$ 1$&$ 1$&      &      &  -1  &  -1  &     &      &      &      &      &    &&&& &&&& &&&&    \\ 
      &      &      &      &      &      &      &      &$ 1$&$ 1$&$ 1$&$ 1$ &      &      &      &   &&&& &&&& &&&&  \\ 
      &      &      &      &       &      &      &      &$ 1$&      &$ 1$&     &  -1  &      & -1   &    &&&& &&&& &&&&   \\
      &      &      &      &      &      &      &      &      &      &      &     &$ 1$&$ 1$&$ 1$&$ 1$ &&&& &&&& &&&& \\
      &      &      &      &      &      &      &      &      &     &      &      &   1 &   1  &      &     &  & && &&&& &&&& \\ 

\multicolumn{4}{c}{$\vdots$} & \multicolumn{4}{c}{$\vdots$} & \multicolumn{4}{c}{$\vdots$} & \multicolumn{4}{c}{$\vdots$} & \multicolumn{4}{c}{$\ddots$} & \multicolumn{4}{c}{$\vdots$}& \multicolumn{4}{c}{$\vdots$} \\

   &   &      &      &     &       &      &      &      &      &      &      &      &      &      &   &&&&  &-1&-1&& &&&&   \\ 
   &   &      &      &     &       &      &      &      &      &      &      &      &      &      &   &&&&  &1&1&1&1 &&&&   \\ 
   &   &      &      &     &       &      &      &      &      &      &      &      &      &      &   &\multicolumn{4}{c|}{$\cdots$}  &1&&1& &-1&&-1&   \\ 
      &   &      &      &     &       &      &      &      &      &      &      &      &      &      &   &&&&  &&&& &1&1&1&1   \\ 
      \cline{1-17}\cline{20-28}
 \end{tabular}}

\medskip

To obtain zeros in all the remaining columns, we need to eliminate the $q$ slack in the $a_2b_1$ cell with linear combinations of other rows. Let's now rename the second and third rows of the table above $X$ and $A$, respectively, and recall that $a_2b_1$ also contains some pivots (by assumption, either it is an in-cell or an out-cell). If $a_2b_1$ is an in-cell, then the $q$ entry in $a_2b_1$++ can only be reduced to 0 by inclusion of $-qA$ in the linear combination, as including the $X$ row in the linear combination will create un-reducible slack in the $a_2b_1$00 entry. Inclusion of $-qA$ transfers a pair of $q$'s to the subsequent cell $a_2b_2$ \emph{with the sign of q unchanged} in the new cell.  On the other hand, if $a_2b_1$ is an out-cell, then reducing the $q$ in 0+$a_2b_1$ requires the inclusion of $-qX+qA$ in the linear combination if we are not to create slack in other entries of $a_2b_1$, thus transferring a pair of $q$'s to the subsequent cell \emph{with the sign flipped}. 

Removing now the $A_z$ rows with support in $a_2b_1$ and replacing them with our running linear combination so far, the table now looks like this:

 \medskip

\hspace{-2.2cm}{\tiny\begin{tabular}{ |cccc|cccc|cccc|cccc|cccc|cccc|cccc| }
\multicolumn{4}{c}{$a_1b_1$} 
&\multicolumn{4}{c}{$a_2b_1$} 
&\multicolumn{4}{c}{$a_2b_2$} 
&\multicolumn{4}{c}{$a_3b_2$} 
&\multicolumn{4}{c}{$\cdots$}
&\multicolumn{4}{c}{$a_nb_n$}
&\multicolumn{4}{c}{$a_1b_n$}  \\

 \multicolumn{1}{c}{++}
 &  \multicolumn{1}{c}{+0}
 &  \multicolumn{1}{c}{0+} 
 &  \multicolumn{1}{c}{00}
  & \multicolumn{1}{c}{++}
 &  \multicolumn{1}{c}{+0}
 &  \multicolumn{1}{c}{0+} 
 &  \multicolumn{1}{c}{00}
  & \multicolumn{1}{c}{++}
 &  \multicolumn{1}{c}{+0}
 &  \multicolumn{1}{c}{0+} 
 &  \multicolumn{1}{c}{00}
  & \multicolumn{1}{c}{++}
 &  \multicolumn{1}{c}{+0}
 &  \multicolumn{1}{c}{0+} 
 &  \multicolumn{1}{c}{00}
 &\multicolumn{4}{c}{$\cdots$}
     & \multicolumn{1}{c}{++}
 &  \multicolumn{1}{c}{+0}
 &  \multicolumn{1}{c}{0+} 
 &  \multicolumn{1}{c}{00}
   & \multicolumn{1}{c}{++}
 &  \multicolumn{1}{c}{+0}
 &  \multicolumn{1}{c}{0+} 
 &  \multicolumn{1}{c}{00} 
\\
\cline{1-17}\cline{20-28}
      &    1  &      &      &     &    &      &      &  $\pm$ q &  $\pm$ q  &     &      &      &      &      &    &&&& &&&& &-p&-p&&    \\ 
      &      &      &      &      &      &      &      &$ 1$&$ 1$&$ 1$&$ 1$ &      &      &      &   &&&& &&&& &&&&  \\ 
      &      &      &      &       &      &      &      &$ 1$&      &$ 1$&     &  -1  &      & -1   &    &&&& &&&& &&&&   \\
      &      &      &      &      &      &      &      &      &      &      &     &$ 1$&$ 1$&$ 1$&$ 1$ &&&& &&&& &&&& \\
      &      &      &      &      &      &      &      &      &     &      &      &   1 &   1  &      &     &  & && &&&& &&&& \\ 

\multicolumn{4}{c}{$\vdots$} & \multicolumn{4}{c}{$\vdots$} & \multicolumn{4}{c}{$\vdots$} & \multicolumn{4}{c}{$\vdots$} & \multicolumn{4}{c}{$\ddots$} & \multicolumn{4}{c}{$\vdots$}& \multicolumn{4}{c}{$\vdots$} \\

   &   &      &      &     &       &      &      &      &      &      &      &      &      &      &   &&&&  &-1&-1&& &&&&   \\ 
   &   &      &      &     &       &      &      &      &      &      &      &      &      &      &   &&&&  &1&1&1&1 &&&&   \\ 
   &   &      &      &     &       &      &      &      &      &      &      &      &      &      &   &\multicolumn{4}{c|}{$\cdots$}  &1&&1& &-1&&-1&   \\ 
      &   &      &      &     &       &      &      &      &      &      &      &      &      &      &   &&&&  &&&& &1&1&1&1   \\ 
      \cline{1-17}\cline{20-28}
 \end{tabular}}

\medskip

\noindent The sign of $q$ is positive if $a_2b_1$ was an out-row and negative if $a_2b_1$ was an in-row. Analyzing now the situation in the $a_2b_2$ cell, we will find once again that in order to reduce the $q$ slack, it will be necessary to include a multiple of the row that transfers $q$ slack to the following cell $a_3b_2$, where the sign of $q$ will be unchanged if $a_2b_2$ is an in-cell and flipped if $a_2b_2$ is an out-cell.

Thus a pattern emerges: as we reduce column entries to zero, we move to the right through the table, all the while shifting some slack of magnitude $q$ to the right but never getting rid of it, and furthermore flipping the sign of $q$ every time we pass through an out-cell and keeping the sign the same every time we pass through an in-cell. At the final step, we find ourselves in the $a_1b_n$ cell in the following situation:

\medskip

\hspace{-2.2cm}{\tiny\begin{tabular}{ |cccc|cccc|cccc|cccc|cccc|cccc|cccc| }
\multicolumn{4}{c}{$a_1b_1$} 
&\multicolumn{4}{c}{$a_2b_1$} 
&\multicolumn{4}{c}{$a_2b_2$} 
&\multicolumn{4}{c}{$a_3b_2$} 
&\multicolumn{4}{c}{$\cdots$}
&\multicolumn{4}{c}{$a_nb_n$}
&\multicolumn{4}{c}{$a_1b_n$}  \\

 \multicolumn{1}{c}{++}
 &  \multicolumn{1}{c}{+0}
 &  \multicolumn{1}{c}{0+} 
 &  \multicolumn{1}{c}{00}
  & \multicolumn{1}{c}{++}
 &  \multicolumn{1}{c}{+0}
 &  \multicolumn{1}{c}{0+} 
 &  \multicolumn{1}{c}{00}
  & \multicolumn{1}{c}{++}
 &  \multicolumn{1}{c}{+0}
 &  \multicolumn{1}{c}{0+} 
 &  \multicolumn{1}{c}{00}
  & \multicolumn{1}{c}{++}
 &  \multicolumn{1}{c}{+0}
 &  \multicolumn{1}{c}{0+} 
 &  \multicolumn{1}{c}{00}
 &\multicolumn{4}{c}{$\cdots$}
     & \multicolumn{1}{c}{++}
 &  \multicolumn{1}{c}{+0}
 &  \multicolumn{1}{c}{0+} 
 &  \multicolumn{1}{c}{00}
   & \multicolumn{1}{c}{++}
 &  \multicolumn{1}{c}{+0}
 &  \multicolumn{1}{c}{0+} 
 &  \multicolumn{1}{c}{00} 
\\
\cline{1-17}\cline{20-28}

   & 1  &      &      &    &       &      &      &      &      &      &      &      &      &      &   &&&&  &&&& &-p$\pm$q&-p&$\pm$q&   \\ 
      &   &      &      &     &       &      &      &      &      &      &      &      &      &      &   &&\multicolumn{2}{c}{$\cdots$}&  &&&& &1&1&1&1   \\ 
      \cline{1-17}\cline{20-28}
 \end{tabular}}
 
\medskip

\noindent In the table above, $q$ is positive if we passed through an even number of out-cells as we methodically worked through cells $a_2b_1$, $a_2b_2$, $a_3b_2$,...., $a_nb_{n-1}$, $a_nb_n$. Recalling that the current argument started with the assumption that the the first cell $a_1b_1$ was an out-cell, this is equivalent to saying that the total number of out-cells among the non-$a_1b_n$ cells is odd. The other possible scenario is that $q$ is negative because the total number of out-cells among the non-$a_1b_n$ cells is even.

Let's examine what happens if the total number of out-cells in the set $\{a_1b_1,a_2b_1,...,a_nb_{n-1}\}$ is odd, and $q$ is therefore positive. Now, if $a_1b_n$ is A) an in-cell, we can reduce the p's and q's in the middle columns of $a_1b_n$ with pivots but must worry about the $-p+q$ factor in $a_1b_n$++. We cannot use the second row in the above table (let's call it $X$) to reduce this column without generating unreducible slack in the 00 column. Therefore the only possibility is for $-p+q = 0$, which can be reconciled with our initial stipulation that $p+q=1$ by setting $p=q=1/2$. Thus we have achieved our original goal of generating the row with $1$ in $a_1b_1$+0 and zeros elsewhere as a linear combination of rows of $A_z$. On the other hand, if $a_1b_n$ is B) an out-cell, we have simultaneous slack in the +0 and 0+ columns that could only be reduced by some multiple $x$ of the $X$ row, which would require $-p+x = 0$ to reduce the +0 column and simultaneously $q+x = 0$ to reduce the 0+ column, which requires that $p+q = 0$ but this is incompatible with the initial requirement that $p+q=1$. Thus it is not possible to complete the task; so in this case it is necessary for the $a_1b_n$ row to be an in-row and thus \emph{the total number of out-rows is odd.}

Alternatively, the total number of out-cells in the set $\{a_1b_1,a_2b_1,...,a_nb_{n-1}\}$ could have been even, in which case $q$ is negative. Here, a similar line of analysis demonstrates that it is only possible to reduce all the columns of $a_1b_n$ to zero in a consistent manner if this cell is an out-cell, thus also reaching the conclusion that the total number of out-rows is odd.

So we have showed that if we start with an out-cell in the first cell, then we can generate the row with 1 in $a_1b_1$ and zeros elsewhere if and only if the total number of out-row cells is odd. It is possible to show with a similar line of argument, by trying to generate a row that has a 1 in $a_1b_1$++ and zeros elsewhere, that the same thing holds if we assume $a_1b_1$ is an in-row; thus, the statement of the lemma holds. $\hfill \Box$

\medskip

In the previous lemma, we studied distribution matrices consisting exclusively of in-rows and out-rows, which implies that the first cell $a_1b_1$ of $A_z$ will always have two pivot rows. In the proof of the lemma, we showed that if the number of in-rows and out-rows are both odd, we can create a third pivot row in the $a_1b_1$ cell as a linear combination of $A_z$ rows. Now note that if we have three pivots in a cell, we can always obtain the fourth pivot by subtracting the three pivots from the ``$X$'' row (the row with four 1s all in the same cell). Furthermore, if we can generate four pivots in the $a_1b_1$ cell, then it is straightforward to check that we can generate four pivots in the adjacent cell $a_1b_2$ under the assumption that it already has either the in-cell or out-cell pivots. (One starts by reducing the $a_1b_1$ components of the row that ``links'' the $a_1b_1$ and $a_2b_1$ cells.) This process can be repeated from cell to cell, and so all pivots can be obtained. Thus $A_z$ has rank $8n$, which means any in-row/out-row mix $z$ will be a extremal point of the polytope, so long as $A_zz=b$ is satisfied (i.e., $z$ is actually an element of the polytope).

\begin{alemma}\label{l:anotherlemma}
Let $z$ be a distribution matrix consisting of an odd number of in-rows and an odd number of out-rows. Then $z$ is a extremal point of the $n$-fold chained Bell polytope if and only if all nonzero entries are 1/2; that is, p=q=1/2 in each row.
\end{alemma}
\emph{Proof.} The probabilities must be 1/2 for the following reason. Considering the outcome table, start from the $a_1b_1$ row and if it is not of the form $(p,0,0,q)$, move down until encountering the first such row -- i.e., the first in-row, at least one of which must occur by the odd/odd requirement. Now, if we continue to move down the outcome table, we can either encounter 1) an even number (possibly zero) of consecutive out-rows followed by an in-row, which also must contain $(p,0,0,q)$ by no-signaling, or 2) an odd number of consecutive out-rows followed by an in-row, which leaves us with the flipped distribution $(q,0,0,p)$ by no-signaling. 

Note that option (1) leaves us in a situation where we cannot be back to the initial starting point in-row, because this option passes an even number of consecutive out-rows, and if we were back at the beginning, this would mean that the total number of out-rows is even. So if option (1) occurs, we are in an in-row of the form $(p,0,0,q)$ that is not the original row, and we have passed an even number of out-rows. If option (2) occurs, we \emph{can} be back at the beginning cell when we arrive at the first in-row, which would require that $(p,0,0,q) = (q,0,0,p)$ because of the distribution flip, and hence $p=q=1/2$. If option (2) occurs and we do not make it back to the beginning cell, we are in an in-row of the form $(q,0,0,p)$ having passed an odd number of out-rows.

As we continue to move down the outcome table (eventually looping back from $a_1b_n$ to $a_1b_1$), every time option (2) occurs, the distribution is flipped from $(q,0,0,p)$ to $(p,0,0,q)$ or vice versa, and every time that option (1) occurs, the distribution remains unchanged. Since the total number of times that (2) must occur before we return to the original row must be odd (by the oddness of the total number of in-rows), we must have $(p,0,0,q) = (q,0,0,p)$, so $p=q=1/2$.

This shows that $p=q=1/2$ is a necessary condition for $z$ to be an extremal point. To see that it is sufficient, note that such a $z$ is indeed an element of the polytope -- the no-signaling conditions and probability conditions are easily verified, and the comments immediately preceding Lemma \ref{l:anotherlemma} demonstrate that $A_z$ is full rank. $\hfill \Box$

Recalling Lemma \ref{l:extremal}, there are two candidates for extremal points of the polytope: in-row/out-row combinations, and distributions containing only 1s and 0s. We have shown that the extremal points of the polytope that are in-row/out-row combinations are precisely the generalized PR boxes in the statement of Theorem \ref{t:chainedBelltheorem}. Furthermore, Lemma \ref{l:mustbeLD} tells us that any extremal point whose elements are only 1s and 0s is necessarily a LD distribution. For any given LD distribution $z$, it is clear that $z$ is an element of the polytope (i.e., it satisfies the no-signaling conditions and probability conditions), and $A_z$ is rank $8n$ because each cell automatically contains three pivots and thus the fourth can always be obtained with linear combinations of $X$-type rows. Hence the extremal points consisting only of 1s and 0s are precisely the local deterministic distributions. This completes the proof of Theorem \ref{t:chainedBelltheorem}.

\end{document}